\documentclass[aps,prb,twocolumn,amsmath,amssymb,superscriptaddress,floatfix,scrartcl]{revtex4-1}

\usepackage{amssymb}
\usepackage{color}
\usepackage{graphicx}
\usepackage{srcltx}
\usepackage{mathrsfs}
\usepackage[unicode=true,pdfusetitle,
 bookmarks=false,colorlinks=true,citecolor=blue,urlcolor=blue,linkcolor=red]{hyperref}

\begin{document}

\title{
Bulk-boundary correspondence in (3+1)-dimensional topological phases
   }

\author{Xiao Chen}
\email{chenxiao.phy@gmail.com}
\affiliation{
Institute for Condensed Matter Theory and
Department of Physics, University of Illinois
at Urbana-Champaign, 
1110 West Green St, Urbana IL 61801
            }
\thanks{The first two authors contributed equally to the work.}

\author{Apoorv Tiwari} 
\email{t.apoorv@gmail.com}
\affiliation{
Institute for Condensed Matter Theory and
Department of Physics, University of Illinois
at Urbana-Champaign, 
1110 West Green St, Urbana IL 61801
            }

\author{Shinsei Ryu}
\email{ryuu@illinois.edu}
\affiliation{
Institute for Condensed Matter Theory and
Department of Physics, University of Illinois
at Urbana-Champaign, 
1110 West Green St, Urbana IL 61801
            }

\date{\today}

\begin{abstract}
We discuss (2+1)-dimensional gapless surface theories of bulk (3+1)-dimensional topological phases, such as 
the BF theory at level $\mathrm{K}$, and its generalization. 
In particular, we put these theories on a flat (2+1) dimensional torus $T^3$ parameterized by its modular parameters,
and compute the partition functions obeying various twisted boundary conditions. 
We show 
the partition functions are transformed into each other under $SL(3,\mathbb{Z})$ modular transformations, 
and furthermore establish the bulk-boundary correspondence in (3+1) dimensions
by matching  
the modular $\mathcal{S}$ and $\mathcal{T}$ matrices 
computed from the boundary field theories 
with those computed in the bulk. 
We also propose the three-loop braiding statistics can be studied by constructing
the modular $\mathcal{S}$ and $\mathcal{T}$ matrices from an appropriate boundary field theory. 
\end{abstract}

\pacs{72.10.-d,73.21.-b,73.50.Fq}

\maketitle

\tableofcontents

\section{Introduction}

{\it The bulk-boundary correspondence} is one of the most salient features of topologically ordered phases of matter.
In topologically ordered states in (2+1) dimensions [(2+1)d],
all essential topological properties in their bulk can be derived and understood from their edge theories,
such as quantized transport properties, 
properties of bulk quasiparticles (fractional charge and braiding statistics thereof),
and 
the topological entanglement entropy, etc.  
\cite{Halperin1982, Witten1989, Wen1992, Hatsugai1993, Cappelli01, Cappelli96, Cappelli10, Cappelli11}
Edge or surface theories also play an important role in symmetry-protected and symmetry-enriched topological phases. 
\cite{Ryu2012, Sule13, Cho2014,Hsieh2014a, Lu2012, Cappelli13, Hsieh2015}

The purpose of this paper is to study the 
bulk-boundary correspondence in the simplest (3+1)d topological field theory, 
the BF topological field theory
\cite{Horowitz1989,
Horowitz1990, 
Blau:1989bq,
Blau:1989dh,
Birmingham1991,
Oda1989,
BergeronSemenoffSzabo1995},
and its generalizations. 
The BF theory describes, among others, 
the long wave-length limit of BCS superconductors, and 
the deconfined phase of the $\mathbb{Z}_{\mathrm{K}}$ gauge theory. 
It is also relevant to the hydrodynamic description of (3+1)d 
symmetry-protected topological (SPT) phases
including topological insulators and related systems. 
\cite{HanssonOganesyanSondhi2004,
Banks2011,
ChoMoore2011,
Vishwanath2013,
Chan2013,
Tiwari2014,
Ye2015,
Gaiotto2015, Cirio2014}

To put our purpose in the proper context, let us 
give a brief overview of the bulk-boundary correspondence in (2+1)d topologically ordered phases. 
For (2+1)d topological phases, 
bulk topological phases can be characterized by the modular $\mathcal{S}$ and $\mathcal{T}$ matrices. 
The $\mathcal{S}$ and $\mathcal{T}$ transformations generate the basis transformation 
in the space of degenerate ground states, which appear when the system is put on a {\it spatial} two-dimensional torus.  
Combined together, the $\mathcal{S}$ and $\mathcal{T}$ transformations form the group $SL(2,\mathbb{Z})$, 
the mapping class group of the two-dimensional torus $T^2$. 
Their geometric meanings are the $90^{\circ}$ rotation and Dehn twist defined on the torus, respectively. 
In the basis in which the $\mathcal{T}$ matrix is diagonal (the so-called quasi-particle basis), 
the diagonal entries of the $\mathcal{T}$ matrix encode the information on the topological spin of quasi-particles. 
On the other hand, the $\mathcal{S}$ matrix contains the information of the braiding and fusion. 
For an Abelian topological phase, the elements of the $\mathcal{S}$ matrix are given by braiding phases between 
quasiparticles, 
up to an over all normalization factor $1/\mathcal{D}$, 
where $\mathcal{D}$ is the total quantum dimension. 

On the other hand, 
at their boundary (edge), 
gapless boundary excitations supported by
a (2+1)d topological phase 
can be described by  
a (1+1)d conformal field theory (CFT).\cite{FMS-CFT} 
There is one-to-one correspondence between quasi-particle excitations in the bulk and primary fields living on the edge. 
On the (1+1)d {\it spacetime} torus, 
one can form the character $\chi_j(\tau)$ from 
the tower of states built upon a primary field $O_j$:
\begin{align}
\chi_j(\tau)=\mbox{Tr}_{\mathcal{H}_j}\left[e^{2\pi i\tau_1P_0-2\pi\tau_2H_0}\right]
\end{align}
where $H_0$ and $P_0$ are the Hamiltonian and the momentum operators, respectively,
the complex parameter $\tau=\tau_1 + i\tau_2$ is the modular parameter parameterizing the spacetime torus, 
and the trace is taken over all states in the Hilbert space $\mathcal{H}_j$   
that is built upon the highest weight state associated with the primary field $O_j$. 
The characters $\chi_j$ transform into each other under the modular transformations of the spacetime torus.
Under the modular $\mathcal{T}$ and $\mathcal{S}$ transformations, the characters $\chi_j(\tau)$ transform as
\begin{align}
\chi_j(\tau+1)&=e^{2\pi i(h_j-\frac{c}{24})}\chi_j(\tau)=\mathcal{T}_{jj}\chi_j(\tau),
\nonumber\\
\chi_j(-{1}/{\tau})&=\sum_{j^{\prime}}\mathcal{S}_{jj^{\prime}}\chi_{j^{\prime}}(\tau), 
\end{align}
where the matrices $\mathcal{T}$ and $\mathcal{S}$ represent the action of 
the $\mathcal{T}$ and $\mathcal{S}$ modular transformations on
the characters, respectively. 
The matrix $\mathcal{T}$ is a diagonal matrix and 
includes the conformal dimension $h$ for each character and the central charge $c$ for the CFT. 
The $\mathcal{T}$ and $\mathcal{S}$ matrices for the characters in the edge theory 
have the direct correspondence (and are essentially identical) to 
the the $\mathcal{T}$ and $\mathcal{S}$ matrices defined for the corresponding $(2+1)$d bulk topological theory.

Coming back to our main focus, i.e., $(3+1)$d topologically ordered phases, 
the bulk topological system can be defined on a spatial torus $T^3$ (while other choices are of course possible). 
The mapping class group 
of the three-dimensional torus is $SL(3,\mathbb{Z})$,
and, as in the case of (2+1)d,
is also generated by two transformations, 
which we also call the modular $\mathcal{S}$ and $\mathcal{T}$ transformations 
(see Sec.\ \ref{Modular transformations on $T^3$} for details).
For $(3+1)$d topological phases defined on a spatial torus $T^3$, 
$\mathcal{S}$ and $\mathcal{T}$ matrices can be introduced to 
describe the basis transformation of degenerate ground states.
As in (2+1)d, 
the $\mathcal{S}$ and $\mathcal{T}$ matrices encode the topological data of the bulk topological phase,
such as the braiding and spin statistics of excitations.
\cite{MoradiWen2015, JiangMesarosRan2014} 
In (3+1)d, the exchange statistics of particles has to be either fermionic or bosonic. 
On the other hand, a particle and a loop-like excitation, 
or two loop-like excitations in the presence of an additional background loop,
can have non-trivial braiding and can obey non-trivial statistics.  
For the Abelian topological phase described by the BF theory, 
the $\mathcal{S}$ matrix describes the braiding phase 
between 
particle and 
loop excitations,
while the $\mathcal{T}$ matrix has the physical meaning of a $(3+1)$d analogue of topological spins.
\cite{MoradiWen2015} 
It has been also proposed that
there exist (3+1)d topological phases that are characterized by their  
three-loop braiding statistics. 
\cite{WangLevin2014, JiangMesarosRan2014,WangWen2015,JianQi2014,WangLevin2015, LinLevin2015, Wan2015}

We will demonstrate that these results, obtained and discussed previously from the bulk point of view, 
can be obtained solely from gapless boundary field theories.
More specifically,
taking various examples of (3+1)d topologically ordered phases and their surface states, 
which we put on the $(2+1)$d spacetime torus $T^3$,
we compute the modular $\mathcal{S}$ and $\mathcal{T}$ matrices explicitly, and show 
that they agree with the $\mathcal{S}$ and $\mathcal{T}$ matrices obtained from the bulk considerations. 
We thereby establish 
the bulk-boundary correspondence
in these (3+1)d topologically ordered phases. 
Along the course, we also propose a bulk continuum field theory which realizes non-trivial 
three-loop braiding statistics.

N.B.
Our strategy adopted in this paper 
is to utilize boundary field theories to learn about bulk excitations
in (3+1)d topological phases,
by establishing a bulk-boundary correspondence. 
One should however bear in mind that boundaries may have more ``life'' than their corresponding bulk,
in that a given bulk topological phase can be consistently terminated by more than one boundary theory.
Therefore, it would be more appropriate to consider a ``stable equivalent class'' of boundary theories for a given bulk theory. 
(See, for example, Ref.\ \onlinecite{Cano2014}.)
Nevertheless, one can expect universal topological properties of the bulk theories may be extracted from 
any boundary theory which consistently terminates the bulk.

\subsection{Outline of the paper} 


The rest of the paper is organized as follows.

In Sec.\ \ref{Compact free boson in $2+1$ dimensions}, 
we consider the compactified free boson theory in $(2+1)$d 
defined on the 3d flat torus $T^3$ is computed. 
This (2+1)d theory is not necessarily tied to a particular (3+1)d bulk topological order 
but serves as a warm up for later sections.
We will show its partition function is invariant under $SL(3,\mathbb{Z})$.

In Sec.\ \ref{The surface theory of the BF theory}, 
the surface theory of the (3+1)d BF theory at level $\mathrm{K}$
is studied. 
This theory can be subjected to twisted boundary conditions, 
which are induced by introducing quasi-particles in the (3+1)d bulk. 
We will show that the partition functions with different boundary conditions 
are transformed into each other under $SL(3,\mathbb{Z})$, and form a representation of $SL(3,\mathbb{Z})$. 
The extracted $\mathcal{S}$ and $\mathcal{T}$ matrices agree with the known result. 
\cite{MoradiWen2015} 
We will also compute the thermal entropy in Sec.\ \ref{Entanglement Entropy for BF boundary}, 
and show that there is a constant negative contribution to the entropy.  
This contribution to the boundary thermal entropy 
is expected to capture the topological entanglement entropy defined 
in the corresponding (3+1)d bulk. 


In Sec.\ \ref{The surface theory of the (3+1)d BF theory with the Theta term}, 
we introduce an additional term,
the axion term or the theta term, 
to the (3+1)d BF theory.  
The theta angle has a texture (spatial inhomogeneity)
and affects the boundary theory by twisting the quantum numbers.
Being static, the texture in the theta angle is interpreted as a topological defect,
and we will show that the introduction of the defect makes 
the surface theory non-modular invariant, in the sense that 
the action of modular transformations is not closed
within the space of the partition functions. 

This BF theory with the theta term motivates us to consider yet another theory
in Sec.\ \ref{Three-loop braiding statistics}, 
which can be constructed by coupling two copies of the BF theory.
Compared with the 
the defect system (the BF theory with the theta term) discussed in
Sec.\ \ref{The surface theory of the (3+1)d BF theory with the Theta term},
in the coupled system, each copy can be interpreted as 
playing a role of a defect to the other.
In this system, however, there is no externally imposed texture.  
We propose this continuum bulk field theory realizes three-loop braiding statistics
discussed previously.
\cite{WangLevin2015, WangLevin2014, LinLevin2015, JianQi2014}
On the surface, 
we consider two copies of the BF surface theories,
which are coupled together in their zero mode sectors.
We will show that, by computing the modular $\mathcal{S}$ and $\mathcal{T}$ matrices explicitly,
this system exhibits three-loop braiding statistics. 

Finally, we conclude in Sec.\ \ref{Discussion}.

\section{The compactified free boson in $(2+1)$d}
\label{Compact free boson in $2+1$ dimensions}

The compactified real scalar theory in $(2+1)$d is described by the Lagrangian density  
\begin{align}
\mathcal{L}=\frac{1}{(2\pi)^2}\left[(\partial_t\phi)^2-(\partial_x\phi)^2-(\partial_y\phi)^2\right], 
\label{lag_bos}
\end{align}
where, for now, the spacetime is the ``canonical'' flat torus $T^3$ parameterized by $(t,x,y)$. 
(We will consider, momentarily in Sec.\ \ref{partition function on $T^3$}, 
a generic torus parameterized by six modular parameters.)
The boson field obeys the compactification condition on a circle of radius $\mathrm{r}$, i.e.,
\begin{align}
\phi\equiv\phi+2\pi \mathrm{r}. 
\label{compactification cond}
\end{align}
This model can be exactly solved and is dual to the compact $U(1)$ gauge theory. 
Under the duality, the boson field $\phi$ is related to the $U(1)$ gauge field $a_{\mu}$ by  
\begin{align}
\epsilon^{\mu\nu\lambda} \partial_{\nu} a_{\lambda}\leftrightarrow  \partial^{\mu}\phi, \quad
f^{\mu\nu}f_{\mu\nu}\leftrightarrow  \partial^{\mu}\phi \partial_{\mu}\phi. 
\end{align}
Furthermore, 
quantized vortices on the boson side are dual to 
quantized charges in the $U(1)$ gauge theory. 
For the compact $U(1)$ gauge field theory, the monopole (instanton) proliferation leads to a confining phase 
and this process on the scalar boson side  corresponds to adding a $\cos(\phi)$ term.
\cite{Polyakov1975, Polyakov1976}
This process breaks the $U(1)$ symmetry in the compact boson theory,
and the particle number is not conserved anymore. 
If we prohibit the monopoles, on the other hand, the Abelian $U(1)$ gauge theory is stably gapless.

The free boson theory can be canonically quantized:
The corresponding Hamiltonian is 
\begin{align}
H=\frac{1}{(2\pi)^2}
\int\limits_0^{2\pi R_1} dx\int\limits_0^{2\pi R_2} dy
\left[ 
\Pi^{2}+(\partial_x\phi)^2
+(\partial_y\phi)^2
\right],
\label{2d_bos}
\end{align}
where 
$x$ and $y$ are periodic with radius $2\pi R_1$ and $2\pi R_2$, respectively, 
and the canonical momentum is ($\mathsf{r}:=(x,y)$)
\begin{align}
\Pi(\mathsf{r}):=\partial_t \phi(\mathsf{r}).
\end{align}
The canonical commutation relation is given by
\begin{align}
\left[\phi(t,{\mathsf{r}}),\Pi(t,{\mathsf{r'}})\right]&=
\frac{i}{2}
(2\pi)^2 
\delta^{(2)}({\mathsf{r}}-{\mathsf{r'}})
\nonumber \\
&=
\frac{i}{2} 
\frac{1}{R_1 R_2}
\sum_{s_{1,2}\in\mathbb{Z}} 
e^{i \mathsf{k} \cdot (\mathsf{r} -\mathsf{r}')}, 
\label{comm}
\end{align}
where 
$\delta^{(2)}({\mathsf{r}}-{\mathsf{r'}})$
is the periodic delta function 
and 
$\mathsf{k} = (s_1/R_1, s_2/R_1)$
is the 2d momentum $(s_i\in \mathbb{Z})$.  

To specify the Hilbert space, we develop 
the mode expansion of the bosonic field $\phi$. 
Due to the compactification condition (\ref{compactification cond}),
the bosonic field has the following expansion:
\begin{align}
\phi(t,\mathsf{r})=\frac{\mathrm{r}N_1}{R_1}x+ \frac{\mathrm{r}N_2}{R_2}y
+\frac{\phi_0 + \pi_0 t}{2\pi\sqrt{R_1R_2}}
+{\phi}_{osc}(t,\mathsf{r}),
\end{align} 
where $N_{1,2}\in \mathbb Z$ characterize the winding zero modes in the $x$ and $y$ direction, respectively. 
The Fourier decomposition of the oscillator part ${\phi}_{osc}(t,\mathsf{r})$ is given by
\begin{align}
{\phi}_{osc}(t,{\mathsf{r}})&=
\frac{1}{\sqrt{R_1R_2}}
\sum_{{\mathsf{k}}\neq 0}
\frac{1}{2\sqrt{\omega({\mathsf{k}})}}
\nonumber \\ 
&\qquad 
\times 
\left[ 
{a}({\mathsf{k}})e^{-i\omega t-i{\mathsf{k}}\cdot{\mathsf{r}}} 
+{a}^{\dagger}({\mathsf{k}})e^{i\omega t+i{\mathsf{k}}\cdot{\mathsf{r}}}\right]. 
\end{align} 
According to Eq.\ \eqref{comm}, $a(\mathsf{k})$ satisfies
the canonical commutation relation
\begin{align}
\left[
{a}(\mathsf{k}),
{a}^{\dag}(\mathsf{k}' )
\right]=
\delta_{\mathsf{k}, \mathsf{k}^{\prime}},
\end{align}
where $\omega({\mathsf{k}})$ is the dispersion of the free boson on a Euclidean three-torus
and given by
\begin{align}
\omega(\mathsf{k})=\sqrt{\left(\frac{s_1}{R_1}\right)^2+\left(\frac{s_2}{R_2}\right)^2}. 
\end{align}
On the other hand,
the zero mode part satisfies
\begin{align}
\left[\phi_0,\pi_0\right]=2\pi^2 i. 
\end{align}
Owing to the $2\pi \mathrm{r}$ periodicity of $\phi(t,\mathsf{r})$, 
the eigenvalues of $\pi_0$ needs to be quantized according to
\begin{align}
\pi_0=\frac{\pi N_0}{\mathrm{r}\sqrt{R_1R_2}},
\quad
N_0 \in \mathbb{Z}.
\end{align}
To summarize, the boson field $\phi(t,\mathsf{r})$ can be 
mode-expanded as 
\begin{align}
\phi(t,\mathsf{r})&= 
\frac{\phi_0}{2\pi\sqrt{R_1R_2}}+\frac{N_0t}{2\mathrm{r}R_1R_2}
+\frac{\mathrm{r}N_1}{R_1}x+ \frac{\mathrm{r}N_2}{R_2}y 
\nonumber \\
&\quad 
+
\frac{1}{\sqrt{R_1R_2}}
\sum_{{\mathsf{k}}\neq 0}
\frac{1}{2\sqrt{\omega({\mathsf{k}})}}
\nonumber \\ 
&\qquad 
\times 
\left[ 
{a}({\mathsf{k}})e^{-i\omega t-i{\mathsf{k}}\cdot{\mathsf{r}}} 
+{a}^{\dagger}({\mathsf{k}})e^{i\omega t+i{\mathsf{k}}\cdot{\mathsf{r}}}\right]. 
\label{mode_bos}
\end{align} 
The Hilbert space $\mathcal{H}_0$ consists of, 
for each winding sector specified by $N_{1,2}$, 
the zero mode part and the bosonic Fock space for each $\mathsf{k}\neq 0$.
States in the zero mode part are labeled by the eigenvalues of $\pi_0$, and hence by $N_0$. 
Furthermore, different winding sectors are summed over.
In the following, the part of the partition function associated to the summation over $N_{0,1,2}$ 
is called the zero mode sector.

\subsection{Modular transformations on $T^3$}
\label{Modular transformations on $T^3$}

We now consider the theory put on a generic flat torus. 
A flat three-torus is parameterized by six real parameters, 
$R_{0,1,2}$ and $\alpha,\beta,\gamma$. 
For a flat three-torus $T^3$, the dreibein is given by \cite{Hsieh2015}
\begin{align}
 {e^A}_{\mu} 
 &=
 \left(
 \begin{array}{ccc}
 R_0 & 0 & 0 \\
 0 & R_1 & 0 \\
 0 & 0 & R_2 
 \end{array}
 \right)
 \left(
  \begin{array}{ccc}
 1 & 0 & 0 \\
 -\alpha & 1 & 0 \\
 -\gamma & -\beta & 1 
 \end{array}
 \right)
\end{align}
where $R_0$, $R_1$, and $R_2$ are the radii for the directions $\tau$, $x$, and $y$, and $\alpha$, $\beta$, and $\gamma$ 
describe the angles between directions $\tau$ and $x$,  $x$ and $y$, and $\tau$ and $y$, respectively. 
The Euclidean metric is then given by

\begin{align}
 g_{\mu\nu} 
 &=
 {e^A}_{\mu}{e^B}_{\nu}\delta_{AB}
 \nonumber\\
 &=
  \left(
\begin{array}{ccc}
 R_0^2+\alpha^2R_1^2+\gamma^2R_2^2 & -\alpha R_1^2+\beta\gamma R_2^2 & -\gamma R_2^2 \\
 -\alpha R_1^2+\beta\gamma R_2^2 & R_1^2+\beta^2R_2^2 & -\beta R_2^2 \\ 
 -\gamma R_2^2 & -\beta R_2^2 & R_2^2 
 \end{array}
 \right),
\end{align}

The group ${SL}(3,\mathbb{Z})$ is generated by two transformations:
\begin{align}
 U_1 =
 \left(
 \begin{array}{ccc}
 0 & 0 & 1 \\
 1 & 0 & 0 \\
 0 & 1 & 0 
 \end{array}
\right),
\quad
 U_2 =
 \left(
 \begin{array}{ccc}
 1 & 1 & 0 \\
 0 & 1 & 0 \\
 0 & 0 & 1 
 \end{array}
\right).
\label{generator}
\end{align}

Under the $U_2$ transformation,
the metric is transformed as 
\begin{align}
& g_{\mu\nu}  
 \overset{U_2}{\longrightarrow} 
 \  {( U_2 g U^T_2)}_{\mu\nu} 
\end{align}
which corresponds to the changes
\begin{align}
\alpha \rightarrow \alpha-1, \quad
\gamma \rightarrow \gamma+\beta,
\end{align}
while $R_0$, $R_1$, $R_2$, and $\beta$ are unchanged.

On the other hand, $U_1$ can be decomposed as
\begin{align}
\label{U1' and M}
 U_1 
 &= 
 U'_1M,
 \quad 
 U'_1
=
 \left(
 \begin{array}{ccc}
 0 & -1 & 0 \\
 1 & 0 & 0 \\
 0 & 0 & 1 
 \end{array}
\right)
\quad
 M =
 \left(
 \begin{array}{ccc}
 1 & 0 & 0 \\
 0 & 0 & -1 \\
 0 & 1 & 0 
 \end{array}
\right)
\end{align}
where $U_1^{\prime}$ corresponds to the $90^{\circ}$ rotation in the $\tau-x$ plane and $M$ is the $90^{\circ}$ rotation in the $x-y$ plane. The generator $U^{\prime}_1$ acts on the metric as
\begin{align}
& g_{\mu\nu}  
 \overset{U'_1}{\longrightarrow} 
 \  {( U'_1 g U'^T_1)}_{\mu\nu} 
\end{align}
which corresponds to the changes
\begin{align}
&R_0 \rightarrow R_0/|{\tau}|, \quad
R_1 \rightarrow R_1|{\tau}|, \quad
\tau_1 \rightarrow -\tau_1/|{\tau}|^2, 
\nonumber \\
&
\gamma \rightarrow -\beta, \quad
\beta \rightarrow \gamma \quad
\text{(while $R_2$ is unchanged)},
\end{align}
where we have introduced 
\begin{align}
{\tau}\equiv\alpha+ir_{01},
\quad
r_{01}\equiv R_0/R_1.
\end{align} 
Observe also that under $R_0\to R_0/|\tau|$ and $R_1\to R_1|\tau|$, $\tau_2\to \tau_2/|\tau|^2$.
Hence, $U'_1$ induces $\tau \to -1/\tau$.

The two transformations $U'_1$ and $U_2$ correspond respectively to 
modular $S$ and $T^{-1}$ transformations in the ${\tau}-x$ plane, generating the ${SL}(2,\mathbb{Z})$ subgroup of ${SL}(3,\mathbb{Z})$ group. Combining with $M$, they generate the whole ${SL}(3,\mathbb{Z})$ group.
In the following, we call $U^{\prime}_1 M$ as $\mathcal{S}$ transformation 
and $U_2$ as $\mathcal{T}^{-1}$ transformation.

Moreover, the two generators of $SL(3,\mathbb{Z})$, the $U_1$ and $U_2$ transformations defined in Eq.\ \eqref{generator}, 
satisfy
\cite{WangWen2015, Generators}
\begin{align}
&U_1U_1^{\dag}=U_1^3=R^6=(U_1R)^4=(RU_1)^4=1,
\\
&(U_1R^2)^4=(R^2U_1)^4=(U_1R^3)^3=(R^3U_1)^3=1, 
\\
&(U_1R^2U_1)^2R^2=R^2(U_1R^2U_1)^2\ \mod 3,
\label{constraint_3}
\end{align}
where 
\begin{align}
R=(U_2U_1)^2U_2^{-1}U_1^2U_2^{-1}U_1U_2U_1.
\end{align}

\subsection{The partition function on $T^3$}
\label{partition function on $T^3$}

In this section, we calculate the partition function of the compactified free boson theory 
on the three-torus in the presence of the generic flat metric. 
The Euclidean action is given by
\begin{widetext}
\begin{align}
S&= 
\frac{1}{(2\pi)^2}\int\limits_{0}^{2\pi}{{d}^{3}\theta\sqrt{|g|}g^{\mu\nu}\partial_{\mu}\phi\partial_{\nu}\phi} 
\nonumber \\
&=
\frac{1}{(2\pi)^2}
\int\limits_{0}^{2\pi R_{0}}{d}\tau
\int\limits_{0}^{2\pi R_{1}}{d}x
\int\limits_{0}^{2\pi R_{2}}{d}y
\left[\left(\partial_{\tau}\phi\right)^{2}
+\left(\frac{\alpha^{2}R^{2}_{1}}{R_{0}^{2}}+1\right)\left(\partial_{x}\phi\right)^{2}
+\left(\frac{R_2^2}{R_0^2}(\alpha\beta+\gamma)^{2}+\frac{R_2^2}{R_1^2}\beta^2+1\right)\left(\partial_{y}\phi\right)^{2}
\right. 
\nonumber \\
&
\qquad
\left.
+
2\alpha\frac{R_1}{R_0}\left(\partial_{\tau}\phi\right)\left(\partial_{x}\phi\right)+
2\frac{R_2}{R_0}(\alpha\beta+\gamma)\left(\partial_{\tau}\phi\right)\left(\partial_{y}\phi\right)+
2\left(\frac{R_1R_2}{R_0^2}\alpha(\alpha\beta+\gamma)+
{\frac{R_2}{R_1}}\beta\right)\left(\partial_{x}\phi\right)\left(\partial_{y}\phi\right)
\right], 
\label{free boson action}
\end{align}
\end{widetext}
where $0\leq \theta^{\mu}\leq 2\pi$ are angular variables and we noted 
$\sqrt{|g|}=R_{0}R_{1}R_{2}$, 
$\tau=R_{0}\theta^{0}$, 
$x=R_{1}\theta^{1}$ and 
$y=R_{2}\theta^{2}$. 
\footnote{
The Euclidean time coordinate $\tau$ should not be confused with the modular parameter. 
}
The inverse metric $g^{\mu\nu}=(e^{\star \mu}_A)^T\delta^{AB}(e_B^{\star \nu})$ ($e^{\star}_A$ is the inverse of $e^A$) is given by
\begin{align}
g^{\mu\nu}=\begin{pmatrix}\frac{1}{R_0^2}&\frac{\alpha}{R_0^2}&\frac{\alpha\beta+\gamma}{R_0^2}\\
\frac{\alpha}{R_0^2}&\frac{\alpha^2}{R_0^2}+\frac{1}{R_1^2}&\frac{\alpha(\alpha\beta+\gamma)}{R_0^2}+\frac{\beta}{R_1^2}\\
\frac{\alpha\beta+\gamma}{R_0^2}&\frac{\alpha(\alpha\beta+\gamma)}{R_0^2}+\frac{\beta}{R_1^2}&\frac{(\alpha\beta+\gamma)^2}{R_0^2}+\frac{\beta^2}{R_1^2}+\frac{1}{R_2^2}
\end{pmatrix}. 
\end{align}

In the operator formalism, 
the partition function corresponding to the action 
\eqref{free boson action}
is given by the trace of the thermal density matrix  
$\exp(-2\pi R_0 H^{\prime})$
over the Hilbert space
 $\mathcal{H}_0$:  
\begin{align}
\mathcal{Z}=\mbox{Tr}_{\mathcal{H}_{0}}\left[e^{-2\pi R_0 H^{\prime}}\right]. 
\end{align} 
The ``boosted'' Hamiltonian $H'$ and the (untwisted) Hilbert space $\mathcal{H}_0$ 
are specified as follows. 
The boosted 
Hamiltonian $H^{\prime}$ consists of 
the ``unboosted'' Hamiltonian $H$
(with $\alpha=\gamma=0$),
and the momentum $P_{x,y}$,
which induces the boost in $x$ and $y$ directions, 
respectively:
\begin{align}
H' &= H + i \frac{\alpha R_1}{R_0}P_x + i(\alpha\beta +\gamma) \frac{R_2}{R_0} P_y 
\end{align}
where
\begin{align}
H&=
\int \frac{dx dy}{(2\pi)^2}
\left[ 
\Pi^{2}
+
G^{ij} \partial_i \phi \partial_j \phi
\right],
\nonumber \\
P_{i} &=
\int \frac{dx dy}{(2\pi)^2}\, 
2 \Pi\partial_i\phi,
\quad
i=x,y. 
\end{align}
and
$G^{ij}$ is defined as
\begin{align}
g^{ij}&=
\begin{pmatrix}
g_{11}&g_{12}\\g_{21}&g_{22}\end{pmatrix}
=
\begin{pmatrix}\frac{1}{R_1^2}&\frac{\beta}{R_1^2}\\\frac{\beta}{R_1^2}&\frac{\beta^2}{R_1^2}+\frac{1}{R_2^2}
\end{pmatrix},
\nonumber \\
G^{ij} & = R_i R_j g^{ij}
=
\begin{pmatrix}
1 & \beta \frac{R_2}{R_1} \\
\beta \frac{R_2}{R_1} &
\frac{\beta^2 R^2_2}{R^2_1} + 1 
\end{pmatrix}
\end{align}
(where $i,j$ are not summed in $R_i R_j g^{ij}$). 
I.e., 
\begin{align}
H^{\prime}&=
\int \frac{dx dy}{(2\pi)^2}
\left[ 
\Pi^{2}+(\partial_x\phi+\beta\frac{R_2}{R_1}\partial_y\phi)^2
+(\partial_y\phi)^2
\nonumber\right. \\
&\qquad \left.+2i\frac{\alpha R_1}{R_0}\Pi\partial_x\phi
+2i(\alpha\beta+\gamma)\frac{R_2}{R_0}\Pi\partial_y\phi
\right],
\end{align}

The mode expansion for the bosonic field $\phi$ is still given by Eq.\ \eqref{mode_bos}, 
where the energy spectrum $\omega({\mathsf{k}})$ is now given by
\begin{align}
\omega(\mathsf{k})
&=\sqrt{G^{ij}k_i k_j}
\nonumber \\
&
=
\sqrt{g^{ij}s_is_j}=\sqrt{\left(\frac{s_1}{R_1}+\beta\frac{s_2}{R_1}\right)^2+\left(\frac{s_2}{R_2}\right)^2}.
\end{align}
The Hilbert space $\mathcal{H}_0$ is given as a direct product of the bosonic Fock spaces each built out of 
a given zero mode state specified by $N_{0,1,2}$. 

Next, we proceed to compute the partition function and study its properties under modular transformations of the three-torus. 
The partition function can be split into the zero mode part, which we call $Z_0$,  and the oscillator part, which we call $Z_{osc}$. 
The total partition function is 
$\mathcal{Z}=Z_{0}Z_{osc}$. 

\begin{widetext}
The partition function of the zero mode part is
\begin{align}
 Z_0
&=\sum_{N_{0,1,2} \in \mathbb{Z}}
\exp\Big[-
\frac{\pi \tau_2}{2 \mathrm{r}^2 R_2} N^2_0
-2\pi \mathrm{r}^2R_2\tau_2(N_1+\beta N_2)^2
-\frac{2\pi \mathrm{r}^2R_0R_1}{R_2}N_2^2
\nonumber\\
&
\qquad 
+
2\pi i \tau_1 N_0(N_1+\beta N_2)+2\pi i\gamma N_0N_2
\Big]
\label{zero mode part fun free boson}
\end{align}
where we recall $\tau_2=R_0/R_1$, $\tau_1=\alpha$ and $\tau=\tau_1+i\tau_2$.

On the other hand, for the oscillator part, the Hamiltonian is
\begin{align}
H^{\prime}_{osc}&
=\sum_{{\mathsf{k}} \neq 0} 
\left[
\omega(\mathsf{k})
+i\alpha\frac{R_1}{R_0} k_1
+i(\alpha\beta+\gamma)
\frac{R_2}{R_0} k_2
\right] 
{a}^{\dag}({\mathsf{k}})
{a}({\mathsf{k}})+E_0,
\end{align}
where $E_0$ is the ground state energy and needs to be properly regularized: 
\begin{align}
E_0=\sum_{\mathsf{s}\in\mathbb{Z}^2/(0,0)}\frac{1}{2}\sqrt{g^{ij}s_is_j}
=-\frac{\sqrt{\det(g)}}{2}
\sum_{\mathsf{s}\in\mathbb{Z}^2/(0,0)}\frac{1}{|g^{ij}s_is_j|^2}. 
\end{align}
The  partition function of the oscillator part can be decomposed into 
the product of the partition functions of one-dimensional non-compact bosons with ``mass'' given by $s_2$. 
When the ``mass'' $s_2=0$,
\begin{align}
Z_{s_2=0}&=e^{-2\pi R_0E_0(s_2=0)}\prod_{s_1\neq 0\in\mathbb{Z}}\left[1-e^{-2\pi R_0(\omega({\mathsf{k}})+i\alpha\frac{s_1}{R_0})}\right]^{-1}
=\left|\frac{1}{\eta(\tau)}\right|^2
\end{align}
where $\eta(\tau)$ is the Dedekind eta function and $\tau$ is the 2-dimensional modular parameter:
\begin{align}
\eta(\tau):=e^{\frac{\pi i\tau}{12}}\prod_{n=1}^{\infty}(1-q^n),
\quad
q:=e^{2\pi i\tau}.
\end{align}
On the other hand, the other massive part equals to
\begin{align}
\nonumber 
Z_{s_2\neq 0}&=
e^{-2\pi R_0E_0(s_2\neq 0)}
\prod_{s_2\neq 0,s_1\in\mathbb{Z}}\left[1-e^{-2\pi R_0(\omega({\mathsf{k}})+i\alpha\frac{s_1}{R_0}+i(\alpha\beta+\gamma)\frac{s_2}{R_0})}\right]^{-1}\nonumber\\
&=\prod_{s_2\in\mathbb{Z}^+}\Theta^{-1}_{[\beta s_2,\gamma s_2]}\left(\tau,\frac{R_1}{R_2}s_2\right),
\end{align}
where the massive theta function $\Theta_{[\beta s_2,\gamma s_2]}(\tau,\frac{R_1}{R_2}s_2)$ is defined as
\begin{align}
\Theta_{[a,b]}(\tau,m)&\equiv e^{4\pi \tau_2\Delta(m,a)}
 \prod_{n\in\mathbb{Z}}\left|1-e^{-2\pi \tau_{2}\sqrt{m^2+(n+a)^2}+2\pi i\tau_{1}(n+a)+2\pi i b}\right|^2
\end{align}
where 
\begin{align}
\Delta(m,a)\equiv\frac{1}{2}\sum_{n\in\mathbb{Z}}\sqrt{m^2+(n+a)^2}-\frac{1}{2}\int_{-\infty}^{\infty}dk(m^2+k^2)^{1/2}
\end{align}
Thus the partition function for the oscillator part equals to
\begin{align}
Z_{osc}=Z_{s_2=0}Z_{s_2\neq 0}=\left|\frac{1}{\eta(\tau)}\right|^2
\prod_{s_2\in\mathbb{Z}^+}\Theta_{[\beta s_2,\gamma s_2]}^{-1}
\left(\tau,\frac{R_1}{R_2}s_2\right). 
\end{align}
Together with \eqref{zero mode part fun free boson},
we have completed the calculation of the total partition function, $\mathcal{Z}=Z_{0}Z_{osc}$. 
\end{widetext}

It is instructive to compare the above partition function 
with the partition function of the (1+1)d compactified free boson.   
Performing dimensional reduction, 
by taking $R_2=1, N_2=0$ and $s_2=0$, the partition function reduces 
to 
\begin{align}
 \mathcal{Z}&=\frac{1}{|\eta(\tau)|^2}
 \sum_{N_{0,1}\in \mathbb{Z}}
 \exp\Big(
 -\frac{\pi\tau_2}{2\mathrm{r}^2} N_0^2
 -2\pi \mathrm{r}^2\tau_2N_1^2
\nonumber\\
&
\qquad
\qquad 
+2\pi i\tau_1N_0N_1
\Big).
\end{align}
This is the partition function for the compactified free boson in $(1+1)$d.

\subsubsection{Modular invariance}
We now show that the total partition function is invariant under the ${SL}(3, \mathbb{Z})$ transformations.

For $Z_0$, under $U_1^{\prime}$ transformation, by using the Poisson resummation formula twice, we have 
\begin{align}
Z_0(\tau)\overset{U_1^{\prime}}{\longrightarrow}  Z_0(-1/\tau)=|\tau|Z_0(\tau)
\end{align}
where the Poisson resummation formula is
\begin{align}
\sum_{n\in \mathbb{Z}} e^{-\pi an^2+bn}=
\frac{1}{\sqrt{a}}\sum_{k\in \mathbb{Z}}
e^{
-\frac{\pi}{a}\left(k+\frac{b}{2\pi i}\right)^2
}.
\end{align}
For $Z_{osc}$ part, under $U_1^{\prime}$ transformation, the massless $s_2=0$ component will contribute  a $1/|\tau|$ prefactor. 
The massive part is invariant under $U_1^{\prime}$ transformation, since the massive theta function satisfies
\begin{align}
\Theta_{[a,b]}\left(\tau,\frac{R_1}{R_2}s_2\right)
&\overset{U_1^{\prime}}{\longrightarrow}  
\Theta_{[a,b]}\left(-\frac{1}{\tau},\frac{R_1}{R_2}s_2|\tau|\right)
\nonumber\\
&=
\Theta_{[b,-a]}\left(\tau,\frac{R_1}{R_2}s_2\right)
\end{align}
Thus the total partition function is invariant under $U_1^{\prime}$ transformation.

Under the $M$ transformation which is basically a $\pi/2$ rotation in the $x-y$ plane, the partition function for the zero mode part becomes
\begin{align}
Z_0&\overset{M}{\longrightarrow} Z_0=
\sum_{N_{0,1,2}\in \mathbb{Z}}
\exp\left(
-\frac{\pi R_0}{2 \mathrm{r}^2 R_1 R_2}
N^2_0
\right.\nonumber\\
&\qquad\qquad\left.
-2\pi \mathrm{r}^2R_0R_1R_2
\left[\left(
\beta\frac{N_1}{R_1}-\frac{N_2}{R_1}\right)^2+\left(\frac{N_1}{R_2}\right)^2 \right]\right.
\nonumber \\
&
\qquad 
\qquad 
-2\pi i\alpha R_0 [\beta N_1-N_2]N_0-2\pi i\gamma N_1 N_0\Big).
\end{align}
Therefore the invariance of the zero mode part of the partition function 
can be seen from relabeling,
\begin{align}
M: \begin{cases} N_1\\N_2\\N_0\end{cases}\to\begin{cases} -N_2\\N_1\\N_0\end{cases}
\end{align}
It is also straightforward to show that the oscillator part is invariant under $M$ transformation 
and thus the total partition function is invariant under $M$ transformation.

Finally, it is also easy to check that the partition function is invariant under $U_2$ transformation. 
Hence the partition function is invariant under the ${SL}(3,\mathbb{Z})$ transformation.

\section{The surface theory of the (3+1)d BF theory}
\label{The surface theory of the BF theory}

\subsection{Bulk and surface theories}

\paragraph{The bulk field theory}
The (3+1)-dimensional one component BF theory is described by 
the action
\begin{align}
S_{bulk}
&=
\int_{\mathcal M}
\left[
\frac{\mathrm{K}}{2\pi}
b\wedge da
-a\wedge J_{qp}
-b\wedge J_{qv}
\right]
\nonumber \\
&=
\int d^4x\,
\left[
\frac{\mathrm{K}}{4\pi}
\varepsilon^{\mu\nu\lambda\rho}
b_{\mu\nu} \partial_{\lambda} a_{\rho}
-a_{\mu}j^{\mu}_{qp}
-\frac{1}{2}b_{\mu\nu}j^{\mu\nu}_{qv}
\right], 
\end{align}
where 
$\mu,\nu, \ldots=0,\ldots,3$, 
$a=a_{\mu} dx^{\mu}$ and $b= (1/2) b_{\mu\nu} dx^{\mu} dx^{\nu}$ 
are one and two form gauge fields;
$\mathcal{M}$ is the bulk spacetime manifold.
The ``level'' $\mathrm{K}$ is an integer. 
The three form $J_{qp}$ and two form $J_{qv}$ 
represent currents of zero-dimensional (point-like) quasi-particles 
and one-dimensional quasi-vortex lines, respectively. 
The BF theory furnishes the following (bulk) equations of motion
\begin{align}
\frac{\mathrm{K}}{2\pi}da=J_{qv},
\quad
\frac{\mathrm{K}}{2\pi}db=J_{qp}. 
\end{align}

The BF theory implements a non-trivial fractional statistics between
quasiparticles and quasivortices. 
To see this, we consider the following configuration of quasiparticles and quasivortices:
\begin{align} 
J_{qp}&=\delta_{3}(\mathcal{C}),
\quad 
J_{qv}=\delta_{2}(\mathcal{S}).
\label{source config}
\end{align}
Here $\mathcal{C}$ and $\mathcal{S}$ represent the one-dimensional wold-line and the two-dimensional world-sheet 
of quasiparticles and quasivortices, respectively;
$\delta_{D-n}(\mathcal{N})$ is the delta function $(D-n)$-form associated a submanifold $\mathcal{N}\subset \mathcal{M}$,
where $D-n=\mathrm{dim}\, \mathcal{M}-\mathrm{dim}\,\mathcal{N}$.   
By definition, 
for any $n$-form $A_n$,
\begin{align}
\int_{\mathcal{N}} A_n = \int_{\mathcal{M}} \delta_{D-n}(\mathcal{N})\wedge A_n.
\end{align}
Hence, for example, 
\begin{align}
 \int_{\mathcal{M}} J_{qp}\wedge a = \int_{\mathcal{C}} a,
 \quad
 \int_{\mathcal{M}} J_{qv}\wedge b = \int_{\mathcal{S}} b. 
 \label{Wilson op1}
\end{align}
Some useful properties of the delta function forms are summarized in Appendix  \ref{delta-function forms}. 

In the presence of these quasiparticles and quasivortices, 
we now integrate over $a$ and $b$ to derive the effective action for $J_{qp}$ and $J_{qv}$. 
Since the theory is quadratic, this can be done by solving the equations of motion. 
These equations, up to a closed form, are solved by
\begin{align}
b	=\frac{{2\pi}}{\mathrm{K}}d^{-1}J_{qp},
\quad
a	=\frac{{2\pi}}{\mathrm{K}}d^{-1}J_{qv}. 
\end{align}
(If the spacetime is trivial, by the Poincar\'e lemma, a closed form is exact. 
If so, such exact term does not affect our final result since, 
for an arbitrary closed submanifold $\mathcal{N}$,
$
\int\delta(\mathcal{N})(d\phi)\sim\int\delta(\partial \mathcal{N)}\phi=0
$.)
From the formula \eqref{formula delta function exterior derivative}, 
$d^{-1}J_{qp}$ and $d^{-1}J_{qv}$ are determined as
\begin{align}
d^{-1}J_{qp} &= \delta_2 (\mathcal{D}),
\quad
\mbox{where}
\quad 
\partial \mathcal{D}=\mathcal{C}, 
\nonumber \\
d^{-1}J_{qv} &= \delta_1 (\mathcal{V}),
\quad 
\mbox{where}
\quad 
\partial \mathcal{V}= \mathcal{S}.
\end{align}
where the two-dimensional manifold $\mathcal{D}$ and 
the three-dimensional manifold $\mathcal{V}$   
are chosen such that $\partial \mathcal{D}=\mathcal{C}$
and $\partial \mathcal{V} =\mathcal{S}$. 
They are not unique, 
but different choices lead to different $d^{-1}J_{qp, qv}$ 
which differ by closed forms.  

Substituting these solution into the action,
\begin{align}
 S_{bulk} 
 &= -\frac{{2\pi}}{\mathrm{K}}\int (d^{-1}J_{qv})\wedge J_{qp}  
 \nonumber \\
 &= -\frac{{2\pi}}{\mathrm{K}}Lk (\mathcal{S}, \mathcal{C}),
\end{align}
where $Lk$ is the linking number between $J_{qp}$ and $J_{qv}$.
Hence,
\begin{align}
\int \mathcal{D}[a,b] e^{i S_{bulk} } 
&= 
e^{-\frac{{2\pi i}}{\mathrm{K}} Lk(\mathcal{S}, \mathcal{C})}
\nonumber \\
&=
e^{-\frac{{2\pi i}}{\mathrm{K}} 
\sum_{ij}
q_i \lambda_j
Lk(\mathcal{S}_i, \mathcal{C}_j)}. 
\label{stat phase}
\end{align}
In the last line, 
we assume 
the world-line $\mathcal{L}$ consists of trajectories $\mathcal{L}_i$ of many quasiparticles each carrying charge $q_i\in \mathbb{Z}$:
$J_{qp}=\delta_{3}(\mathcal{C})=\sum_{i}q_{i}\delta_{3}(\mathcal{C}_{i})$.
Similarly, 
the world-line $\mathcal{S}$ consists of trajectories $\mathcal{S}_i$ of many quasivortices each carrying charge $\lambda_i\in \mathbb{Z}$:
$J_{qv}=\delta_{2}(\mathcal{S})=\sum_{i}\lambda_{i}\delta_{2}(\mathcal{S}_{i})$.
The fractional phase (when $|\mathrm{K}|>1$)
in Eq. \eqref{stat phase} represents statistical interactions between quasiparticles and quasivortices.

Once the coupling of the gauge fields to the currents is prescribed, 
it also specifies the set of Wilson loops and Wilson surfaces included in theory
(see Eq.\ \eqref{Wilson op1}). 
If the theory is canonically quantized on $\mathcal{M}=\mathbb{R}\times \Sigma$, 
the set of the Wilson loop and Wilson surface operators of our interest is
\begin{align}
 \exp i m \int_L a,
 \quad
 \exp in \int_S b, 
\end{align}
where $m,n$ are integers, and $L$ and $S$ are arbitrary closed loops and surfaces in $\Sigma$.  
These operators satisfy the commutation relations,
\begin{align}
&
 \Big[ {\textstyle \int_L a}, {\textstyle \int_S b}\Big]=\frac{2\pi i}{\mathrm{K}} I(L,S),
 \nonumber \\
 &
 e^{im \int_L a} e^{ in\int_S b}
 =
 e^{\frac{2\pi i}{\mathrm{K}} I(L,S)}
e^{ in\int_S b}
 e^{im \int_L a}. 
\end{align}
where $I(L,S)$ is the intersection number of
the loop $L$ and the surface $S$.

\paragraph{The boundary theory}

On a closed manifold, the BF theory is invariant under gauge transformations 
$a \to a+d\varphi$, where $\varphi$ is zero form, 
and $b\to b+d\zeta$, where $\zeta$ is one form.
In the presence of a boundary (surface), 
there may appear gapless degrees of freedom localized on the surface. 
The action describing the boundary degrees of freedom can be inferred by
adopting the temporal gauge
$a_0=b_{i0}=b_{0i}=0$ ($i=1,2,3$), 
solving the Gauss law constraints
$\varepsilon^{ijk0}\partial_k b_{ij} = \varepsilon^{0ijk}\partial_j a_k=0$
by
$
a_k = \partial_k \varphi
$,
$
b_{ij}=\partial_i \zeta_j-\partial_j \zeta_i
$,
and 
then plugging these back to the action.  
The resulting action is 
\cite{Wu1991,Balachandran1993,Amoretti2012}
\begin{align}
S_{\partial \mathcal M}=
\int_{\partial \mathcal M}
dt dxdy
\left[
\frac{\mathrm{K}}{2\pi}
\epsilon_{ij} \partial_i \zeta_j \partial_t \varphi
-V(\varphi,\zeta)
\right]
\label{surf theory}
\end{align}
where $i,j=1,2$.
Here we have added the potential $V(\varphi,\zeta)$,
which originates from microscopic details of the boundary
and is non-universal. 
This boundary action can be obtained from the 
the free scalar and the $U(1)$ Maxwell theories 
by imposing a self-dual (or an anti-self-dual) constraint,  
$\epsilon^{\mu\nu\lambda}\partial_{\nu}\zeta_{\lambda}=\pm\partial^{\mu}\varphi$. 
\cite{Balachandran1993}

The appearance of the gapless degrees of freedom on the surface deserves more comments.
In particular they should be contrasted with the gapless edge theory of the (2+1)-dimensional Chern-Simons theory.
For the single-component Chern-Simons theory in $(2+1)$ dimensions, 
the boundary is described by the single-component chiral boson theory,
which is stable and cannot be gapped out. 
The appearance of the gapless edge theory is necessary 
since the bulk theory is anomalous and the anomaly must be 
compensated by the degrees of freedom living on the edge.  

On the other hand, 
the surface theory of the BF theory in $(3+1)$ dimensions (also in $(2+1)$ dimensions) 
can be gapped out by adding suitable perturbations (if we do not require any symmetry). 
In other words, there is no anomaly protecting the gapless nature of the surface theory.
Nevertheless, the appearance of the surface theory \eqref{surf theory}
can still be understood in terms of an anomaly.
While the BF theory (both in (2+1) and (3+1) dimensions) 
is equivalent to the topological $\mathbb{Z}_{\mathrm{K}}$ gauge theory,
which does not have gauge anomaly on a manifold with boundary,
the continuum action of the BF theory artificially preserves $U(1)$ symmetry.  
The BF theory is anomalous under $U(1)$ in the presence of a boundary,
and this anomaly must be canceled by gapless degrees of freedom 
living on the boundary.
As in the bulk theory, 
the boundary theory is invariant under the artificial $U(1)$ symmetry;
this symmetry translates the boson field $\varphi\to \varphi+const$.
If the $U(1)$ symmetry is strictly preserved, the gapless boundary 
theory cannot be gapped, as can be inferred easily from the fact that any 
gapping term of cosine type $\cos (n \varphi+\alpha)$ violates the $U(1)$ symmetry. 
(If we use the dual picture of the compactified boson, i.e., the compact $U(1)$ gauge theory,   
the $U(1)$ symmetry is equivalent to prohibiting monopoles.)
On the other hand, once we relax the $U(1)$ symmetry, 
which is, from our point of view, an artificial symmetry after all, 
this gapless boundary theory can be easily gapped out by adding some relevant perturbation and is not stable at all. 

While this gapless surface theory is not stable at all, 
it does encode topological data of the bulk, as we will demonstrate later. 
Let us for now discuss, in more detail,  
the connection between the bulk excitations and the fields living on the boundary.
In the following we choose $\mathcal M=S^1\times \Sigma$, where the spatial manifold $\Sigma$ is a solid torus, 
$\Sigma=D^2\times {S^1}$, and hence $\partial \mathcal M=T^3$. 
Let us first consider a quasiparticle current consisting of 
a quasiparticle carrying $n_0$ units of charges ($n_0\in \mathbb{Z}$):
\begin{align}
j_{qp}^{\mu}(x)=n_0 \int_{L}d\tau\frac{{dX^{\mu}(\tau)}}{d\tau}\delta^{(4)}[x-X(\tau)]
\end{align} 
where $L$ is the world-line of the quasiparticle, and the coordinate $X^{\mu}(\tau)$ represents 
the trajectory of the particle. 
For the quasiparticle at rest, 
$X^{1,2,3}(\tau)=const. = X^{1,2,3}$, 
\begin{align}
j_{qp}^{0}(x)=
n_{0}\delta^{(3)}(\vec{{x}}-\vec{{X}}).
\label{config qp}
\end{align}
Integrating the equation of motion over the total space,
\begin{align}
\frac{\mathrm{K}}{4\pi}\int_{\Sigma} d^{3}x\,\varepsilon^{0ijk}\partial_{i}b_{jk}=\int_{\Sigma} d^{3}x\,j_{qp}^{0}=n_{0}. 
\end{align}
Using Stokes' theorem, 
$\int_{\Sigma}db=\int_{\partial\Sigma}b =(1/2)\int_{\partial \Sigma}  b_{ij}\epsilon^{ij}d^{2}x$, 
and
substituting 
$b_{ij}=\partial_i \zeta_j -\partial_j\zeta_i$, 
this reduces to 
\begin{align}
\int_{\partial\Sigma} d^{2}x\,\epsilon_{ij}\partial_{i}\zeta_{j}=\frac{2\pi}{\mathrm{K}}n_0. 
\end{align}
Hence adding a quasiparticle in the bulk corresponds to introducing flux on the surface.

\begin{figure}[t]
\centering
\includegraphics[scale=0.4]{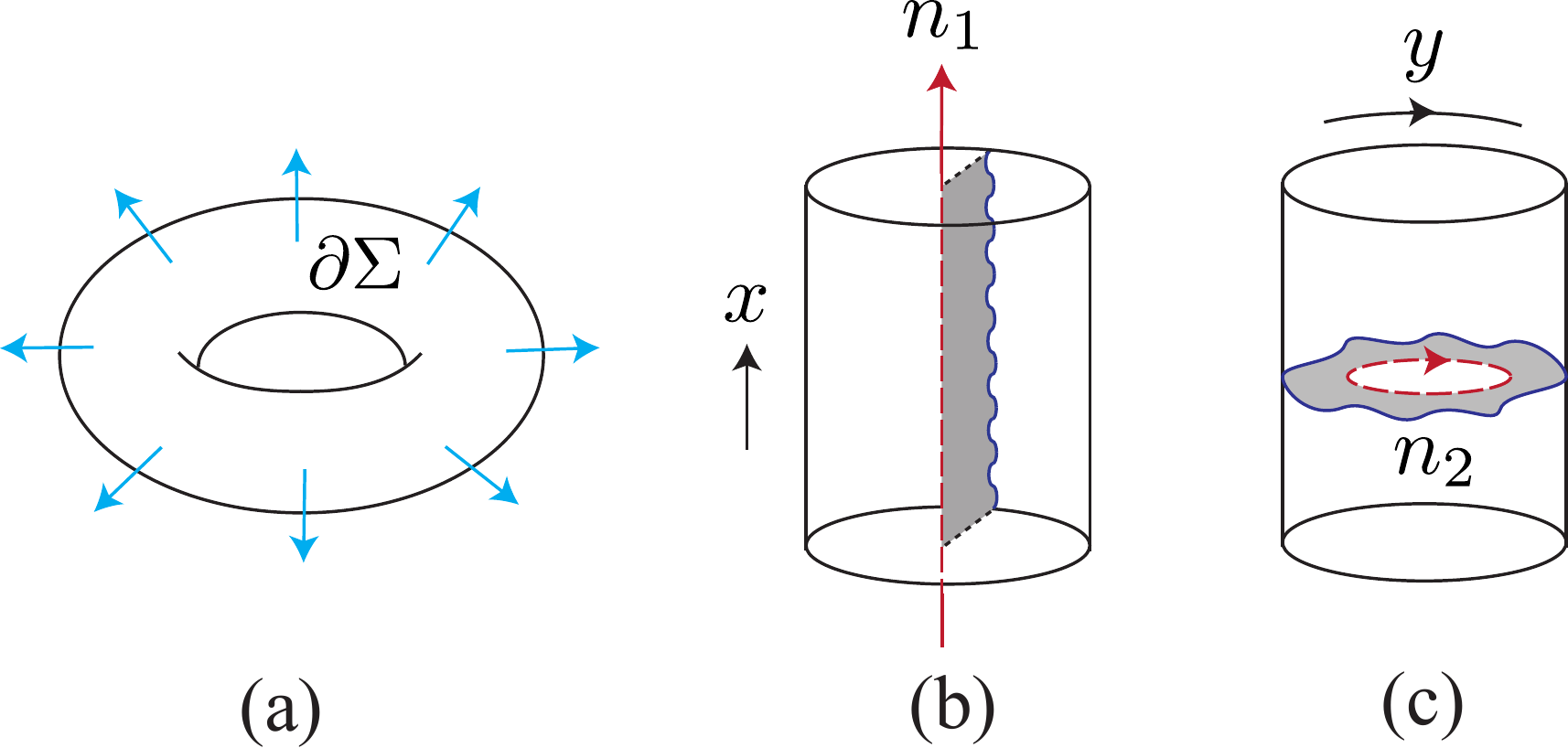}
\caption{
(a) 
The presence of a point-like quasiparticle in the bulk (solid torus) induces 
a fractional flux on the spatial boundary $\Sigma$ (torus). 
(b) 
The presence of a quasivortex line in the bulk twists
the boundary conditions of the surface theory.
Here and in (c), the bulk is presented as a filled cylinder where 
the top and the bottom of the cylinder are identified.
The shaded surface is a sheet of the branch cut which 
emanates from the quasivortex, and intersects with the spatial boundary
(depicted by a wavy line). 
The surface excitations experience a twisted boundary condition as they go through 
the branch cut.
(c)
Similar to (b),
a bulk quasivortex,
which creates a branch cut on the surface 
which now goes along 
a different cycle of the surface, 
is depicted.
\label{fig1}}
\end{figure}

Similarly, let us consider to introduce a quasivortex source: 
\begin{align}
j_{qv}^{\mu\nu}(x)=n
\int_{S}d^{2}\sigma\epsilon^{\alpha\beta}\frac{{\partial X^{\mu}(\sigma)}}{\partial\sigma^{\alpha}}\frac{{\partial X^{\nu}(\sigma)}}{\partial\sigma^{\beta}}
\delta^{(4)}[x-X(\sigma)],
\end{align}
where
 $S$ is the world-surface of the quasivortex, and the coordinate $X^{\mu}(\sigma)$ represents 
the trajectory of the particle in spacetime, and $n$ is an integer.
Let us consider a straight quasivortex at rest, stretching along 
a non-contractible cycle of the bulk solid torus. 
For convenience, this direction is taken as the $x$-direction
(Fig.\ \ref{fig1}).
Then, 
$j_{qv}^{02}=j_{qv}^{03}=0$ and 
\begin{align}
j_{qv}^{01}(x)	
&=n_2 \delta(x^{2}-X^2)\delta(x^{3}-X^3).
\label{config qv}
\end{align}
where $X^{2,3}(\sigma)=const.$ and we have renamed the integer $n$ as $n_2$. 
Integrating the equation of motion over space,
\begin{align}
L_1
\times 
\frac{\mathrm{K}}{2\pi}
\int dydz\,
\varepsilon^{01ij}
\partial_{i}a_{j}
= n_2 \times L_{1}, 
\end{align}
where $i,j=1,2$ and $L_{1}=2\pi R_1$
is the length of the quasivortex stretching in the $x$-direction,
and we noted the flux $\varepsilon^{01ij}\partial_i a_j$ is independent of $x_{1}$.
Using Stokes' theorem, 
and substituting $a_i = \partial_i \varphi$, 
\begin{align}
\oint dy\partial_{y}\varphi=\frac{{2\pi}}{\mathrm{K}}n_{2}. 
\label{winding and qv}
\end{align} 
Hence introducing a quasivortex (quasivortices) 
along the non-contractible loop in the bulk 
corresponds to introducing winding of the scalar boson on the surface.  

One may wish to develop a similar argument for a quasivortex (quasivortices) stretching in the $y$-direction.
(Fig.\ \ref{fig1} (c)). 
It should however be noted that once we fix our geometry as above
(Fig.\ \ref{fig1} (b)), 
loops running in the $y$-direction are contractible in the bulk.
In other words, 
if one constructs a solid torus by filling ``inside'' of a two-dimensional torus, 
one needs to specify one of non-contractible cycles on the two-dimensional torus, 
such that after filling, this cycle now is contractible in the sold torus.

\subsection{The surface theory and quantization}

We now proceed to the canonical quantization of the surface theory.
We start from the surface Lagrangian density 
\begin{align}
\mathcal{L}&=
\frac{\mathrm{K}}{2\pi} (\epsilon_{ij} \partial_i \zeta_j)(\partial_t \varphi) 
\nonumber \\
&\quad 
-\frac{1}{2\lambda_1} (\epsilon_{ij} \partial_i \zeta_j)^2
-\frac{1}{2\lambda_2} 
 G^{ij} \partial_i \varphi \partial_j \varphi.
 \label{surf lag}
\end{align}
The boson field $\varphi$ is compact and satisfy
\begin{equation}
\varphi\equiv\varphi+2\pi.
\end{equation}
I.e., physical observables are made of bosonic exponents
\begin{align}
\exp [i m \varphi(t,\mathsf{r})],
\quad
m\in \mathbb{Z},
\label{bosonic exp}
\end{align}
and the derivative of the boson fields (current operators). 
The winding number of $\varphi$ is quantized, 
in the absence of bulk quasiparticles, according to 
\begin{align}
 \oint dx^i \partial_i \varphi = 2\pi N_i,
\quad 
N_i\in \mathbb{Z},
\end{align}
where $i=1,2$ and $i$ is not summed on the right hand side.
On the other hand,
the gauge field $\zeta_i$ is compact, meaning that physical observables are
Wilson loops,
\begin{align}
 \exp i m \int_C dx^i \zeta_i(t,\mathsf{r}) ,
 \quad
 m\in \mathbb{Z}, 
 \label{Wilson loops}
\end{align}
where $C$ is a closed loop on $\partial\Sigma=T^2$. 
The flux associated to $\zeta_i$ is quantized,
in the absence of bulk quasiparticles, according to 
\begin{align}
\int dxdy\, \epsilon_{ij} \partial_i \zeta_j = 2\pi N_0 
\label{compact cond for zeta}
\end{align}
where $N_0$ is an integer. 
The canonical commutation relation is 
\begin{align}
\left[\varphi(t,{\mathsf{r}}),\epsilon_{ij}\partial_{i}\zeta_{j}(t,{\mathsf{r}^{\prime}})\right]
&=\frac{2\pi i}{\mathrm{K}}\delta^{(2)}({\mathsf{r}}-{\mathsf{r}^{\prime}})
\end{align}

In the following, we fix $\lambda_1$ and $\lambda_2$ according to
\begin{align}
 \frac{(2\pi)^2}{\mathrm{K}^2 \lambda_1 \lambda_2} = 1. 
\end{align}
This choice is convenient since it gives rise to the same energy dispersion 
as the compactified free boson discussed in the previous section. 

To proceed, we consider the mode expansion of the fields.
The equations of motion are
\begin{align}
&
\frac{-\mathrm{K}}{2\pi} \epsilon_{ij} \partial_i \partial_t \zeta_j +
\frac{1}{\lambda_2} G^{ij} \partial_i \partial_j \varphi =0,
\nonumber \\
&
\frac{-\mathrm{K}}{2\pi} \epsilon_{lk} \partial_l \partial_t \varphi +
\frac{1}{\lambda_1}
\epsilon_{lk} \partial_l (\epsilon_{ij} \partial_i \zeta_j) =0. 
\end{align}
The mode expansion consistent with the equations of motion are
\begin{align}
&\varphi(\mathsf{r}) = 
\alpha_0 
+
\frac{\beta_1 x}{R_1}
+
\frac{\beta_2 y}{R_2}
\nonumber \\
&\qquad \quad 
+
\frac{1}{ \sqrt{R_1 R_2}}
\sqrt{\frac{1}{2\mathrm{K}^2 \lambda_1}}
\sum_{\mathsf{k}\neq 0}
\frac{1}{\omega(\mathsf{k})^{1/2}}
\nonumber \\
&\qquad \qquad\qquad 
\times 
\left[
 a(\mathsf{k}) 
 e^{-i \mathsf{k}\cdot \mathsf{r}}
+
a^{\dag}(\mathsf{k}) e^{+i\mathsf{k}\cdot \mathsf{r}}
\right],
\nonumber \\
&
 \zeta_j (\mathsf{r})
=
\frac{\alpha_j}{2\pi R_j}  
+
\frac{\beta_0}{2\pi R_1 R_2}
x \delta_{j,2}
\nonumber \\
&\qquad \quad
+
\frac{1}{ \sqrt{R_1 R_2}}
\sqrt{
\frac{\lambda_1}{8\pi^2}
}
\sum_{\mathsf{k}\neq 0} 
\frac{-1 }{ \omega(\mathsf{k})^{3/2} }
\epsilon_{jm}G^{ml}k_l
\nonumber \\
&\qquad\qquad \qquad
\times 
\left[
 a(\mathsf{k}) e^{ -i \mathsf{k}\cdot \mathsf{r}}
+
a^{\dag}(\mathsf{k}) e^{+i\mathsf{k}\cdot \mathsf{r}}
\right],
\label{BF mode expansion}
\end{align}
where 
the eigenvalues of 
$\beta_{0}$ and
$\beta_{1,2}$
describes the flux (associated with the gauge field $a$ in the bulk),  
and the winding of the $\varphi$ field, respectively. 
The quantization conditions of these variables will be discussed momentarily. 
Reflecting the compact nature of the $\varphi$ and $\zeta_j$ fields, 
the zero modes are compact variable $\alpha_{\mu} \equiv \alpha_{\mu} +2\pi$ ($\mu=0,1,2$); 
For $\alpha_0$, the compactification condition comes from the fact that physical observables are given 
as bosonic exponents \eqref{bosonic exp}. 
Similarly, for $\alpha_{1,2}$, 
that physical observables are given in terms of Wilson loops \eqref{Wilson loops},
and 
that these Wilson loop operators must be invariant under large gauge transformations  
imposes the compactification condition,
$\alpha_{1,2}\equiv \alpha_{1,2}+2\pi$.

From the commutator $[\varphi(t,\mathsf{r}), \epsilon_{ij} \partial_i \zeta_j (t, \mathsf{r}')]$, 
we read off
\begin{align}
&
 [a(\mathsf{k}), a^{\dag}(\mathsf{k}')] = 
 \delta_{\mathsf{k},\mathsf{k}'}, 
 \nonumber \\
 &
 [ \alpha_0, \beta_0] =
 \frac{2\pi i}{\mathrm{K} } \frac{1}{2\pi}. 
\end{align}
From the compactification condition $\alpha_0 \equiv \alpha_0 + 2\pi$,  
$\beta_0$ is quantized according to
\begin{align}
 \beta_0 &=
 \frac{M_0}{  \mathrm{K} },
 \quad
 M_0\in \mathbb{Z}. 
\end{align}
This quantization condition translates into 
\begin{align}
\int dxdy\, \epsilon_{ij} \partial_i \zeta_j &= 
\frac{2\pi M_0}{\mathrm{K} }. 
\end{align}
Compared with the quantization condition \eqref{compact cond for zeta}, 
the flux is now quantized in the fractional unit. 
We will separate $M_0$ into its non-fractional and fractional parts as 
\begin{align}
M_0 = \mathrm{K} N_0 + n_0,
\quad
N_0 \in \mathbb{Z}, 
\quad
n_0 = 0,\ldots, \mathrm{K}-1,
\end{align}
and write the quantization condition of $\beta_0$ as 
\begin{align}
 \beta_0 = 
N_0 + n_0/\mathrm{K}.
\end{align}

The quantization condition of $\beta_{1,2}$ can be discussed similarly. 
From the commutator $[\varphi(t,\mathsf{r}), \epsilon_{ij} \partial_i \zeta_j (t, \mathsf{r}')]$, 
we infer 
\begin{align}
[\epsilon_{ij}\partial_i \varphi(t,\mathsf{r}), \zeta_j (t, \mathsf{r}')] 
=
-\frac{2\pi i}{\mathrm{K} }  \delta^{(2)}(\mathsf{r}-\mathsf{r}'), 
\end{align}
which implies
\begin{align}
 \left[\beta_1, \alpha_2\right]
- \left[\beta_2, \alpha_1\right]
= 
-
\frac{2\pi i}{\mathrm{K}}
\frac{1}{2\pi}.
\end{align}
One can choose, for example, 
\begin{align}
 \left[\beta_1, \alpha_2\right]=
 0,
\quad 
\left[\beta_2, \alpha_1\right]
=
\frac{2\pi i}{\mathrm{K}} \frac{1}{2\pi}.
\end{align}
This choice may be consistent with 
the previous consideration from the bulk point of view, 
and in particular with the comment below \eqref{winding and qv}. 
I.e.,
this choice may correspond to 
choosing which non-contractible loops on the surface 
are contractible in the bulk, 
when forming a solid torus starting from the two-dimensional torus 
by filling its ``inside''. 

From the compactness of the gauge field $\zeta_i$, 
the zero modes satisfy $\alpha_i\equiv \alpha_i + 2\pi$,  
which imposes the quantization condition
\begin{align}
\beta_{2} = \frac{ M_2 }{\mathrm{K}},
\quad
M_2\in \mathbb{Z}.
\end{align}
As before, we split $M_{2}$ into 
the fractional and non-fractional parts, 
\begin{align}
 M_2 = \mathrm{K}N_2 + n_2, 
 \quad
 N_2 \in \mathbb{Z},
 \quad
 n_2 = 0, \ldots, \mathrm{K}-1.
\end{align}
With this, 
the boson field obeys the twisted boundary condition
\begin{align}
\varphi(t,x,y+2\pi R_2)&=
\varphi(t,x,y) + 
2\pi
\left(
N_2 +\frac{n_2}{\mathrm{K}}
\right). 
\end{align}

While above consideration allows winding in the $y$-direction but not in the $x$-direction,  
in computing the partition functions of the surface theory in the next section, 
we consider winding in both directions, 
\begin{align}
\varphi(t,x+2\pi R_1,y)&= \varphi(t,x,y) + 
2\pi \left(
N_1 +\frac{n_1}{\mathrm{K}}
\right),
\nonumber \\
\varphi(t,x,y+2\pi R_2)&=
\varphi(t,x,y) + 
2\pi
\left(
N_2 +\frac{n_2}{\mathrm{K}}
\right), 
\end{align}
That is
\begin{align}
\beta_{i=1,2} = N_{1,2} + n_{1,2}/\mathrm{K}. 
\end{align}

To summarize, in the presence of twisted boundary conditions,
the mode expansion of the fields are given by 
\begin{align}
&
\partial_i \varphi(t,\mathsf{r}) = 
\frac{N_i + n_i/\mathrm{K}}{R_i}
\nonumber \\
&\qquad \quad 
+
\frac{-i}{ \sqrt{R_1 R_2}}
\sqrt{\frac{1}{2\mathrm{K}^2 \lambda_1}}
\sum_{\mathsf{k}\neq 0}
\frac{k_i}{\sqrt{\omega(\mathsf{k}) }}
\nonumber \\
&\qquad\qquad 
\times 
\left[
- a(\mathsf{k}) e^{-i \omega (\mathsf{k})t -i \mathsf{k}\cdot \mathsf{r}}
+
a^{\dag}(\mathsf{k}) e^{+i \omega (\mathsf{k}) t+i\mathsf{k}\cdot \mathsf{r}}
\right],
\nonumber \\
&
\epsilon_{ij} \partial_i \zeta_j (t,\mathsf{r})
=
\frac{N_0+n_0/\mathrm{K}}{2\pi R_1 R_2}
\nonumber \\
&\qquad \quad 
+
\frac{-i}{ \sqrt{R_1 R_2}}
\sqrt{
\frac{\lambda_1}{8\pi^2}
}
\sum_{\mathsf{k}\neq 0} 
 \sqrt{\omega(\mathsf{k})}
\nonumber \\
&\qquad\qquad 
\times 
\left[
 a(\mathsf{k}) e^{-i \omega (\mathsf{k})t -i \mathsf{k}\cdot \mathsf{r}}
-
a^{\dag}(\mathsf{k}) e^{+i \omega (\mathsf{k}) t+i\mathsf{k}\cdot \mathsf{r}}
\right],
\end{align}

The above consideration is somewhat analogous to the quantization of the chiral boson theory 
that appears at the edge of the (2+1)d Chern-Simons theory at level $\mathrm{K}$. 
The (1+1)d chiral boson theory defined on a spatial circle of radius $2\pi$ is described by  
the Lagrangian density
\begin{align}
\mathcal{L} = 
\frac{\mathrm{K}}{4\pi} 
\partial_x \Phi (\partial_t -\partial_x)\Phi, 
\end{align}
where $\Phi$ is a single component boson theory  
compactified as $\Phi \equiv \Phi + 2\pi$,
and obeys the canonical commutation relation
$[\Phi(x), \partial_x \Phi(x')] = (2\pi i/\mathrm{K})\delta(x-x')$. 
The zero mode part of $\Phi$, defined by the mode expansion
\begin{align}
 \Phi(t, x) = \Phi_0 - p (t+x) + i \sum_{n\neq 0} b_n e^{-in (t+x)},  
\end{align}
satisfies 
$
[ \Phi_0, p] = i /\mathrm{K}
$.
This then suggests the quantization rule, 
$ p = (\mbox{integer})/\mathrm{K}$,
and the boundary condition of the chiral boson field 
\begin{align}
 \Phi(t, x+2\pi) = \Phi(t,x) - \frac{2\pi (\mbox{integer})}{\mathrm{K} }. 
\end{align}
Thus, the canonical quantization naturally leads
to the twisted boundary condition of the chiral boson field.

Quantization of the surface theory with the above twisted boundary conditions 
gives the spectrum of local as well as nonlocal (quasiparticles) excitations, 
which obey untwisted and twisted boundary conditions, respectively. 
Once we specify the boundary condition (with some integer vector $n_{\mu=0,1,2}$), 
the theory is quantized within one sector 
(labeled by the equivalence class $[\vec{n}]$ 
with the relation $\vec{n} \equiv \vec{n} + \mathrm{K}\vec{\Lambda}$
where $\vec{\Lambda}$ is a vector with integer entries) 
of the original spectrum. For this surface theory, 
there are $\mathrm{K}^3$ sectors in this compactified theory and is consistent with the $\mathrm{K}^3$ ground states of single component BF theory defined on $T^3$. 

\subsection{The partition functions and modular transformations}

Now we compute the partition function (coupled to the $T^3$ metric):
\begin{align}
\mathcal{Z}^{n_0n_1n_2}
=
\mathrm{Tr}_{\mathcal{H}_{n_0 n_1 n_2}}\left[e^{-2\pi R_{0}H'}\right]
\end{align}
where $\mathcal{H}_{n_0 n_1 n_2}$ is the Hilbert space twisted by $n_0,n_1,n_2 $ 
fractional quantum numbers, and
\begin{align}
H'&= H + i \frac{\alpha R_1}{R_0} P_x + i(\alpha\beta +\gamma)\frac{R_2}{R_0}P_y,
\nonumber \\
H&=
\int dxdy\,\frac{\mathrm{K}^2\lambda_1}{8\pi^2} 
\left[
\frac{4\pi^2}{\mathrm{K}^2\lambda^2_1}
(\epsilon_{ij} \partial_i \zeta_j)^2
+
G^{ij}\partial_i\varphi\partial_j\varphi  
\right],
\nonumber \\
P_i &=
\int dxdy \frac{\mathrm{K}}{2\pi} (\epsilon_{lm}\partial_l \zeta_m) (\partial_i \varphi). 
\end{align}

The calculation of the partition function
goes in parallel with the calculation 
presented in the previous section for the free boson theory.  
To see this, we note, from the equation of motion, 
\begin{align}
\epsilon_{ij}\partial_i \zeta_j
=
 \frac{\mathrm{K}\lambda_1}{2\pi} \partial_t \varphi
\end{align}
up to a constant term.
Thus, in terms of $\varphi$, the Hamiltonian density and the
commutation relation are given by
\begin{align}
&
 \mathcal{H}
 =
 \frac{1}{2} \frac{\mathrm{K}^2 \lambda_1}{(2\pi)^2} 
 \left[
 (\partial_t \varphi)^2 + G^{ij} \partial_i\varphi \partial_j\varphi
 \right],
 \nonumber \\
 &
[\varphi(t,\mathsf{r}), \partial_t \varphi(t,\mathsf{r}') ]
=
\frac{ (2\pi)^2 i }{\mathrm{K}^2 \lambda_1} 
\delta^{(2)} (\mathsf{r}-\mathsf{r}'). 
\end{align}
By introducing the rescaled field,
\begin{align}
 \tilde{\phi} =
\mathrm{K}
 \sqrt{\frac{ \lambda_1}{2}}\varphi
\end{align}
the Hamiltonian and the commutation relation 
can be made isomorphic to those of the free boson theory. 
The compactification condition of the rescaled boson field is
\begin{align}
 \mathrm{r} = \mathrm{K} \sqrt{\frac{\lambda_1}{2}}. 
\end{align}

The partition function $\mathcal{Z}^{n_0n_1n_2}$
can now be computed from 
the partition function of the free boson theory.
The zero mode part of the partition function for each excitation sector is
obtained from $Z_0$ (Eq.\ \eqref{zero mode part fun free boson})
by making replacement 
$N_0 \to \mathrm{K}N_0+n_0$
and 
$N_i \to N_i+n_i/\mathrm{K}$ ($i=1,2$):
\begin{widetext}
\begin{align}
Z^{n_{0}n_{1}n_{2}}&=
\sum_{N_{0,1,2}\in\mathbb{{Z}}}
\exp\Big\{-\frac{{\pi}\mathrm{K}^2\tau_2}{2 \mathrm{r}^2 R_2}
\tilde{N}_0^2
-2\mathrm{r}^2 \pi R_{2}\tau_{2}\left[\tilde{N}_{1}+\beta\tilde{N}_2 \right]^{2}
-\frac{{2\mathrm{r}^2\pi R_{0}R_{1}}}{R_{2}} \tilde{N}_2^2 
\nonumber \\
&\qquad 
+2\pi i\tau_{1}\mathrm{K}\tilde{N}_0 \left[ \tilde{N}_1 +\beta\tilde{N}_2 \right]
+2\pi i\gamma \mathrm{K}\tilde{N}_0 \tilde{N}_2 \Big\},
\label{part_bf}
\end{align}
\end{widetext}
where we have introduced the notation
\begin{align}
\tilde{N}_{\mu}:=N_{\mu} + n_{\mu}/\mathrm{K}. 
\end{align}
For the oscillator part, since the Hamiltonian is the same as the oscillator part for the compact boson, the partition function is exactly the same as the free boson case presented above. Thus we have
\begin{align}
Z_{osc}&=Z_{s_2=0}Z_{s_2\neq 0}
\nonumber \\
&=\left|\frac{1}{\eta(\tau)}\right|^2\prod_{s_2\in\mathbb{Z}^+}
\Theta_{[\beta s_2,\gamma s_2]}^{-1}\left(\tau,\frac{R_1}{R_2}s_2\right). 
\end{align}
The total partition function for each sector is $\mathcal{Z}^{n_0n_1n_2}=Z^{n_0n_1n_2}Z_{osc}$. 

Although 
the surface theory of the (3+1)d BF theory resembles
the compactified free boson discussed in the previous section,
these theories are physically different. 
For the compactified boson, 
the partition function is invariant under the $\mathcal{S}$ and $\mathcal{T}$
modular transformations: It is anomaly-free and a well-defined theory on the $(2+1)$d spacetime torus. 
On the other hand, 
for the surface theory, 
the partition function for each sector is not modular invariant and thus it is not a well-defined theory on the $(2+1)$d torus. 
It should be regarded as the boundary theory of a higher-dimensional topological phase. 
There are $\mathrm{K}^3$ sectors determined by three quantum number $n_{0,1,2}$ 
and they form a complete basis under $S$ and $T$ modular transformations,
as we will show now.

Under $M$ transformation,
quantum numbers are transformed as
\begin{align}
M: \begin{cases} 
N_1+\frac{n_1}{\mathrm{K}}\\
N_2+\frac{n_2}{\mathrm{K}}\\
N_0+\frac{n_0}{\mathrm{K}}
\end{cases}
\to\begin{cases} 
-N_2-\frac{n_2}{\mathrm{K}}\\
N_1+\frac{n_1}{\mathrm{K}}\\
N_0+\frac{n_0}{\mathrm{K}}\end{cases}
\end{align}

To discuss $U_1^{\prime}$ transformation,
we use the Poisson resummation to rewrite the summation over $N_0$ and $N_1$ in
$Z^{n_{0}n_{1}n_{2}}$
and rewrite the zero-mode partition function as 
\begin{widetext}
\begin{align}
Z^{n_{0}n_{1}n_{2}}&=\frac{{1}}{\mathrm{K}|\tau|}\sum_{N_{2},M_{0,1}\in \mathbb{Z}}
\exp\Bigg\{
-\frac{{\pi\tau_{2}}}{2r^{2}R_{2}|\tau|^{2}}M_{1}^{2}
-\frac{{2r^{2}\pi R_{2}\tau_{2}}}{|\tau|^{2}}\left[\frac{{M_{0}}}{\mathrm{K}}-\gamma\tilde{{N}}_{2}\right]^{2}
-\frac{{2\pi r^{2}R_{0}R_{1}}}{R_{2}}\tilde{{N}}_{2}^{2}
\nonumber \\
&
\quad 
-\frac{{2\pi i\tau_{1}\mathrm{K}}}{|\tau|^{2}}\frac{{M_{1}}}{\mathrm{K}}
\left[\frac{{M_{0}}}{\mathrm{K}}-\gamma\tilde{{N}}_{2}\right]
+2\pi i\beta\tilde{{N}}_{2}M_{1}+\frac{{2\pi in_{1}}}{\mathrm{K}}M_{1}
+\frac{{2\pi in_{0}}}{\mathrm{K}}M_{0}
\Bigg\}. 
\end{align}
Let us introduce
\begin{align}
M_{1}:=\mathrm{K}N'_{0}+n'_{0,}\quad M_{0}:=\mathrm{K}N'_{1}+n'_{1}
\end{align}
Then, the partition function can be written as 
\begin{align}
Z^{n_{0}n_{1}n_{2}}	&=
\frac{{1}}{K|\tau|}\sum_{N_{2},M_{0,}\in \mathbb{Z}}
\sum_{n_{0,1}^{\prime}\in \mathbb{Z}_\mathrm{K}}\exp\Bigg[
-\frac{{\pi\tau_{2}K^{2}}}{2r^{2}R_{2}|\tau|^{2}}\tilde{{N}}_{0}^{\prime2}
	-\frac{{2r^{2}\pi R_{2}\tau_{2}}}{|\tau|^{2}}\left[\tilde{{N}}_{1}^{\prime}-\gamma\tilde{{N}}_{2}\right]^{2}
	-\frac{{2\pi r^{2}R_{0}R_{1}}}{R_{2}}\tilde{{N}}_{2}^{2}
	\nonumber \\
	&\quad 
	-\frac{{2\pi i\tau_{1}\mathrm{K}}}{|\tau|^{2}}\tilde{{N}_{0}^{\prime}}\left[\tilde{{N}_{1}^{\prime}}-\gamma\tilde{{N}}_{2}\right]
	+2\pi i\mathrm{K}\beta\tilde{{N}}_{2}\tilde{{N}}_{0}^{\prime}
	+\frac{{2\pi in_{1}n_0^{\prime}}}{\mathrm{K}}+\frac{{2\pi in_{0}n_1^{\prime}}}{\mathrm{K}}\Bigg],
	\label{part bf after poisson}
 \end{align}
\end{widetext}
where $\sum_{n\in \mathbb{Z}_{\mathrm{K}}}:= \sum_{n=0}^{\mathrm{K}-1}$. 
From these expressions, under $U_1^{\prime}$ transformation,
\begin{align}
(U_1^{\prime} \mathcal{Z})^{n_0n_1n_2}
&
=
\frac{1}{\mathrm{K}}
\sum_{n^{\prime}_{0,1} \in \mathbb{Z}_\mathrm{K}}
e^{
\frac{2\pi i}{\mathrm{K}} (n_0 n'_1+n_1 n'_0)
}
\mathcal{Z}^{n^{\prime}_0n^{\prime}_1n_2}.
\label{S_trans}
\end{align}
Combined with the $M$ transformation, we can write down the modular $\mathcal{S}$ 
and  $\mathcal{T}$ matrices:
\begin{align}
\mathcal{S}_{n_i,n^{\prime}_i}&=
\frac{1}{\mathrm{K}}\delta_{n_1,n_2^{\prime}}e^{-\frac{2\pi i}{\mathrm{K}}(n^{\prime}_0n_2-n_0n_1^{\prime})},
\nonumber\\
\mathcal{T}_{n_i, n^{\prime}_i}&=
\delta_{n_0,n_0^{\prime}}\delta_{n_1,n_1^{\prime}}\delta_{n_2,n_2^{\prime}}e^{\frac{2\pi i}{\mathrm{K}}n_0n_1}.
\label{stmatrix}
\end{align}
This result is consistent with previous works, 
Refs.\ \onlinecite{MoradiWen2015, JiangMesarosRan2014},  
and also \onlinecite{WangWen2015}, where the action of 
the modular transformations are calculated in the bulk. 
(See also other related works: Refs. \onlinecite{WangLevin2015, WangLevin2014, LinLevin2015, JianQi2014}.)
In terms of the bulk physics, 
the $\mathcal{S}$ matrix describes the braiding phase between particle and loop excitations,
whereas the $\mathcal{T}$ matrix 
encodes information related to $(3+1)$d analogue of topological spins.
\cite{MoradiWen2015} 
(See also Refs.\ \onlinecite{WangLevin2015, WangLevin2014, LinLevin2015, JianQi2014}.)
The exact agreement between 
the $\mathcal{S}$ and $\mathcal{T}$ matrices calculated in the bulk and the boundary 
suggests there is one-to-one correspondence,
the bulk-boundary correspondence in (3+1)d.

The computed $\mathcal{S}$ and $\mathcal{T}$ matrices \eqref{stmatrix} are expected to be consistent with
the algebraic relations in Eq.\ \eqref{constraint_3}: 
As in (1+1)d CFTs, together with the charge conjugation matrix $\mathcal{C}$,
$\mathcal{S}$ and $\mathcal{T}$ matrices should obey essentially the same algebraic relations as Eq.\ \eqref{constraint_3}.
Assuming the charge conjugation matrix is unity, $\mathcal{C}=1$, 
we have checked, for the case of $\mathrm{K}=2,3,4,5$, 
the $\mathcal{S}$ and $\mathcal{T}$ matrices satisfy all the above constraints 
except the last equation in Eq.\ \eqref{constraint_3}. 

Before we leave this section, 
as we have done in the previous section,
it is instructive to dimensionally reduce the partition functions
of the surface theory of the (3+1)d BF theory.
For each given sector, after dimensional reduction, the partition function is given by
\begin{align}
\mathcal{Z}^{n_0,n_1}
&=\frac{1}{|\eta(\tau)|^2}
\sum_{N_{0,1}\in \mathbb{Z}}
\exp\left\{-\pi\tau_2 \mathrm{K}
\left(N_0+\frac{n_0}{\mathrm{K}}\right)^2
\right.
\nonumber\\
&
\quad
-\pi \tau_2\mathrm{K}\left(N_1+\frac{n_1}{\mathrm{K}}\right)^2
\nonumber \\
&
\left.
\quad 
+2\pi i\tau_1\mathrm{K}\left(N_0+\frac{n_0}{\mathrm{K}}\right)
\left(N_1+\frac{n_1}{\mathrm{K}}\right)
\right\}. 
\label{part_red}
\end{align}
Here, we made a convenient choice $\lambda_1 = 1/\mathrm{K}$, i.e., $2\mathrm{r}^2 = \mathrm{K}$. 
This is the same as the character of the edge theory of the (2+1)d $\mathbb{Z}_{\mathrm{K}}$ gauge theory in its topological phase. 
The effective Lagrangian density of the edge CFT is by
\begin{align}
\mathcal{L}=\frac{1}{4\pi}\partial_t\vec{\Phi}^T\textbf{K}\partial_x\vec{\Phi}
-\partial_x\vec{\Phi}^T \textbf{V}\partial_x\vec{\Phi}, 
\label{lutt}
\end{align}
where $\textbf{K}=\mathrm{K}\sigma_x$ and $\textbf{V}$ 
is a symmetric and positive definite matrix that accounts for the interaction on the edge and is non-universal. 
The characters defined in Eq.\ \eqref{part_red} can be simplified as
\begin{align}
\chi_{ab}(\tau)=
\frac{1}{|\eta(\tau)|^2}\sum_{s,t}
q^{\frac{1}{4\mathrm{K}}(\mathrm{K}s+a+\mathrm{K}t+b)^2}
\bar{q}^{\frac{1}{4\mathrm{K}}(\mathrm{K}s+a-\mathrm{K}t-b)^2}
\end{align}
where $a=n_0$ and $b=n_1$.
There are $\mathrm{K}^2$ characters in total. 
Under the $\mathcal{S}$ and $\mathcal{T}$ modular transformations,
they are transformed as
\begin{align}
\chi_{ab}(\tau+1)&=e^{2\pi i\frac{ab}{\mathrm{K}}}\chi_{ab}(\tau),
\nonumber \\
\chi_{ab}(-{1}/{\tau})&=\frac{1}{\mathrm{K}}\sum_{a^{\prime} ,b^{\prime}}
\chi_{a^{\prime}b^{\prime}}(\tau)
e^{-2\pi i\frac{a^{\prime}b+b^{\prime}a}{\mathrm{K}}
}. 
\end{align}

\subsection{Entropy of the boundary theory}
\label{Entanglement Entropy for BF boundary}

In this section, 
we compute the thermal entropy 
\begin{align}
S_{T}&:=
\frac{{d}}{{d}T}\left[T \ln\chi_{a}\right],
\end{align}
obtained from the partition functions of the boundary theory discussed above. 
Here, 
$\chi_a$ is the partition function in the sector labeled by $a=(n_0,n_1,n_2)$, and 
\begin{align}
1/T = 2\pi R_0 
\end{align}
is the inverse temperature. 

While $S_{T}$ is defined for a system with a real (physical) boundary, 
it is expected to carry information on the universal topological part of the entanglement entropy
(the topological entanglement entropy). 
The latter is defined for the bulk system (the BF theory) defined on a manifold without a physical boundary,
and obtained by integrating out (tracing over) 
a subregion B (compliment to, say, subregion A). 
\cite{
KitaevPreskill2006, Fendley2007, Cappelli11, QiKatsuraLudwig2012}
%
%
%
%

We are interested in the entropy $S_{T}$
in the limit $R_1/R_0 \to \infty$ and $R_1/R_2 \to \infty$. 
(We could also equivalently take the limit with $R_1$ and $R_2$ exchanged, 
in which case, we have to resum differently but the result would be the same.)
To evaluate the entropy in this limit,  
we first make use of the $\mathcal{S}$-modular transformation, 
\cite{Affleck:1991tk}
and write
\begin{align}
S_{T}&=
\frac{{d}}{{d}T}\left[T
\ln\left(\mathcal{S}^{b}_{a}\chi_{b}(-1/\tau) \right)\right]. 
\end{align}
In the above limit,
only the identity character gives rise to the dominant contribution,
$\lim_{R_{1}/R_0\to \infty,R_1/R_{2}\to \infty} \chi_{a}(\tau)
=\mathcal{S}^{0}_{a}\chi_{0}({-1}/{\tau})$,
as seen from Eq.\ \eqref{part bf after poisson}.
Hence 
\begin{align}
S_{T} |_{ R_1/R_0 \to \infty, R_1/R_2 \to \infty} 
&=
\frac{{d}}{{d}T}\left[T\ln\left(\mathcal{S}^{0}_{a}\chi_{0}\right)\right]. 
\end{align}
Then using the modular $\mathcal{S}$ matrix computed in the previous section, 
\begin{align}
S_{T} |_{ R_1/R_0 \to \infty, R_1/R_2 \to \infty} 
&=
\frac{{d}}{{d}T}\left[T \ln\left(\frac{1}{\mathrm{K}}\chi_{0}\right)\right] 
\nonumber \\
&= -\ln \mathrm{K}+ \frac{{d}}{{d}T}\left[T\ln\chi_{0}\right]. 
\end{align}

The first term is the subleading term, 
and identical to the bulk topological entanglement entropy, although $S_{T}$ and the entanglement entropy are defined differently. 
The second term is the extensive piece, which basically corresponds to the entropy of the free boson and is the usual leading order term.
(When $S_{T}$ is interpreted as the entanglement entropy, the second term corresponds to the area law term.)

\section{The surface theory of the $(3+1)$d BF theory with the $\Theta$ term}
\label{The surface theory of the (3+1)d BF theory with the Theta term}

Recall that in the surface theory of the BF theory discussed in the previous section,  
there are three quantum numbers $M_{0,1,2}$,
which we wrote in terms of their non-fractional and fractional parts as
\begin{align}
M_{\mu} = N_{\mu} + \frac{n_{\mu}}{\mathrm{K}},
\quad 
\mu = 0,1,2.
\end{align}
These quantum numbers in the surface theory can be interpreted as 
arising from the presence of bulk quasi-particles or quasi-vortices;
$M_{1,2}$ represents the fractional winding number of
the $\varphi$ field induced by a bulk quaxi-vortex, whereas $M_0$
represents a fractionalized flux threading the surface induced by
a bulk quasi-particle. 

In this section, we consider the following ``twist'' of the quantum number 
\begin{align}
M_0
&\to M_0+ \frac{Q_1 M_2 - Q_2 M_1}{\mathrm{K}} 
\nonumber \\
&= M_0+ \frac{Q\times M}{\mathrm{K}}, 
\label{theta_twist}
\end{align}
in the surface theory of the BF theory, 
where $Q_{1,2}$ are fixed integers,
and we have introduced the notation
\begin{align}
Q\times M :=
Q_{1}M_{2}-Q_{2}M_{1}.
\end{align}
This twist can be induced by considering 
a modification of the BF theory by introducing 
the $\Theta$ term (axion term).
In the next section, we will consider a 
similar twist to discuss three-loop braiding statistics. 

\subsection{The BF theory with the $\Theta$-term in (3+1)d}


We motivate 
the twist
\eqref{theta_twist}
by considering the following modification of the bulk BF theory 
by adding a $\Theta$-term:
\begin{align}
S_{bulk}=&\;\int_{\mathcal M}
\left[
\frac{\mathrm{K}}{2\pi}b\wedge da
-
\frac{\mathrm{p}}{8\pi^2}d\Theta\wedge a\wedge d a \right. 
\nonumber \\
&\quad \qquad 
-a\wedge J_{qp} -b\wedge J_{qv}
\Big].
\label{bf plus theta}
\end{align}
In the second term (the-$\Theta$ term or axion term), 
$\mathrm{p}$ is a parameter, specific value of which will be discussed later,  
$\Theta$ is a non-dynamical background field, 
and we consider an inhomogeneous but time-independent configuration of $\Theta$, which will be specified later. 
Compared to the standard form of the $\Theta$ term, $\Theta da\wedge da$, 
we have done an integration by part and put the derivative acting on $\Theta$.
Since the $\Theta$ field is non-dynamical, we will interpret the presence of the $\Theta$ term as an introduction of a static 
{\it defect}. 
In Ref.\ \onlinecite{Lopes2015}, a similar effective action has been proposed to describe 
the thermal and gravitational response of topological defects in superconducting topological insulators. 
\cite{Chan2013}
We also note that the BF theory with the $\Theta$-term, $\Theta da\wedge da$, 
has been proposed to describe the fermonic and  bosonic topological insulators.
In Ref.\ \onlinecite{JianQi2014}, the BF theory with the $\Theta$ term was used to discuss three-loop braiding processes.

To see the $\Theta$-term 
induces the twist \eqref{theta_twist}, 
we assume the following configuration of the $\Theta$-field:
\begin{align}
\Theta(x,y,z)=\frac{Q_1x}{R_1}+\frac{Q_2y}{R_2}.
\end{align}
where $Q_{1,2}$ are fixed integers. 
From the equation of motion,
\begin{align}
&\frac{\mathrm{K}}{2\pi}\varepsilon^{\mu\nu\lambda\rho}\partial_{\lambda}a_{\rho}=j_{qv}^{\mu\nu}, 
\nonumber \\
&\frac{\mathrm{K}}{4\pi}\varepsilon^{\mu\nu\lambda\rho}\partial_{\nu}b_{\lambda\rho}
+\frac{\mathrm{p}}{4\pi^2} \varepsilon^{\mu\nu\lambda\rho}\partial_{\nu}\Theta\partial_{\lambda}a_{\rho}=j_{qp}^{\mu}.
\end{align}
By plugging the first equation into the second, 
these equations of motion reduce to
\begin{align}
&
\frac{\mathrm{K}}{2\pi}\varepsilon^{\mu\nu\lambda\rho}\partial_{\lambda}a_{\rho}=j_{qv}^{\mu\nu},
\nonumber \\
&
\frac{\mathrm{K}}{4\pi}\varepsilon^{\mu\nu\lambda\rho}\partial_{\nu}b_{\lambda\rho}
=
-
\frac{{\mathrm{p}}}{2\pi \mathrm{K}}\partial_{\nu}\Theta j_{qv}^{\mu\nu}+j_{qp}^{\mu}.  
\label{mod eom}
\end{align}
In the presence of quasiparticle and quasivortex sources,
\eqref{config qp} and \eqref{config qv}, 
the equations of motion
integrated over space
are
\begin{align}
&
\frac{\mathrm{K}}{2\pi}
\int dydz\,
\varepsilon^{01ij}
\partial_{i}a_{j}
= M_1,
\nonumber \\
&
\frac{\mathrm{K}}{2\pi}
\int dxdz\, 
\varepsilon^{02ij}\partial_{i}a_{j}=M_{2},
\nonumber\\ 
&
\frac{\mathrm{K}}{4\pi}\int_{\Sigma} d^{3}x\,\varepsilon^{0ijk}\partial_{i}b_{jk}
=
-
\frac{\mathrm{p}}{\mathrm{K}} 
 Q_{i}M_{i}
+
N_{0}, 
\end{align}
where $M_{1,2},N_0\in \mathbb{Z}$ and we noted
\begin{align}
\partial_{i}\Theta\int d^{3}x\,j_{qv}^{0i}=
 \frac{{Q_{i}}}{R_i} M_{i}\times(2\pi R_{i})=
 2\pi Q_{i}M_{i}
\end{align} 
($i$ is not summed over). 
These can be reduced to, by using Stokes' theorem, 
\begin{align}
& 
\int dy\partial_{y}\varphi=\frac{{2\pi}}{\mathrm{K}}N_{2},
\nonumber \\
& 
\int dx\partial_{x}\varphi=\frac{{2\pi}}{\mathrm{K}}N_{1},
\nonumber \\
&
\int d^{2}x\,\epsilon_{ij}\partial_{i}\zeta_{j}
=
\frac{2\pi}{\mathrm{K}}N_{0}
+
\frac{{2\pi}\mathrm{p}}{\mathrm{K}^2} (Q\times N), 
\label{q condition}
\end{align}
where $M_1= -N_2$ and $M_2=N_1$.
Hence,
upon choosing $\mathrm{p}=1$, 
in the presence of the defect field $\Theta$, 
the quantum number $N_0$ in the surface theory is ``twisted''
as in Eq.\ \eqref{theta_twist}.

We observe that 
the following action
\begin{align}
S'_{bulk}&=\int d^{4}x\,\frac{\mathrm{K}}{4\pi}\varepsilon^{\mu\nu\lambda\rho}\partial_{\nu}b_{\lambda\rho}a_{\mu}
\nonumber \\
&
-j_{qp}^{\mu}a_{\mu}-j_{qv}^{\mu\nu}\frac{{1}}{2}\left[b_{\mu\nu}-\frac{\mathrm{p}}{2\pi \mathrm{K}}(a_{\mu}\partial_{\nu}\Theta-a_{\nu}\partial_{\mu}\Theta)\right]
\label{modified theory}
\end{align}
shares the same equations of motion,
Eq.\ \eqref{mod eom},
as the BF theory with $\Theta$ terms, \eqref{bf plus theta}. 
Hence, the boundary theory derived from $S'_{bulk}$ 
has the same quantization rules of the zero modes
as the boundary theory of $S_{bulk}$.  
In the next section, we will consider the boundary theory derived from $S'_{bulk}$,
and its partition functions. 

To contrast the two theories $S_{bluk}$ and $S'_{bulk}$,
we note, in $S_{bulk}$, 
that the coupling to the currents are ``normal'' while the commutators are ``abnormal'',
in the sense that the commutators among fields $a,b$ are modified due to the presence of the theta term. 
On the other hand, in $S'_{bulk}$, 
the commutators are normal (the same as the ordinary BF theory)
while the coupling to the current is ``abnormal''.
(Since that the commutators are the same as the ordinary BF theory, 
$S'_{bulk}$ and the corresponding boundary theory
can be analyzed in a complete parallel with the BF theory 
-- a practical reason why we will consider on $S'_{bulk}$ in the following --
expect for the zero mode part.)

In spite of these differences, 
these theories lead to the same quantization conditions (the same ``lattice'' of quantum numbers) of zero modes. 
To see how this is possible, we note that the quantization rule of
the zero modes are determined both by 
(a) the canonical commutation relations
and 
(b) the compactification conditions.
The compactification condition
is determined by declaring physically observable Wilson loop operators.
This in term is determined from the coupling of the theory to the
current. 
Therefore, in the original theory, (a) is abnormal
but (b) is normal. In the modified theory, (a) is normal but (b) is
abnormal. 
In the next section, 
we demonstrate this by deriving the quantization conditions
\eqref{q condition}, derived from the bulk point of view here,
in terms of the boundary theory of $S'_{bulk}$. 
In Appendix \ref{The surface theory of the BF theory with theta term}, 
we quantize the boundary theory of the original theory, $S_{bulk}$, 
to derive the quantization rule \eqref{q condition}.

\subsection{The surface theory and partition functions}

\subsubsection{The compactification conditions and quantization rules}
\label{The compactification conditions and quantization rules}

We now proceed to consider the surface theory of the bulk theory \eqref{modified theory}.
Without sources, the surface theory is described by the same Lagrangian density as the surface of the BF theory,
\eqref{surf lag}, 
and hence has the same canonical commutation relations. 
This immediately means that the oscillator part of the surface theory can be treated in exactly the same as before. 
On the other hand, reflecting the abnormal coupling of the gauge fields to the currents in the bulk action $S'_{bulk}$, 
the compactification conditions of the boundary fields $\varphi$ and $\zeta$ are modified, as we will now discuss.  

As we noted earlier, the coupling to the current can be written, e.g.,
$\int d^{4}x\,j_{qp}^{\mu}a_{\mu}=\oint_{L}a$.
Thus, introducing a proper current corresponds to introducing a Wilson
loop. If we now consider a Wilson line $L$ that is spatial, and that ends at the boundary,
\begin{align}
\int_{L}a=\int_{L}d\varphi=\int_{\partial L}\varphi=\varphi(\partial L)
\end{align}
where we noted $\partial L$ is a point, and we have used the solution
to the Gauss law constraint, $a_{a}=\partial_{a}\varphi$ ($a=1,2,3$).
Thus, 
\begin{align}
\exp im\int_{L}a=\exp \left[im\varphi(\partial L)\right] 
\end{align}
This means that $\varphi$ is compactified with the radius $2\pi.$

Let us repeat the same exercise for
the coupling to the quasivortex current:
\begin{align}
&
\int d^{4}x\,j_{qv}^{\mu\nu}\frac{{1}}{2}\left[b_{\mu\nu}
-\frac{{ \mathrm{p}}}{2\pi\mathrm{K}}(a_{\mu}\partial_{\nu}\Theta-a_{\nu}\partial_{\mu}\Theta)\right]
\nonumber \\
&=
\int_{S}\left[b-\frac{{\mathrm{p}}}{2\pi \mathrm{K}}a\wedge d\Theta\right],
\end{align}
where $S$ is the world surface of a quasivortex (quasivortices). 
In the presence of a boundary and using $b=d\zeta$, 
this is evaluated as
\begin{align}
  =\int_{S}\left[d\zeta-\frac{{ \mathrm{p}}}{2\pi \mathrm{K}}d\varphi\wedge d\Theta\right]
  =\int_{\partial S}\left[\zeta-\frac{{ \mathrm{p}}}{2\pi \mathrm{K}}\varphi\wedge d\Theta\right]
\end{align}
where the boundary of the world sheet is on the surface. 
We thus have a Wilson line on the surface:
\begin{align}
\exp im\int_{\partial S}\left[\zeta-\frac{{\mathrm{p}}}{2\pi \mathrm{K}}\varphi\wedge d\Theta\right]
\end{align}
We now consider the case where $\partial S$ is along the $x$- or $y$- cycles. 
Recalling the mode expansion
Eq.\ \eqref{BF mode expansion},
and noting  
$({\mathrm{p}}/{2\pi \mathrm{K}})\int_{L_{i}}\varphi d\Theta
=({\mathrm{p}}/{\mathrm{K}})\alpha_{0}Q_{i}$
the zero modes enter into the integral 
$\int_{\partial S}\left[\zeta-\frac{{ \mathrm{p}}}{2\pi \mathrm{K}}\varphi\wedge d\Theta\right]$
through the following combinations
\begin{align}
\alpha_1-\frac{\mathrm{p}}{\mathrm{K}}Q_{1}\alpha_{0},
\quad
\alpha_2-\frac{{\mathrm{p}}}{\mathrm{K}}Q_{2}\alpha_{0}.
\end{align}

Together with $\alpha_0$, the following three linear combinations 
\begin{align}
 v^{a}_{\mu}\alpha_{\mu}, \quad a=0,1,2
\end{align}
are angular variables,
where
\begin{align}
& v^0 = (1,0,0)^T,
 \nonumber \\
&
 v^1 = \left(-\frac{\mathrm{p}Q_1}{\mathrm{K}}, 1, 0\right)^T ,
\quad  
 v^2 = \left(-\frac{\mathrm{p}Q_2}{\mathrm{K}}, 0, 1\right)^T. 
\end{align}
Noting the commutation relations among zero modes, 
\begin{align}
[\alpha_{0,}\beta_{0}]=[\alpha_{1},-\beta_{2}]
=[\alpha_{2},\beta_{1}]=\frac{{i}}{\mathrm{K}},
\end{align}
we consider the linear combinations
\begin{align}
 w_{a}^{\mu}\bar{\beta}_{\mu}, 
 \quad
\bar{\beta}=(\beta_0, -\beta_2, \beta_1), 
 \quad a=0,1,2,
\end{align}
where $w^a$ are translation vectors reciprocal to $v^a$:
\begin{align}
 w_a^{\mu} v^b_{\mu} = \delta^{a}_{b}.
\end{align}
Explicitly, they are given by
\begin{align}
w_0=\left( 1, \frac{\mathrm{p}Q_1}{\mathrm{K}}, \frac{\mathrm{p}Q_1}{\mathrm{K}}\right),
\,\,
w_1 = (0,1,0),
\,\,
w_2 = (0,0,1). 
\end{align}
Then, in the ``rotated'' basis, the commutation relation takes the following canonical form:
\begin{align}
 [v^a_{\mu} \alpha_{\mu}, w^{\nu}_b \bar{\beta}_{\nu} ]=
 \frac{i}{\mathrm{K}} v^a_{\mu} w_{b}^{\mu}
 =
 \frac{i}{\mathrm{K}} \delta^a_b.
\end{align}
Due to the compacticity of $v^a_{\mu}\alpha_{\mu}$, 
$w^{\nu}_b\bar{\beta}_{\nu}$ takes on values
\begin{align}
w^{\nu}_b\bar{\beta}_{\nu} = \frac{1}{\mathrm{K}} \times m_b,
\quad
m_{b=1,2,3}\in \mathbb{Z}. 
\end{align}
Inverting this relation, 
\begin{align}
 \bar{\beta}_{\mu}= \frac{1}{\mathrm{K}} u_{\mu}^a m_a,\quad 
 u_{\nu}^aw^{\mu}_a = \delta^{\mu}_{\nu}
\label{recplocal lattice}
\end{align}
where 
\begin{align}
&
u_0=\left( 1, 0,0\right)^T,
\nonumber \\
&
u_1 = \left(\frac{-\mathrm{p}Q_1}{\mathrm{K}},1,0\right)^T,
\quad
u_2 = \left(\frac{-\mathrm{p}Q_2}{\mathrm{K}},0,1\right)^T. 
\end{align}
Renaming the integers as
$m_0\to N_0$, $m_1 \to -N_2$, and $m_2\to N_1$,
Eq.\ \eqref{recplocal lattice}
is nothing but the quantization rule \eqref{q condition}. 
%

%
%

\subsubsection{The partition functions}
\label{The partition functions}

%
%
%

With the twist (\ref{theta_twist}),
we can now write down the zero mode partition function. 
Let us recall
the partition function of the BF surface without the theta term, 
$Z^{n_{0}n_{1}n_{2}}$, 
defined in Eq.\ \eqref{part_bf}. 
For later use, we write $Z^{n_{0}n_{1}n_{2}}$ as
\begin{align}
Z^{n_{0}n_{1}n_{2}} 
 & =\sum_{N_{0,1,2}\in \mathbb{Z} }f_{\mathrm{K}}\left(M_{0,}M_{i}\right)
\nonumber \\
& 
=\sum_{N_{0,1,2}\in\mathbb{{Z}}}
f_{\mathrm{K}}\left(\mathrm{K}N_{0}+n_{0},\mathrm{K}N_{i}+n_{i}\right), 
\end{align}
where $f_{\mathrm{K}}$ is defined by the summand in Eq.\ \eqref{part_bf},
and recall $M_{\mu}=\mathrm{K}N_{\mu} + n_{\mu}$. 
We will call the partition function resulting from the twist 
$Z_{Q_{1}Q_{1}}^{n_{0}n_{1}n_{2}}$.
It is given by
\begin{align}
Z_{Q_{1}Q_{2}}^{n_{0}n_{1}n_{2}}=\sum_{N_{0,1,2}\in\mathbb{{Z}}}
f_{\mathrm{K}}\left(\mathrm{K}N_{0}+n_{0}+\frac{{Q\times M}}{\mathrm{K}},
\mathrm{K}N_{i}+n_{i}\right).
\end{align}

To proceed, we write
\begin{align}
 M_{i}&=\mathrm{K}N_{i}+n_{i}=\mathrm{K}^{2}\bar{{N}}_{i}+\mathrm{K}t_{i}+n_{i},
 \nonumber \\
 Q_{i}&=\mathrm{K}R_{i}+r_{i}=\mathrm{K}^{2}\bar{{R}}_{i}+\mathrm{K}s_{i}+r_{i},
 \label{Mi and Qi}
\end{align}
where new integers $\bar{N}_i, \bar{R}_i, R_i$ and 
$\mathbb{Z}_{\mathrm{K}}$ variables $t_i, s_i, r_i$ are introduced.
In the following, 
we will show that 
the zero mode partition function depends on $Q_i$ only through $r_i$,  
and hence can be denoted as 
$Z^{n_0n_1n_2}_{r_1r_2}$,
and that 
the partition function can be written as
\begin{align}
Z^{n_0n_1n_2}_{r_1r_2}
=\sum_{t_{1,2}\in \mathbb{Z}_{\mathrm{K}}}
X^{\bar{n}_0\bar{n}_1\bar{n}_2},
\label{goal}
\end{align}
where we have introduced
\begin{align}
&
\underbar{n}_0 : =n_{0}+s\times n+r\times t \mod \mathrm{K},
\nonumber \\
&
\begin{cases}
\bar{n}_0:= \mathrm{K}\underbar{n}_0+(r\times n) \\
\bar{n}_1:= \mathrm{K} t_1+n_1\\ 
\bar{n}_2:= \mathrm{K}t_2+n_2,
\end{cases}
\label{bar variables}
\end{align}
(i.e., $\underbar{n}_0  =(n_{0}+s\times n+r\times t)\% \mathrm{K}$)
and $X^{\bar{n}_0\bar{n}_1\bar{n}_2}$ is defined by 
\begin{align}
X^{\bar{{n}}_{0}\bar{{n}}_{1}\bar{{n}}_{2}}:=
\sum_{A_0,\bar{{N}}_{1,2}\in\mathbb{{Z}}}
f_{\mathrm{K}}
\left(\mathrm{K}A_0+\frac{\bar{{n}}_{0}}{\mathrm{K}},
\mathrm{K}^{2}\bar{{N}}_{i}+\bar{{n}}_{i}\right). 
\end{align}

To show Eq.\ \eqref{goal}, 
we start by writing the partition function in terms of
variables introduced in Eq.\ \eqref{Mi and Qi}:
\begin{align}
&Z_{Q_{i}=\mathrm{K}^{2}\bar{{R}}_{i}+\mathrm{K}s_{i}+r_{i}}^{n_{0}n_{1}n_{2}}
=\sum_{N_0, \bar{{N}}_{1,2}\in\mathbb{{Z}}}
\sum_{t_{1,2}\in\mathbb{{Z}}_{\mathrm{K}}}
\nonumber \\
&
\times 
f_{\mathrm{K}}\left(\mathrm{K}N_{0}+n_{0}+\frac{{Q\times M}}{\mathrm{K}},\mathrm{K}^{2}\bar{{N}}_{i}+\mathrm{K}t_{i}+n_{i}\right). 
\end{align}
By further introducing
\begin{align}
a_0 & =n_{0}+s\times n+r\times t,
\nonumber \\
A_0 & =N_{0}+(\mathrm{K}\bar{{R}}+s)\times N+\bar{{R}}\times n+r\times\bar{{N}}, 
\end{align}
and noting the equality 
\begin{align}
\mathrm{K}N_{0}+n_{0}+\frac{{Q\times M}}{\mathrm{K}}
=\mathrm{K}A_0+a_0+\frac{{r\times n}}{\mathrm{K}}
\end{align}
Then,
\begin{align}
&
Z_{Q_{i}=\mathrm{K}^{2}\bar{{R}}_{i}+\mathrm{K}s_{i}+r_{i}}^{n_{0}n_{1}n_{2}}
=\sum_{N_0,\bar{{N}}_{1,2}\in\mathbb{{Z}}}
\sum_{t_{1,2}\in\mathbb{{Z}}_{\mathrm{K}}}
\nonumber \\
&\quad
\times
f_{\mathrm{K}}
\left(\mathrm{K}A_0+a_0+\frac{{r\times n}}{\mathrm{K}},
\mathrm{K}^{2}\bar{{N}}_{i}+\mathrm{K}t_{i}+n_{i}\right).
\end{align}

We now fix $t_{1,2}$ and consider
\begin{align}
&
X_{Q_{i}=\mathrm{K}^{2}\bar{{R}}_{i}+\mathrm{K}s_{i}+r_{i}}^{n_{0}n_{1}n_{2},t_{1}t_{2}}
=\sum_{N_0,\bar{{N}}_{1,2}\in\mathbb{{Z}}}
\nonumber \\
&\quad
\times 
f_{\mathrm{K}}\left(\mathrm{K}A_0+a_0+\frac{{r\times n}}{\mathrm{K}},
\mathrm{K}^{2}\bar{{N}}_{i}+\mathrm{K}t_{i}+n_{i}\right).
\end{align}
Note that once $t_{1,2}$ are fixed, $a_0$ is fixed. 
Converting the summation over $N_{0}$ to a summation over $A_0$,
\begin{align}
&
X_{Q_{i}=\mathrm{K}^{2}\bar{{R}}_{i}+\mathrm{K}s_{i}+r_{i}}^{n_{0}n_{1}n_{2},t_{1}t_{2}}
=
\sum_{A_0,\bar{{N}}_{1,2}\in\mathbb{{Z}}}
\nonumber \\
&\quad
\times 
f_{\mathrm{K}}
\left(\mathrm{K}A_0+a_0+\frac{{r\times n}}{\mathrm{K}},
\mathrm{K}^{2}\bar{{N}}_{i}+\mathrm{K}t_{i}+n_{i}\right).
\end{align}
Note that the $s$ and $n_{0}$ dependence of the right hand side comes only from $a_0$.
Also, after converting the sum $\sum_{N_{0}}\to\sum_{A_0},$
the $\bar{{R}}_{i}$ dependence is gone. So, we write
$
X_{Q_{i}=\mathrm{K}^{2}\bar{{R}}_{i}+\mathrm{K}s_{i}+r_{i}}^{n_{0}n_{1}n_{2},t_{1}t_{2}}
$
simply as
$X_{r_{i}}^{a_0n_{1}n_{2},t_{1}t_{2}}$.
Observe that 
$X_{Q_{i}=\mathrm{K}^{2}\bar{{R}}_{i}+\mathrm{K}s_{i}+r_{i}}^{n_{0}n_{1}n_{2},t_{1}t_{2}}$
appears to depend on nine $\mathbb{{Z}}_{\mathrm{K}}$-valued parameters,
$n_{0,1,2},t_{1,2},s_{1,2},r_{1,2}$
After the reorganization we have just done, 
we lost $s_{1,2}$, and we now only have six $\mathbb{{Z}}_{\mathrm{K}}$ parameters, 
$n_{1,2},t_{1,2},r_{1,2}$ and $a_0$.
While $a_0$ is not $\mathbb{Z}_{\mathrm{K}}$-valued, 
we can shift $a_0$ such that
\begin{align}
a_0=\mathrm{K}\times (\mbox{{integer}})+[a_0]
\end{align}
where the second term takes values $0,\ldots \mathrm{K}-1$.
Then, 
\begin{align}
&
X_{r_{1}r_2}^{[a_0]n_{1}n_{2},t_{1}t_{2}}=\sum_{A_0,\bar{{N}}_{1,2}\in\mathbb{{Z}}}
\nonumber\\
&
\quad\times 
f_{\mathrm{K}}\left(\mathrm{K}A_0+[a_0]+\frac{{r\times n}}{\mathrm{K}},
\mathrm{K}^{2}\bar{{N}}_{i}+\mathrm{K}t_{i}+n_{i}\right). 
\label{X intermidiate expression}
\end{align}
Observing that 
$X_{r_{1}r_2}^{[a_0]n_{1}n_{2},t_{1}t_{2}}$
depends on $[a_0],n_{1,2},t_{1,2}$ 
only through $\bar{n}_{0,1,2}$
defined in Eq.\ \eqref{X intermidiate expression},
rewriting 
Eq.\ \eqref{X intermidiate expression}
in terms of
$\bar{n}_{0,1,2}$
completes the derivation of Eq.\ \eqref{goal}.

\subsubsection{Modular transformations}
We now discuss the modular properties of the partition functions. 
%
Under the $U_2$ transformation, the zero-mode partition functions are transformed according to
\begin{align}
(U_2X)^{\bar{n}_0\bar{n}_1\bar{n}_2}
=
e^{-\frac{2\pi i}{\mathrm{K}^2}\bar{n}_0\bar{n}_1}X^{\bar{n}_0\bar{n}_1\bar{n}_2}. 
\end{align}
%
On the other hand, under the $U_1^{\prime}$ transformation, the partition functions are transformed as
\begin{align}
(U^{\prime}_1X)^{\bar{n}_0\bar{n}_1\bar{n}_2}
&=\frac{1}{|\tau|\mathrm{K}^2}
\sum_{\bar{n}^{\prime}_{0,1},\bar{r}^{\prime}_{0,1}}
e^{\frac{2\pi i}{\mathrm{K}^2} ( \bar{n}_0\bar{n}^{\prime}_1 
+ \bar{n}_1\bar{n}^{\prime}_0)
}
{X}^{\bar{n}^{\prime}_0\bar{n}^{\prime}_1\bar{n}_2}
\end{align}
where 
$\bar{n}_0^{\prime}\equiv \mathrm{K}\underbar{n}_0^{\prime}
+r_1^{\prime}n_2-r_2^{\prime}n_1^{\prime}$. 

Observe that,
upon the $U_1^{\prime}$ transformation,
partition functions with new parameters $r_1^{\prime}$ and $r_2^{\prime}$ generated.
Since $r_1$ and $r_2$ are the given quantum numbers from the $\Theta$ term,
the action of modular transformations is not closed.

\section{Coupling two BF theories -- three-loop braiding statistics} 
\label{Three-loop braiding statistics}

In the twist \eqref{theta_twist},
the integers $Q_{1,2}$ are fixed and treated as a background. 
I.e.,  $\Theta$ is a non-dynamical field.
We have seen that the surface partition functions do not form a complete basis under modular transformations. 
To circumvent this issue, one may consider to treat $Q_1$ and $Q_2$ as dynamical variables,
which may come from another copy of the BF theory.
In this section, we will discuss  
two copies of the BF surface theories which are coupled via cubic terms.

Let us start from two decoupled copies of the BF surface theories.
Let $M_{0,1,2}$ and $Q_{0,1,2}$ label different twisted sectors of the first 
and second copy, respectively. 
We consider to twist these quantum numbers by 
\begin{align}
M_0&\to M_0+
\frac{Q\times M}{\mathrm{K}},\nonumber\\
Q_0&\to Q_0+
\frac{M\times Q}{\mathrm{K}}.
\label{cubic_twist}
\end{align}
Here, unlike Eq.\ \eqref{theta_twist}, both $M$ and $Q$ are dynamical variables.

In the next section, we start by introducing an (3+1)d bulk field theory, Eq.\ \eqref{action 3loop}, 
or its alternative form \eqref{alternative action 3-loop}, 
which realizes precisely the twist \eqref{cubic_twist}.
We will analyze the modular properties of the resulting zero mode partition functions at the surface.
The oscillator part of the partition function is simply given by the partition function of the two decoupled copies of free boson theories. 
By computing the $\mathcal{S}$ and $\mathcal{T}$ matrices acting on the zero mode partitions,
we argue that the action \eqref{action 3loop} realizes three loop braiding statistics. 
While we were finalizing the draft,
a preprint \onlinecite{YeGu2015} appeared where the similar bulk actions were discussed 
and conjectured to realize three-loop braiding statistics.


\subsection{The bulk field theory}

\paragraph{The cubic theory}
Let us motivate the twist \eqref{cubic_twist}.
We propose to work with the following bulk action:
\begin{align}
S_{bulk}&=
\int_{\mathcal M}
\Bigg[\frac{\mathrm{K}}{2\pi}\delta_{IJ}b^{I}\wedge {d}a^{J}
\nonumber \\
&\qquad 
+
\frac{ \mathrm{p}_1}{4\pi^2} a^{1}\wedge a^{2}\wedge da^{2}
+
\frac{\mathrm{p}_2}{4\pi^2}  a^{2}\wedge a^{1}\wedge da^{1}
 \nonumber \\
&\qquad 
-\delta_{IJ}b^{I}\wedge J^J_{qv}
-\delta_{IJ}a^{I}\wedge J^J_{qp}
\Big], 
\label{action 3loop}
\end{align}
where $I,J=1,2$ and $\mathrm{p}_{1,2}$ are, as the level $\mathrm{K}$, constant parameters of the theory.  Similar action has been discussed in Ref.\ \onlinecite{Kapustin2014d, wang2015_prl, wan2015twisted}. 
The equations of motion are
\begin{align}
&\frac{\mathrm{K}}{2\pi}da^{I}=J^{I}_{qv},
\nonumber \\
&
\frac{\mathrm{K}}{2\pi}db^1+
\frac{\mathrm{p}_1}{4\pi^2}a^2\wedge da^2
\nonumber \\
&\qquad 
-
\frac{\mathrm{p}_2}{2\pi^2} 
a^2\wedge da^1
+
\frac{\mathrm{p}_2}{4\pi^2} da^2\wedge a^1= J_{qp}^1,
\nonumber \\
&
\frac{\mathrm{K}}{2\pi}db^2+
\frac{\mathrm{p}_2}{4\pi^2} a^1\wedge da^1
\nonumber \\
&\qquad 
-
\frac{\mathrm{p}_1}{2\pi^2}a^1\wedge da^2
+
\frac{\mathrm{p}_1}{4\pi^2}da^1\wedge a^2= J_{qp}^2. 
\label{3loop EOM}
\end{align} 

As in our previous discussion in the BF theory 
with and without the theta term, 
let us consider a fixed, static quasiparticle and quasivortex configuration and integrate the equation of motion over space.
By solving the first equation of motion 
as $a^I = (2\pi/\mathrm{K}) (d^{-1}J^I_{qv})$,
plugging the solution to the second and the third equations of motion, 
and integrating over space, 
\begin{align}
\frac{\mathrm{K}}{2\pi} \int_{\Sigma} db^1
&=
-\frac{ \mathrm{p}_1}{\mathrm{K}^2} \int_{\Sigma} (d^{-1}J^2_{qv}) \wedge J^2_{qv}
\nonumber \\
&\quad  
+\frac{ \mathrm{p}_2}{\mathrm{K}^2} \int_{\Sigma} (d^{-1}J^2_{qv})\wedge J^1_{qv}
+ \int_{\Sigma} J_{qp}^1,
\nonumber \\
\frac{\mathrm{K}}{2\pi} \int_{\Sigma} db^2
&=
-\frac{\mathrm{p}_2}{\mathrm{K}^2} \int_{\Sigma} (d^{-1}J^1_{qv})\wedge J^1_{qv} 
\nonumber \\
&\quad  
+\frac{\mathrm{p}_1}{\mathrm{K}^2} \int_{\Sigma} (d^{-1}J^1_{qv})\wedge J^2_{qv}
+ \int_{\Sigma} J_{qp}^2, 
\end{align} 
where note that in the static configurations considered here,
$J_{qv}$ is a delta function one form supporting a spatial loop,
whereas
$J_{qp}$ is a delta function three form supporting a spatial point. 
Correspondingly, $d^{-1}J_{qv}$ is a delta function 0 form supporting 
a three dimensional manifold. 
The contributions to $\int_{\Sigma}db^I$ is coming 
from quasivortex loops,
$\int_{\Sigma} (d^{-1}J^I_{qv}) \wedge J^J_{qv}$, 
are given in terms of their linking number.  

Considering now the specific geometry $\Sigma= D^2\times S^1$ 
with the boundary (surface) $\partial \Sigma=T^2$, 
we can derive the quantization rule of the zero modes of the boundary fields. 
Using the Gauss law constraint to write the boundary conditions in terms of $\varphi^{I}$ and $\zeta^{I}$,
the bulk equations of motion translate in to 
\begin{align}
&
\frac{\mathrm{K}}{2\pi}\int_{S^{1}_i}d\varphi^{1}=M_i, 
\quad
\frac{\mathrm{K}}{2\pi}\int_{S^{1}_i}d\varphi^{2}=Q_i, 
\nonumber \\
&
\frac{\mathrm{K}}{2\pi}
\int_{\partial \Sigma} 
d \zeta^{1}
-
\frac{\mathrm{p}_2}{4\pi^2} \int_{\partial \Sigma} d\varphi^{2}\wedge d\varphi^{1}=M_0, 
\nonumber \\
&
\frac{\mathrm{K}}{2\pi}
\int_{\partial\Sigma} 
d \zeta^{2}-\frac{\mathrm{p}_1}{4\pi^2} \int_{\partial \Sigma} d\varphi^{1}\wedge d\varphi^{2}= Q_0.
\label{3loopbcond}
\end{align}
With $\mathrm{p}_1=\mathrm{p}_2=\mathrm{K}$,
these correspond precisely to the twist \eqref{cubic_twist}.

Note that if we naively gauge transform as $b^I\to b^I+d\zeta^I$ and $a^I\to a+d\varphi^I$,
we find that the theory is not gauge invariant. Moreover these gauge transformations are not generated by Gauss constraints. 
We propose the following alternative gauge transformations:
\begin{align}
b^{1}\to &\; b^{\prime 1}
= b^{1} +{d}\zeta^{1}
-\frac{\mathrm{p}_2}{2\pi \mathrm{K}} 
\left(
a^{2}\wedge {d}\varphi^{1}
+
{d}\varphi^{2}\wedge a^{1} 
\right), 
\nonumber \\
b^{2}\to &\; b^{\prime 2}
= b^{2} +{d}\zeta^{2}
-\frac{\mathrm{p}_1}{2\pi \mathrm{K}}
\left(
a^{1}\wedge {d}\varphi^{2}
+
{d}\varphi^{1}\wedge a^{2}
\right),
\nonumber \\
a^{I}\to &\; a^{\prime I}
= a^{I}+{d}\varphi^{I}. 
\label{gauge3loop}
\end{align}
Therefore the action with cubic terms in Eq.\ \eqref{action 3loop} is gauge invariant. 
On the other hand, as for the coupling to the sources, 
by demanding the gauge invariance, 
we can read off the conserved currents, which are modified
due to the presence of the cubic terms and the modified gauge transformations.

On an open manifold, the action picks up a gauge anomaly on the boundary under these gauge transformations
\begin{align}
&S_{bulk}[b^{\prime},a^{\prime}]=S_{bulk}[b,a]
\nonumber \\
&\qquad
+
\frac{\mathrm{K}}{2\pi}
\int_{\partial\mathcal M}
\delta_{IJ}d\zeta^{I}\wedge a^{J}
\nonumber \\
&\qquad 
+
\frac{1}{4\pi^2}
\int_{\partial \mathcal M}{
\left(\mathrm{p}_1d\varphi^1\wedge d \varphi^2\wedge a^2
+\mathrm{p}_2d\varphi^2\wedge d\varphi^1 \wedge a^1 \right)}.
\end{align}
This anomaly then must be compensated by 
an appropriate boundary field theory. 

\paragraph{The alternative quadratic theory}

Instead of tackling
the cubic theory \eqref{action 3loop}
and the corresponding surface theory,
as in our discussion in the BF theory with theta term,  
we consider an alternative form of the theory.
We note that 
the equations of motion \eqref{3loop EOM}
can be derived from the following alternative action: 
\begin{align}
S'_{bulk} &=
\frac{\mathrm{K}}{2\pi} \int \delta_{IJ}b^I\wedge da^J
-\int
\delta_{IJ} a^J\wedge J_{qp}^I
\nonumber\\
&\quad
-\int
\left[b^1
+\frac{\mathrm{p}_2}{2\pi\mathrm{K}}a^1\wedge a^2\right]
\wedge J^1_{qv}
\nonumber\\
&\quad
-\int 
\left[b^2
+\frac{\mathrm{p}_1}{2\pi\mathrm{K}}a^2\wedge a^1\right]
\wedge J^2_{qv}.
\label{alternative action 3-loop}
\end{align}
Unlike $S_{bulk}$, this theory is quadratic.
Integrating over $a^I$ and $b^I$, 
one obtains
the effective action of the currents
\begin{align}
 \int\mathcal{D}[a^I, b^I] e^{i S'_{bulk}}
 =
 e^{i S_{eff}}
\end{align}
where
\begin{align}
S_{eff} &=
-\frac{2\pi}{\mathrm{K}}
\int (d^{-1}J^I_{qv})\wedge J^I_{qp}
\nonumber \\
\quad 
&+
\left(
\frac{2\pi}{\mathrm{K}}
\right)^3
\mathrm{p}_{1}
\int (d^{-1}J^1_{qv})\wedge (d^{-1}J^2_{qv})\wedge J^2_{qv}
\nonumber \\
\quad 
&+
\left(
\frac{2\pi}{\mathrm{K}}
\right)^3
\mathrm{p}_{2}
\int (d^{-1}J^2_{qv})\wedge (d^{-1}J^1_{qv})\wedge J^1_{qv}.   
\end{align}
The first term in the effective action
describes, as in the ordinary BF theory, 
the quasparticle-quasivortex braiding statistics
while 
the second and third terms 
include interactions among three quasivortex lines.

From the coupling to the currents, 
we read off the Wilson loop and Wilson surface operators
in the theory:
\begin{align}
\exp \left[im \int_L a^I\right],
\quad 
\exp in 
\int_S \left[
b^I 
+
\frac{\mathrm{p}_{\bar{I}}}{2\pi \mathrm{K} } a^I\wedge a^{\bar{I}}
\right], 
\label{surface op. quadratic theory}
\end{align}
where 
$n$ and $m$ are integers, 
$L$ and $S$ are arbitrary closed loop and surfaces, respectively, 
and we introduced the notation $\bar{1}=2$ and $\bar{2}=1$, and 
the repeated capital Roman indices are not summer over here. 
These operators (or rather their exponents) satisfy 
\begin{align}
 \Big[ {\textstyle \int_C a^I}, {\textstyle \int_S B^J} \Big] &= 
 \frac{2\pi i}{\mathrm{K}}
 \delta^{IJ} I(C,S),
 \nonumber \\
 \Big[ {\textstyle \int_S B^I}, {\textstyle \int_{S'} B^J} \Big] &= 
 -
 \frac{2i \mathrm{p}_{\bar{J}}}{\mathrm{K}^2} \delta^{IJ} {\textstyle \int_{S\#S'} a^{\bar{J}}}
 +
 \frac{2i \mathrm{p}_{\bar{J}}}{\mathrm{K}^2} \delta^{I\bar{J}} \textstyle{\int_{S\#S'} a^{J}},
\end{align}
where 
\begin{align}
B^I:=
b^I 
+
\frac{\mathrm{p}_{\bar{I}}}{2\pi \mathrm{K} } a^I\wedge a^{\bar{I}},
\end{align}
and as before the repeated capital Roman indices are not summed over. 
Note also the triple commutator among $\int_S B^I$ is computed as 
\begin{align}
&
 \Big[
 \Big[ {\textstyle \int_S B^I}, {\textstyle \int_{S'} B^J} \Big],
{\textstyle \int_{S^{\prime\prime}}B^K}
 \Big]
 \nonumber \\
 &\quad 
 =
 \frac{4\pi \mathrm{p}_{\bar{J}} }{\mathrm{K}^3} 
\left( 
 \delta^{IJ}\delta^{\bar{J}K}
- \delta^{I\bar{J}}\delta^{JK}
 \right)
 I(S\#S', S^{\prime\prime}). 
\end{align}

To make a comparison between
the cubic and quadratic theories,
in the cubic theory,
the canonical commutation relations 
differ from the ordinary BF theory,
while they remain the same in the quadratic theory.
In fact, in the cubic theory, 
the commutator among fields generates another field,
$[b,b]\sim a$, schematically. 
On the other hand, the set of Wilson loop and surface operators
in the cubic theory is  conventional (i.e., identical to the ordinary BF theory)
while it is modified in the quadratic theory
as in \eqref{surface op. quadratic theory}.
In spite of these differences, 
the algebra of Wilson loop and surface operators of 
the two theories appear to be identical. 
Therefore, we argue that the two theories are equivalent. 
In the following, we will proceed with the quadratic theory. 

\paragraph{the quantization rule of the zero modes}

We now derive the compactification condition of the boundary fields 
from Eq.\eqref{surface op. quadratic theory}. 
In the presence of a boundary and using $b^I=d\zeta^I$ and $a^I=d\varphi^I$, 
the surface operators reduce to 
\begin{align}
&
\exp im
\int_S 
\left[
d\zeta^I +\frac{\mathrm{p}_{\bar{I}}}{2\pi \mathrm{K}}
d\varphi^I \wedge d\varphi^{\bar{I}}
\right]
\nonumber \\
&=
\exp im
\int_{\partial S} 
\left[
\zeta^I +\frac{\mathrm{p}_{\bar{I}}}{2\pi \mathrm{K}}
\varphi^I \wedge d\varphi^{\bar{I}}
\right]
\end{align}
where the boundary of the world sheet is on the surface. 
We now consider the case where $\partial S$ is along the $x$- or $y$- cycles
on the surface.
Recalling the mode expansion
\begin{align}
\varphi^I(\mathsf{{r}}) & =\alpha_{0}^I+\frac{{\beta_{1}^Ix}}{R_{1}}+\frac{{\beta_{2}^Iy}}{R_{2}}+\cdots, 
\nonumber \\
\zeta_{j}^I(\mathsf{{r}}) & =\frac{{\alpha_{j}^I}}{2\pi R_{j}}+\frac{{\beta_{0}^I}}{2\pi R_{1}R_{2}}x\delta_{j,2}+\cdots,
\end{align}
the zero modes enter into the integral 
$
\int_{\partial S}
[
\zeta^I
+( \mathrm{p}_{\bar{I}} /2\pi\mathrm{K})\varphi^I\wedge d\varphi^{\bar{I}}
]
$
through the combinations
\begin{align}
\alpha^I_j+\frac{\mathrm{p}_{\bar{I}} }{\mathrm{K}} \alpha^{I}_0 \beta^{\bar{I}}_j.
\end{align}
We thus conclude
\begin{align}
 v^{1a}_{\mu}\alpha^1_{\mu},
 \quad
 v^{2a}_{\mu}\alpha^2_{\mu},
\end{align}
are angular variables, where 
\begin{align}
&
v^{I0}  =(1,0,0)^T,
\nonumber \\
&
v^{I1}  =(\mathrm{p}_{\bar{I}}\beta_{1}^{\bar{I}}/\mathrm{K},1,0)^T,
\quad
v^{I2}  =(\mathrm{p}_{\bar{I}}\beta_{2}^{\bar{I}}/\mathrm{K},0,1)^T. 
\end{align}
The rest of the discussion 
is essentially identical to
the analysis made in Sec.\ \ref{The compactification conditions and quantization rules}. 
We recall the commutation relations among zero modes 
\begin{align}
[\alpha^I_0, \beta^J_0]
=
[\alpha^I_1, -\beta^J_2]
=
[\alpha^I_2, \beta^J_2]
=
\frac{i}{\mathrm{K}}\delta_{IJ},
\end{align}
the following linear combinations of the zero modes are integer-valued
\begin{align}
\mathrm{K} w^{I \nu}_{b}\bar{\beta}_{\nu}^I= m^I_b, 
\quad
m^I_b\in \mathbb{Z}, 
\end{align}
where 
\begin{align}
w_{0}^I & =
\left(1,-\mathrm{p}_{\bar{I}}\beta^{\bar{I}}_{1}/\mathrm{K},-{p}_{\bar{I}}\beta^{\bar{I}}_{2}/\mathrm{K}\right),
\nonumber \\
w_{1}^I & = \left(0,1,0\right),
\quad
w_{2}^I =\left(0,0,1\right).
\end{align}
Inverting this relation, the eigenvalues are given by 
\begin{align}
&\mathrm{K}\beta_0^1=M_{0}-\frac{\mathrm{p}_2}{\mathrm{K}}(Q\times M),\nonumber\\
&\mathrm{K}\beta_1^1=M_{2},
\quad
\mathrm{K}\beta_2^1=-M_{1},\nonumber\\
&\mathrm{K}\beta_0^2=Q_{0}-\frac{\mathrm{p}_1}{\mathrm{K}}(M\times Q),\nonumber\\
&\mathrm{K}\beta_1^2=Q_{2},
\quad
\mathrm{K}\beta_2^2=-Q_{1},
\end{align}
where $M_{\mu}$ and $Q_{\mu}$ are integers.

\subsection{The surface partition functions}

With the twist \eqref{cubic_twist}, the two copies of the surface theories are coupled together.
The partition functions are given by 
\begin{align}
Z^{n_0n_1n_2}_{r_1r_2}Z^{r_0r_1r_2}_{n_1n_2}
\end{align}
where, as before, we decompose the quantum numbers as 
\begin{align}
&M_{\mu} = \mathrm{K} N_{\mu}+ n_{\mu},
\quad
n_{\mu} = 0,1,\ldots, \mathrm{K}-1,
\quad
N_{\mu} \in \mathbb{Z}, \nonumber\\
&
Q_{\mu} = \mathrm{K} R_{\mu} + r_{\mu},
\quad
r_{\mu} = 0,1,\ldots, \mathrm{K}-1,
\quad
R_{\mu} \in \mathbb{Z}, 
\end{align}
and noted,
following the discussion in Sec.\ \ref{The partition functions}, 
the partition functions depend only on the fractional parts of $M_{\mu}/\mathrm{K}$
and $Q_{\mu}/\mathrm{K}$.
Following Sec.\ \ref{The partition functions} further, 
we can write the partition functions as 
\begin{align}
Z^{n_0n_1n_2}_{r_1r_2}Z^{r_0r_1r_2}_{n_1n_2}
&=\sum_{t_{1,2},s_{1,2}\in \mathbb{Z}_\mathrm{K}}
X^{\bar{n}_0\bar{n}_1\bar{n}_2}
X^{\bar{r}_0\bar{r}_1\bar{r}_2}, 
\end{align}
where 
\begin{align} 
&
\underbar{n}_0 : =n_{0}+s\times n+r\times t \mod \mathrm{K},
\nonumber \\
&
\underbar{r}_0 : =r_{0}+t\times r+n\times s \mod \mathrm{K},
\nonumber \\
&
\begin{cases}
\bar{n}_0\equiv \mathrm{K}\underbar{n}_0+(r\times n) \\
\bar{n}_1\equiv \mathrm{K}t_1+n_1\\ 
\bar{n}_2\equiv \mathrm{K}t_2+n_2
\end{cases}
\quad
\begin{cases}
\bar{r}_0\equiv \mathrm{K}\underbar{r}_0+(n\times r) \\
\bar{r}_1\equiv \mathrm{K}s_1+r_1 \\ 
\bar{r}_2\equiv \mathrm{K}s_2+r_2
\end{cases}
\label{ts}
\end{align}

Under the $U_2$ transformation,
the product 
$X^{\bar{n}_0\bar{n}_1\bar{n}_2}X^{\bar{r}_0\bar{r}_1\bar{r}_2}$ is 
invariant up to a phase, 
\begin{align}
&
(U_2X)^{\bar{n}_0\bar{n}_1\bar{n}_2}
(U_2 X)^{\bar{r}_0\bar{r}_1\bar{r}_2}
\nonumber \\
&=
e^{
-\frac{2\pi i}{\mathrm{K}}(\tilde{n}_0n_1+\tilde{r}_0r_1)
-\frac{2\pi i}{\mathrm{K}^2}(r_1n_2-r_2n_1)(n_1-r_1)}
X^{\bar{n}_0\bar{n}_1\bar{n}_2}
X^{\bar{r}_0\bar{r}_1\bar{r}_2}, 
\label{U2 on XX}
\end{align}
where we have introduced
\begin{align}
\tilde{n}_0&\equiv \underbar{n}_0-r_2t_1+r_2s_1\ \mbox{mod}\ \mathrm{K}, 
\nonumber \\
\tilde{r}_0&\equiv \underbar{r}_0+n_2t_1-n_2s_1\ \mbox{mod}\ \mathrm{K}.
\end{align}
For the above equation, if we write down the phase in terms of $\tilde{n}_0$ and $\tilde{r}_0$, 
it will be independent of $t_i$ and $s_i$. 
In other words, 
for two different $X^{\bar{n}_0\bar{n}_1\bar{n}_2}X^{\bar{r}_0\bar{r}_1\bar{r}_2}$, 
if they have the same $\tilde{n}_0$, $\tilde{r}_0$, $n_1$, $r_1$, $n_2$ and $r_2$, 
the phases they acquire under the $U_2$ transformation are the same. 
This motivates us to combine these partition functions and define,
for fixed $\tilde{n}_0, \tilde{r}_0\in \mathbb{Z}_{\mathrm{K}}$, 
\begin{align}
\chi^{\tilde{n}_0 n_1 n_2}_{\tilde{r}_0r_1r_2}&=
\sum_{t_{1,2},s_{1,2}\in \mathbb{Z}_\mathrm{K}}
X^{\tilde{n}_0 \bar{n}_1 \bar{n}_2}
X^{\tilde{r}_0 \bar{r}_1 \bar{r}_2},
\label{char}
\end{align}
where the sum is taken over all quartets $(t_{1,2},s_{1,2})$ giving rise to given $\tilde{n}_0, \tilde{r}_0$.
Observe that $X^{\bar{n}_0 \bar{n}_1 \bar{n}_2}$
is labeled by two 
$\mathbb{Z}_{\mathrm{K}}\times \mathbb{Z}_{\mathrm{K}}$-valued 
quantum numbers, 
and one $\mathbb{Z}_{\mathrm{K}}$-valued quantum number.  
On the other hand, 
$\chi^{\tilde{n}_0 n_1 n_2}_{\tilde{r}_0r_1r_2}$
depends on six 
$\mathbb{Z}_{\mathrm{K}}\times \mathbb{Z}_{\mathrm{K}}$-valued indices. 
There are $\mathrm{K}^6$ sectors. 
From Eq.\ \eqref{U2 on XX}, 
it is straightforward to  read off the transformation of $\chi$ 
under the $U_2$ transformation:
\begin{align}
&\quad
(U_2 \chi)^{\tilde{n}_0 n_1 n_2}_{\tilde{r}_0 r_1 r_2}
\nonumber \\
&
=e^{
-\frac{2\pi i}{\mathrm{K}}(\tilde{n}_0n_1+\tilde{r}_0r_1)
-\frac{2\pi i}{\mathrm{K}^2}(r_1n_2-r_2n_1)(n_1-r_1)}
\chi^{\tilde{n}_0n_1n_2}_{\tilde{r}_0r_1 r_2}.
\end{align}

As for the $U'_1$ transformation,
the product $X^{\bar{n}_0\bar{n}_1\bar{n}_2}X^{\bar{r}_0\bar{r}_1\bar{r}_2}$
transforms under $U_1^{\prime}$ as
\begin{align}
&\qquad 
(U_1^{\prime}X)^{\bar{n}_0\bar{n}_1\bar{n}_2}
(U_1^{\prime}X)^{\bar{r}_0\bar{r}_1\bar{r}_2}
\nonumber\\
&
=\frac{1}{|\tau|^2\mathrm{K}^4}
\sum_{\bar{n}^{\prime}_{0,1},\bar{r}^{\prime}_{0,1}\in \mathbb{Z}_{\mathrm{K}} }
e^{i\theta^{ \bar{n}'_0\bar{n}'_1 \bar{r}'_0 \bar{r}'_1 } }
{X}^{\bar{n}^{\prime}_0\bar{n}^{\prime}_1\bar{n}_2}
{X}^{\bar{r}^{\prime}_0\bar{r}^{\prime}_1\bar{r}_2}
\end{align}
where the phase $\theta$ is given by
\begin{align}
\theta^{ \bar{n}'_0\bar{n}'_1 \bar{r}'_0 \bar{r}'_1 } 
&=
  \frac{2\pi  \bar{n}_0\bar{n}^{\prime}_1}{\mathrm{K}^2}
+\frac{2\pi  \bar{n}_1\bar{n}^{\prime}_0}{\mathrm{K}^2}
+\frac{2\pi \bar{r}_0\bar{r}^{\prime}_1}{\mathrm{K}^2}
+\frac{2\pi  \bar{r}_1\bar{r}^{\prime}_0}{\mathrm{K}^2}
\nonumber \\
&
=\frac{2\pi}{\mathrm{K}}
(\tilde{n}_0n_1^{\prime}+\tilde{n}_0^{\prime}n_1
+\tilde{r}_0r_1^{\prime}+\tilde{r}_0^{\prime}r_1)
\nonumber\\
&
\quad 
+\frac{2\pi }{\mathrm{K}^2}
\big[
(r\times n) (n_1^{\prime}-r_1^{\prime})
\nonumber \\
&\qquad
\qquad
+
(r_1^{\prime}n_2-r_2n_1^{\prime})
(n_1-r_1)\big].
\label{three_loop}
\end{align}
To derive this result, we note that $n_2$ and $r_2$ 
are invariant under the $U_1^{\prime}$ transformation. 
As in our previous discussion on the $U_2$ transformation, 
it is crucial to observe 
that the phase  
$\theta^{ \bar{n}'_0\bar{n}'_1 \bar{r}'_0 \bar{r}'_1 }$ 
is independent of $t_{1,2},s_{1,2}$.  
We are thus led to consider 
the partition functions
$\chi^{\tilde{n}_0 n_1 n_2}_{\tilde{r}_0r_1r_2}$
defined in Eq.\ \eqref{char}, 
which transform,
under the $U_1^{\prime}$ transformation,
as 
\begin{align}
&\quad
(U_1^{\prime}\chi)^{\tilde{n}_0n_1n_2}_{\tilde{r}_0r_1r_2}
\nonumber\\
&
=\frac{1}{|\tau|^2\mathrm{K}^2}
\sum_{\tilde{n}_0^{\prime},n_{1,2}^{\prime},
\tilde{r}_0^{\prime},r_{1,2}^{\prime}}
e^{ i\theta^{ \bar{n}'_0\bar{n}'_1 \bar{r}'_0 \bar{r}'_1 } }
\chi^{\tilde{n}_0^{\prime}n_1^{\prime}n_2^{\prime}}_{\tilde{r}_0^{\prime}r_1^{\prime}r_2^{\prime}}
\delta_{n_1,n_2^{\prime}}\delta_{r_1,r_2^{\prime}}
\label{U_1_matrix}
\end{align}

Summarizing, 
the modular $\mathcal{S}$ and $\mathcal{T}$ matrices are given by
\begin{align}
\mathcal{S}_{n_{\mu},n^{\prime}_{\mu},r_{\mu},r_{\mu}^{\prime}}
&=\frac{1}{\mathrm{K}^2}
\delta_{n_1,n_2^{\prime}}\delta_{r_1,r_2^{\prime}}
e^{-\frac{2\pi i}{\mathrm{K}}(\tilde{n}^{\prime}_0n_2-\tilde{n}_0n_1^{\prime}+\tilde{r}^{\prime}_0r_2-\tilde{r}_0r_1^{\prime})} 
\nonumber\\
&
\quad 
\times 
e^{-\frac{2\pi i}{\mathrm{K}^2}\left[(n_1+r_1)(n_2r_1^{\prime}+n_1^{\prime}r_2)-2n_2n_1^{\prime}r_1-2n_1r_2r_1^{\prime}\right]}, 
\nonumber \\
\mathcal{T}_{n_{\mu},n^{\prime}_{\mu},r_{\mu},r_{\mu}^{\prime}}&=
\delta_{{n}_{\mu},{n}_{\mu}^{\prime} }
\delta_{r_{\mu},r_{\mu}^{\prime}} 
\nonumber\\
&
\quad 
\times
e^{\frac{2\pi i}{\mathrm{K}}(\tilde{n}_0n_1+\tilde{r}_0r_1)+\frac{2\pi i}{\mathrm{K}^2}(r_1n_2-r_2n_1)(n_1-r_1)}. 
\label{S and T 3-loop braiding}
\end{align} 
where $n_{\mu}=(\tilde{n}_0, n_1,n_2)$.
Observe that a quick way to obtain this three loop braiding phase is to replace 
$
n_0\to n_0+(r\times n)/\mathrm{K}
$,
$
r_0\to r_0+(n\times r)/\mathrm{K},
$
in the $\mathcal{S}$ and $\mathcal{T}$ matrices for the surface of the BF theory (Eq.\ \eqref{stmatrix}).

The first exponential in the $\mathcal{S}$ matrix, 
$e^{ -2\pi i ( \tilde{n}'_0 n_2  - \cdots)/\mathrm{K}}$, 
and the first term in Eq.\ \eqref{three_loop} 
represents the particle-loop braiding phase, which exists also in the ordinary BF theory.
On the other hand, 
the second exponential in the $\mathcal{S}$ matrix, 
$e^{ -2\pi i [ (n_1 + r_1) (n_2 r'_1 +n'_1 r_2) - \cdots]/\mathrm{K}^2}$, 
and the second term in Eq.\ \eqref{three_loop} 
describes a topological invariant which can be considered 
as the higher dimensional generalization of the linking number of closed lines (in three dimensions),
and is also related with the three-loop braiding process.
\cite{JiangMesarosRan2014,WangLevin2014,WangWen2015,JianQi2014}
More precisely, from the second term in Eq.\ \eqref{three_loop}, 
one can extract three-loop braiding statistical phases. 
For example, 
the phase factor 
$e^{2\pi i r_1 n_2 n'_1/\mathrm{K}^2}$ included in Eq.\ \eqref{three_loop} 
can be interpreted as the three-loop braiding statistical phases associated to 
two loops running in the $x$-direction with quantum numbers $r_1$ and $n'_1$
in the presence of a base loop running in the $y$-direction with quantum number $n_2$ 
(Table \ref{table1}).
The three-loop braiding statistics encoded in the $\mathcal{S}$-matrix 
can be further understood through dimensional reduction discussed below.

\begin{table}[t]
  \begin{center}
    \begin{tabular}{c|c|c|c}
      $(n_1^{\prime}, r_1;n_2)$ & $(n_1^{\prime},r_1;r_2)$ & $(r_1^{\prime},r_1;n_2)$ & $(n_1^{\prime},n_1;r_2)$\\
      \hline
      $\frac{2\pi n_2n_1^{\prime}r_1}{\mathrm{K}^2}$ 
      & $\frac{2\pi r_2n_1^{\prime}r_1}{\mathrm{K}^2}$ 
      & $-\frac{4\pi n_2 r_1^{\prime}r_1}{\mathrm{K}^2}$ & $-\frac{4\pi r_2n_1^{\prime}n_1}{\mathrm{K}^2}$\\
    \end{tabular}
 \caption{
The braiding statistical phases (the second line) for 
the braiding processes between loop $a$ and loop $b$ with base loop $c$ linking both of them (denoted by $(a,b;c)$ in the first line). 
Here, $a$, $b$ and $c$ are the quantum numbers for loop excitations. 
}
    \label{table1}
  \end{center}
\end{table}

As for the $\mathcal{T}$ matrix, 
the first phase factor $e^{2\pi i (\tilde{n}_0 n_1 + \tilde{r}_0 r_1)/\mathrm{K}}$ 
is proposed to be the topological spin for the composite particle-loop excitations in the BF theory. 
On the other hand, 
the second phase factor 
$e^{2\pi i (r_1n_2-r_2n_1)(n_1-r_1)/\mathrm{K}^2}$
can be considered as the topological spin for the loop excitations with a base loop threading through it.  
For instance, $e^{2\pi i r_1n_2n_1/\mathrm{K}^2}$ 
represents the topological spin for the loop excitation with quantum number $(r_1,n_1)$ 
threaded by the loop excitation carrying quantum number $n_2$.

These results extracted from
the boundary $\mathcal{S}$ and $\mathcal{T}$ matrices, \eqref{S and T 3-loop braiding},
are consistent with the previous bulk calculations in the literature. 
\cite{JiangMesarosRan2014,WangLevin2014,WangWen2015,JianQi2014}
In particular, 
in Ref.\ \onlinecite{JiangMesarosRan2014}
the $\mathcal{S}$ and $\mathcal{T}$ matrices in the bulk  
are calculated in the basis that is constructed from 
the so-called minimum entropy states (MESs) on the bulk spatial three torus.
In Ref.\ \onlinecite{WangWen2015},
the bulk $\mathcal{S}$ and $\mathcal{T}$ matrices
were constructed for $Z_{N_1}\times Z_{N_2}\times Z_{N_3}$ gauge theories.

Several comments are in order:

%

(i)
The entropy $S_T$ computed from these characters and the modular $\mathcal{S}$ matrix shows, 
in the limit $ R_1/R_0 \to \infty$ and $R_1/R_2 \to \infty$, 
the asymptotic behavior 
$S_T = -2\ln \mathrm{K}+\cdots$,
where $\cdots$ is the term proportional to the area of the surface.   
I.e., the constant piece in the (entanglement) entropy is the same as 
the two decoupled copies of the BF theories.

(ii) For $(3+1)$d topological phases (gauge theories) with $\mathbb{Z}_{\mathrm{K} }\times \mathbb{Z}_{\mathrm{K}}$ gauge symmetry,
we expect there are (at least) 
$\mathrm{K}^2$ different topological phases that are differentiated by their three-loop braiding statistical phases. 
This is expected from the group cohomology classification (construction) of SPT phases;
from $H^4[\mathbb{Z}_{\mathrm{K}}\times\mathbb{Z}_{\mathrm{K}},U(1)]=\mathbb{Z}_{\mathrm{K}}\times\mathbb{Z}_{\mathrm{K}}$, 
we expect there are at least $\mathrm{K}^2$ different SPT phases in (3+1)d protected by
unitary on-site symmetry $G=\mathbb{Z}_{\mathrm{K}}\times \mathbb{Z}_\mathrm{K}$.
Once the global symmetry is gauged, 
these different SPT phases give rise to $\mathrm{K}^2$ different topologically ordered phases
which are differentiated by the three-loop braiding phases.\cite{dijkgraaf1990topological,Chen2013} 
The model we studied in this section, the two copies of coupled BF surface theories,  
corresponds to the surface theory of one of the $\mathrm{K}^2$ topological phases.  
The surface theories of all the other topological phases can be obtained 
by tuning the coefficient in the coupling terms.
In our model, the coefficient $\mathrm{p}_1$ and $\mathrm{p}_2$ in front of the cubic terms 
are chosen to be $\mathrm{K}$. 
In general, they can take value $\mathrm{q}_1\mathrm{K}$ and $\mathrm{q}_2\mathrm{K}$ 
with $\mathrm{q}_{1,2}=0,1,\ldots,\mathrm{K}-1$
\cite{Gaiotto2015}
which will lead to $\mathrm{K}^2$ different topological phases with different $\mathcal{S}$ and $\mathcal{T}$ matrices. 
The three-loop braiding phases will be slightly modified and are shown in Table \ref{table2}, which are consistent with 
Ref.\ \onlinecite{WangLevin2014}. 

\begin{table}[t]
  \begin{center}
    \begin{tabular}{c|c|c|c}
      $(n_1^{\prime}, r_1;n_2)$ & $(n_1^{\prime},r_1;r_2)$ & $(r_1^{\prime},r_1;n_2)$ & $(n_1^{\prime},n_1;r_2)$\\
      \hline
         $\frac{2\pi \mathrm{q}_2 n_2 n_1^{\prime}r_1}{\mathrm{K}^2}$ 
      & $\frac{2\pi \mathrm{q}_1 r_2 n_1^{\prime}r_1}{\mathrm{K}^2}$ 
      & $-\frac{4\pi \mathrm{q}_1 n_2 r_1^{\prime}r_1}{\mathrm{K}^2}$ 
      & $-\frac{4\pi \mathrm{q}_2 r_2 n_1^{\prime}n_1}{\mathrm{K}^2}$\\
    \end{tabular}
 \caption{
 Same as Table \ref{table1}, 
but for generic values of the parameter $\mathrm{q}_{1,2}=0,1,2,\ldots,\mathrm{K}-1$.
    \label{table2}
}
  \end{center}
\end{table}

Observe also that for $G=\mathbb{Z}_{\mathrm{K}}$, $H^4[\mathbb{Z}_{\mathrm{K}},U(1)]=0$,
i.e., there is no non-trivial SPT phase protected by $G=\mathbb{Z}_{\mathrm{K}}$ symmetry.
Hence, there is essentially only one topologically ordered phase 
with $\mathbb{Z}_{\mathrm{K}}$ gauge group, 
whose surface is described by the one-component surface theory studied in Sec.\ \ref{The surface theory of the BF theory}. 
On the other hand,
the two-component surface theory studied in this section allows richer possibilities.   

(iii)
An insight on the three-loop braiding statistics phase can be obtained from dimensional reduction. 
For the trivial two-component BF theory, there is only a non-trivial particle and loop braiding phase described 
in Eq.\ \eqref{stmatrix}.
This model, after dimensional reduction, reduces to 
the $D(\mathbb{Z}_{\mathrm{K}}\times \mathbb{Z}_{\mathrm{K}})$ 
quantum double model with the $\mathrm{K}$-matrix given by $\mathrm{K}\sigma_x\oplus \mathrm{K}\sigma_x$. 

For the topological phase with non-trivial three-loop braiding statistics phase, 
the dimensional reduction is more interesting. 
Here, we consider the simplest non-trivial example with $\mathrm{K}=2$. 
We perform dimensional reduction on $U_1^{\prime}$ and $\mathcal{T}$ defined 
in Eq.\ \eqref{U_1_matrix} and Eq.\ \eqref{S and T 3-loop braiding}.
When we do so, we need to fix the quantum numbers $n_2$ and $r_2$.
For example, if we take $n_2=0$ and $r_2=0$, 
i.e., there is no third loop connecting the first and second loops, 
the $\mathcal{S}$ and $\mathcal{T}$ matrices after dimensional reduction are the same as 
those for the two copies of the toric code model.

On the other hand, 
if we take $n_2=1$ and $r_2=1$, the dimensional reduction results in 
the $\mathcal{S}$ and $\mathcal{T}$ matrices given by 
\begin{align}
\mathcal{S}_{n_i,r_i}^{n_i^{\prime},r_i^{\prime}}
&=
\frac{1}{4}
e^{\pi i(n_0n_1^{\prime}+n_0^{\prime}n_1+r_0r_1^{\prime}+r_0^{\prime}r_1)
+\pi i(r_1-n_1)(n_1^{\prime}-r_1^{\prime})},
\nonumber\\
\mathcal{T}_{n_i,r_i}^{n_i^{\prime},r_i^{\prime}}
&=\delta_{n_i,n_i^{\prime}}\delta_{r_i,r_i^{\prime}}
e^{\pi i(n_0n_1+r_0r_1)-\frac{\pi i}{2}(n_1-r_1)^2}. 
\end{align}
This indicates that the (2+1)d topological order described by the K-matrix 
\begin{align}
\mathbf{K}=\begin{pmatrix}
2&2&-2&0\\
2&0&0&0\\
-2&0&2&2\\
0&0&2&0
\end{pmatrix}.
\end{align}
By an $SL(2,\mathbb{Z})$ similarity transformation, 
this $\mathrm{K}$-matrix is equivalent to $\mathbf{K}=2\sigma_z\oplus 2\sigma_z$,
which represents two copies of the double semion model.

Similarly, if we choose $(n_2,r_2)=(1,0)$ and $(0,1)$, 
the corresponding $(2+1)$d topological order is described by the K-matrix 
\begin{align}
\mathbf{K}=\begin{pmatrix}
0&2&-1&0\\
2&0&0&0\\
-1&0&2&2\\
0&0&2&0
\end{pmatrix},
\end{align}
and 
\begin{align}
\mathbf{K}=\begin{pmatrix}
2&2&-1&0\\
2&0&0&0\\
-1&0&0&2\\
0&0&2&0
\end{pmatrix}, 
\end{align}
respectively.
To summarize,
after dimensional reduction, 
the original $(3+1)$d topological order with non-trivial three-loop braiding statistics  
``splits'' into four different $(2+1)$d topological order, 
which are controlled by the quantum numbers $n_2$ and $r_2$.
This result seems to 
be related with the group cohomology classification of symmetry-protected topological (SPT) phases in $(2+1)$d
with $G=\mathbb{Z}_2\times \mathbb{Z}_2$ symmetry, 
i.e., $H^3[\mathbb{Z}_2\times\mathbb{Z}_2,U(1)]=\mathbb{Z}_2\times\mathbb{Z}_2$.

\section{Discussion}
\label{Discussion}

Let us summarize our main results.

-- In the (3+1)d BF theory, 
we have demonstrated, through explicit calculations in the boundary field theories
and by comparisons with known bulk results, 
there is a bulk-boundary correspondence in (3+1)d topological phases.
In particular the modular $\mathcal{S}$ and $\mathcal{T}$ matrices 
are calculated from the gapless boundary field theory 
and shown to match with the bulk results. 

-- The surface theory of the (3+1)d BF theory with the theta term is introduced and solved.  
The action of the modular $\mathcal{S}$ and $\mathcal{T}$ transformations on the partition
functions is calculated. It is shown that the partition functions do not form the complete basis 
under the modular $\mathcal{S}$ and $\mathcal{T}$ transformations. 

-- Finally, we propose a (3+1)d bulk field theory with cubic coupling
that may realize three-loop braiding statistics.
We discuss the twist that the cubic term of the field theory adds to the zero modes. 
By considering the alternative form of the bulk and boundary field theories, in which 
the quantization rule of the zero modes is twisted, 
we computed the surface partition functions,
and 
the $\mathcal{S}$ and $\mathcal{T}$ matrices are  constructed. 

These results extend the well-established bulk-boundary correspondence in (2+1)d topological phases and their (1+1)d edge theories. 
Our approach from the surface field theories provide an alternative point of view to (3+1)d topological phases,
and to recently discussed, novel braiding properties, such as 
three-loop braiding statistics.

There are, however, still some aspects  in the (2+1)d-(1+1)d correspondence,
which we do not know if have an analogue in the (3+1)d-(2+1)d correspondence.
For example, 
in the case of the bulk-boundary correspondence connecting
(2+1)d topological phases and (1+1)d edge theories, 
that the edge theories are invariant under an infinite-dimensional algebra 
seems to play a significant role: 
the Virasoro algebra or an extended chiral algebra of (1+1)d CFTs  
faithfully mirrors bulk topological properties of (2+1)d bulk phases. 
On the practical side, that edge theories enjoy an infinite-dimensional symmetry algebra
provides many non-trivial solvable examples. 
For our example of (2+1)d surface theories of (3+1)d topological phases, on the other hand, 
we did not make use of such infinite-dimensional symmetry. 
In fact, the surface theories studied in this paper are not conformal field theories.
For example, 
the two-point correlation function of the boson field
$\langle \phi(t,\mathsf{r}) \phi(t',\mathsf{r}')\rangle$
in the free boson theory in (2+1)d
decays algebraically.
This should be contrasted 
with the logarithmic decay 
of the corresponding correlator in the (1+1)d compactified boson theory.
As a consequence, the correlation functions of the bosonic exponents
$\exp[ i m \phi(t, \mathsf{r})]$ ($m\in \mathbb{Z}$)
do not decay algebraically in the (3+1)d free boson theory. 
Whether or not 
there exists a unified field theory framework in (2+1)d field theories
that strongly resonates with topological properties of (3+1)d bulk topological phases
requires further investigations.

\acknowledgements

We thank Chang-Tse Hsieh for sharing his notes with us. We thank
Chien-Hung Lin,
Michael Levin, 
C.\ W.\ von Keyserlingk and Peng Ye for useful discussion.
XC and AT acknowledge  
Prospects in Theoretical Physics 2015-Princeton Summer School on Condensed Matter Physics.
This work was supported in part by the National Science Foundation grant DMR-1408713 (XC)
and DMR-1455296 (SR) at the University of Illinois, 
and by Alfred P. Sloan foundation.

\appendix

\section{The surface theory of the BF theory with theta term}
\label{The surface theory of the BF theory with theta term}

In this appendix, we go through canonical analysis of the 
surface theory of the BF theory with theta term.
It is described by the Lagrangian density 
\begin{align}
\mathcal{{L}}&=\left[\frac{\mathrm{K}}{2\pi}\epsilon_{ij}\partial_{i}\zeta_{j}
+\frac{\mathrm{p}}{4\pi^{2}}\epsilon_{ij}\partial_{i}\Theta\partial_{j}\varphi\right]\partial_{t}\varphi
\nonumber \\
&\quad
-\frac{{1}}{2\lambda_{1}}(\epsilon_{ij}\partial_{i}\zeta_{j})^{2}-\frac{{1}}{2\lambda_{2}}G^{ij}\partial_{i}\varphi\partial_{j}\varphi.
\end{align}
In this theory, 
physical observables are bosonic exponents
$\exp [i m \varphi(t,\mathsf{r})]$,
and 
Wilson loops
$ \exp i m \int_C dx^i \zeta_i(t,\mathsf{r})$,
where 
$ m\in \mathbb{Z}$,
and $C$ is a closed loop.
The boson field $\varphi$ and gauge fields $\zeta_i$ are compactified accordingly.

The canonical commutators are 
\begin{align}
&
[\epsilon_{ij}\partial_i\zeta_j(t,\mathsf{r}),\epsilon_{lm}\partial_l\zeta_m(t,\mathsf{r}')]  =
\frac{{-2i\mathrm{p}}}{\mathrm{K}^{2}}\epsilon_{ij}\partial_{i}\Theta\partial_{j}^{\mathsf{{r}}}
\delta^{(2)}(\mathsf{{r}}-\mathsf{{r}}'), 
\nonumber \\
&
[\varphi(t,\mathsf{r}),\epsilon_{ij}\partial_i\zeta_j(t,\mathsf{r}')]  =
\frac{{2\pi i}}{\mathrm{K}}\delta^{(2)}(\mathsf{{r}}-\mathsf{{r}}').
\end{align}
The mode expansions consistent with the equations of motion  
are given by 
(only the oscillator parts are shown)
\begin{align}
&
\varphi(t,\mathsf{{r}})  =
\frac{1}{ \sqrt{2\mathrm{K}^2 \lambda_1 R_1 R_2 } }
\sum_{\mathsf{{k}}\neq0}
\frac{1}{ \sqrt{ \Lambda(\mathsf{k})} } 
e^{i\mathsf{{k}}\cdot\mathsf{{r}}}
\nonumber \\
&\quad\quad  
\times 
\left[a^{\dag}(\mathsf{{k}})
e^{i\Delta_{+}(\mathsf{{k}})t}+a(-\mathsf{{k}})e^{i\Delta_{-}(\mathsf{{k}})t}\right],
\nonumber \\
&\epsilon_{ij}\partial_i\zeta_{ij}(t,\mathsf{{r}})  =
\sqrt{\frac{\lambda_1}{8\pi^2 R_1 R_2}}
\sum_{\mathsf{{k}}\neq0}
\frac{1}{ \sqrt{\Lambda(\mathsf{k})} }
e^{i\mathsf{{k}}\cdot\mathsf{{r}}}
\nonumber \\
&\quad\quad  
\times 
\left[
i\Delta_{+}(\mathsf{{k}})a^{\dagger}(\mathsf{{k}})e^{i\Delta_{+}(\mathsf{{k}})t}
+i\Delta_{-}(\mathsf{{k}})a(-\mathsf{{k}})e^{i\Delta_{-}(\mathsf{{k}})t}
\right].
\end{align}
where
\begin{align}
\Omega^{2}(\mathsf{{k}}) & =\frac{{(2\pi)^{2}}}{\mathrm{K}^{2}\lambda_{1}\lambda_{2}}G^{ij}k_{i}k_{j},
\quad 
\Xi(\mathsf{{k}})  =\frac{\mathrm{p}}{\mathrm{K}^{2}\lambda_{1}}\epsilon_{ij}\partial_{i}\Theta k_{j}, 
\nonumber \\
\Lambda(\mathsf{k})^2 &=
\Xi(\mathsf{k})^2 + \Lambda(\mathsf{k})^2 
\nonumber \\
&=
\frac{{1}}{\mathrm{K}^{2}\lambda_{1}}\left(\frac{{(2\pi)^{2}}}{\lambda_{2}}G^{ij}
+\frac{{\mathrm{p}^{2}}}{\mathrm{K}^{2}\lambda_{1}}H^{ij}\right)k_{i}k_{j}
\nonumber \\
\Delta_{\pm}(\mathsf{{k}})&=
-\Xi(\mathsf{k})+\Lambda(\mathsf{k}),
\end{align}
where 
\begin{align}
H^{ij} & =\begin{pmatrix}Q_{2}^{2}/R_{2}^{2} & -Q_{1}Q_{2}/R_{1}R_{2}\\
-Q_{1}Q_{2}/R_{1}R_{2} & Q_{1}^{2}/R_{1}^{2}
\end{pmatrix}
\end{align}

We now attempt to derive the zero mode quantization rule solely from the surface theory. 
The canonical commuators among the zero modes are
\begin{align}
[\alpha_{\mu},\beta_{\nu}]=iM_{\mu\nu}^{-1}
\end{align}
where 
\begin{align}
 M& =
\begin{pmatrix}
\mathrm{K} & 0 & 0\\
-\mathrm{p}Q_{2} &  0& +\mathrm{K}\\
+\mathrm{p}Q_{1} & -\mathrm{K} &0
\end{pmatrix},
\quad
M^{-1}=
\begin{pmatrix}
\frac{{1}}{\mathrm{K}} & 0 & 0\\
\frac{{\mathrm{p}Q_{1}}}{\mathrm{K}^{2}} & 0 & -\frac{{1}}{\mathrm{K}}\\
\frac{\mathrm{p}Q_{2}}{\mathrm{K}^{2}} & \frac{1}{\mathrm{K}} &0 
\end{pmatrix}.
\end{align}
The commutators take 
a canonical form in the rotated basis 
\begin{align}
\tilde{{\beta}}_{\lambda}=\beta_{\nu}M_{\nu\lambda}
\quad
\Rightarrow 
\quad 
[\alpha_{\mu},\tilde{{\beta}}_{\lambda}]=i\delta_{\mu\lambda}. 
\end{align}
Thus, from the compactification condition on $\alpha_{\mu}$, 
the eigenvalues of $\tilde{\beta}$ are integers: 
\begin{align}
\tilde{{\beta}}_{\lambda}=m_{\lambda}
\in \mathbb{Z}.
\end{align}
This means the eigenvalues of $\beta$ are
\begin{align}
\beta_{\mu}&=\tilde{\beta}_{\lambda}M_{\lambda\mu}^{-1}
=m_{\lambda}M_{\lambda\mu}^{-1},
\nonumber \\
&=
\left(
\frac{{m_{0}}}{\mathrm{K}}+\frac{\mathrm{p}}{\mathrm{K}^{2}}(m_{1}Q_{1}+m_{2}Q_{2}),
\frac{{m_{2}}}{\mathrm{K}},
\frac{{-m_{1}}}{\mathrm{K}}
\right). 
\end{align}
Renaming integers by $m_0\to n_0$,
$m_1\to -n_2$, $m_2\to n_1$,
\begin{align}
(
\beta_{0}, \beta_{1}, \beta_{2}
)
=
\left(
\frac{{n_{0}}}{\mathrm{K}}-
\frac{\mathrm{p}}{\mathrm{K}^{2}}
(Q\times n),
\frac{{n_{1}}}{\mathrm{K}}, \frac{{n_{2}}}{\mathrm{K}}
\right). 
\end{align}
This is consistent with the 
quantization rules derived from the equations of motion 
in the bulk field theory.

\section{$\delta$-function forms}
\label{delta-function forms}

In this appendix, we summarize the properties of $\delta$-function forms. 
For an $n$-dimensional submanifold $\mathcal{N}$ of a $D$-dimensional maniofold $\mathcal{M}$,
we define a $(D-n)$-form $\delta_{D-n}(\mathcal{N})$ by
\begin{align}
\int_{\mathcal{N}} A_n = \int_{\mathcal{M}} \delta_{D-n}(\mathcal{N})\wedge A_n,
\quad
\forall A_n,
\end{align}
where $A_n$ is an arbitrary $n$-form on $\mathcal{M}$. 
If we flip the orientation of $\mathcal{N}$, 
\begin{align}
 \delta_{D-n}(-\mathcal{N}) = -\delta_{D-n}(\mathcal{N}).
\end{align}
More generally, for oriented submanifolds $\mathcal{N}_i$,
\begin{align}
\delta\big( \sum_i c_i \mathcal{N}_i\big)
=
\sum_i c_i \delta(\mathcal{N}_i)
\end{align}
where $c_i$ is a coefficient. 

The exterior derivative acts on the delta function form as 
\begin{align}
\delta_{D-n+1}(\partial \mathcal{N}) = (-1)^{D-n+1} d \delta_{D-n}(\mathcal{N}).
\label{formula delta function exterior derivative}
\end{align}

Let $\mathcal{N}_1$ and $\mathcal{N}_2$ be a submanifold of $\mathcal{M}$ with 
dimensions $n_1$ and $n_2$, respectively. 
Define $d$
as 
\begin{align}
d = n_1 + n_2 -D.
\end{align}
When $d\ge 0$, $\mathcal{N}_1$ and $\mathcal{N}_2$ can have a $d$-dimensional intersection within $\mathcal{M}$. 
By properly defining an orientation, we define
the intersection of $\mathcal{N}_1$ and $\mathcal{N}_2$, 
$I= \mathcal{N}_1 \# \mathcal{N}_2$.
The orientation of $I$ is defined to be consistent with
\begin{align}
\delta_{D-d}(I) = \delta_{D-n_1} (\mathcal{N}_1)\wedge \delta_{D-n_1}(\mathcal{N}_2). 
\end{align}
If the two submanifolds have complementary dimensions, $n_{1}+n_{2}=D$,
they intersect at points. 
Then, the intersection number 
\begin{align}
I(\mathcal{N}_{1},\mathcal{N}_{2})=\int\delta_{n}(\mathcal{N}_{1})\wedge\delta_{D-n}(\mathcal{N}_{2})\in\mathbb{{Z}}
\end{align}
counts the number of intersection points. 

The linking number of two submanifolds, $\mathcal{L}$ and $\mathcal{N}$, 
is defined when $\mathrm{dim}\,\mathcal{L}+\mathrm{dim}\, \mathcal{N} +1 =D$. 
By considering an auxiliary manifold satisfying $\partial \mathcal{N}_{1}=\mathcal{L}$, 
the linking number is given by 
\begin{align}
Lk(\mathcal{L},\mathcal{N})
=
I(\mathcal{N}_{1}, \mathcal{N}).
\end{align}

\bibliography{reference}

\begin{thebibliography}{59}%
\makeatletter
\providecommand \@ifxundefined [1]{%
 \@ifx{#1\undefined}
}%
\providecommand \@ifnum [1]{%
 \ifnum #1\expandafter \@firstoftwo
 \else \expandafter \@secondoftwo
 \fi
}%
\providecommand \@ifx [1]{%
 \ifx #1\expandafter \@firstoftwo
 \else \expandafter \@secondoftwo
 \fi
}%
\providecommand \natexlab [1]{#1}%
\providecommand \enquote  [1]{``#1''}%
\providecommand \bibnamefont  [1]{#1}%
\providecommand \bibfnamefont [1]{#1}%
\providecommand \citenamefont [1]{#1}%
\providecommand \href@noop [0]{\@secondoftwo}%
\providecommand \href [0]{\begingroup \@sanitize@url \@href}%
\providecommand \@href[1]{\@@startlink{#1}\@@href}%
\providecommand \@@href[1]{\endgroup#1\@@endlink}%
\providecommand \@sanitize@url [0]{\catcode `\\12\catcode `\$12\catcode
  `\&12\catcode `\#12\catcode `\^12\catcode `\_12\catcode `\%12\relax}%
\providecommand \@@startlink[1]{}%
\providecommand \@@endlink[0]{}%
\providecommand \url  [0]{\begingroup\@sanitize@url \@url }%
\providecommand \@url [1]{\endgroup\@href {#1}{\urlprefix }}%
\providecommand \urlprefix  [0]{URL }%
\providecommand \Eprint [0]{\href }%
\providecommand \doibase [0]{http://dx.doi.org/}%
\providecommand \selectlanguage [0]{\@gobble}%
\providecommand \bibinfo  [0]{\@secondoftwo}%
\providecommand \bibfield  [0]{\@secondoftwo}%
\providecommand \translation [1]{[#1]}%
\providecommand \BibitemOpen [0]{}%
\providecommand \bibitemStop [0]{}%
\providecommand \bibitemNoStop [0]{.\EOS\space}%
\providecommand \EOS [0]{\spacefactor3000\relax}%
\providecommand \BibitemShut  [1]{\csname bibitem#1\endcsname}%
\let\auto@bib@innerbib\@empty
\bibitem [{\citenamefont {{Halperin}}(1982)}]{Halperin1982}%
  \BibitemOpen
  \bibfield  {author} {\bibinfo {author} {\bibfnamefont {B.~I.}\ \bibnamefont
  {{Halperin}}},\ }\href {\doibase 10.1103/PhysRevB.25.2185} {\bibfield
  {journal} {\bibinfo  {journal} {\prb}\ }\textbf {\bibinfo {volume} {25}},\
  \bibinfo {pages} {2185} (\bibinfo {year} {1982})}\BibitemShut {NoStop}%
\bibitem [{\citenamefont {{Witten}}(1989)}]{Witten1989}%
  \BibitemOpen
  \bibfield  {author} {\bibinfo {author} {\bibfnamefont {E.}~\bibnamefont
  {{Witten}}},\ }\href {\doibase 10.1007/BF01217730} {\bibfield  {journal}
  {\bibinfo  {journal} {Communications in Mathematical Physics}\ }\textbf
  {\bibinfo {volume} {121}},\ \bibinfo {pages} {351} (\bibinfo {year}
  {1989})}\BibitemShut {NoStop}%
\bibitem [{\citenamefont {Wen}(1992)}]{Wen1992}%
  \BibitemOpen
  \bibfield  {author} {\bibinfo {author} {\bibfnamefont {X.-G.}\ \bibnamefont
  {Wen}},\ }\href {\doibase 10.1142/S0217979292000840} {\bibfield  {journal}
  {\bibinfo  {journal} {Int. J. Mod. Phys.}\ }\textbf {\bibinfo {volume}
  {B6}},\ \bibinfo {pages} {1711} (\bibinfo {year} {1992})}\BibitemShut
  {NoStop}%
\bibitem [{\citenamefont {{Hatsugai}}(1993)}]{Hatsugai1993}%
  \BibitemOpen
  \bibfield  {author} {\bibinfo {author} {\bibfnamefont {Y.}~\bibnamefont
  {{Hatsugai}}},\ }\href {\doibase 10.1103/PhysRevLett.71.3697} {\bibfield
  {journal} {\bibinfo  {journal} {Physical Review Letters}\ }\textbf {\bibinfo
  {volume} {71}},\ \bibinfo {pages} {3697} (\bibinfo {year}
  {1993})}\BibitemShut {NoStop}%
\bibitem [{\citenamefont {Cappelli}\ \emph {et~al.}(2002)\citenamefont
  {Cappelli}, \citenamefont {Huerta},\ and\ \citenamefont
  {Zemba}}]{Cappelli01}%
  \BibitemOpen
  \bibfield  {author} {\bibinfo {author} {\bibfnamefont {A.}~\bibnamefont
  {Cappelli}}, \bibinfo {author} {\bibfnamefont {M.}~\bibnamefont {Huerta}}, \
  and\ \bibinfo {author} {\bibfnamefont {G.~R.}\ \bibnamefont {Zemba}},\
  }\href@noop {} {\bibfield  {journal} {\bibinfo  {journal} {Nucl.\ Phys.\ B}\
  }\textbf {\bibinfo {volume} {636}},\ \bibinfo {pages} {568} (\bibinfo {year}
  {2002})}\BibitemShut {NoStop}%
\bibitem [{\citenamefont {Cappelli}\ and\ \citenamefont
  {Zemba}(1997)}]{Cappelli96}%
  \BibitemOpen
  \bibfield  {author} {\bibinfo {author} {\bibfnamefont {A.}~\bibnamefont
  {Cappelli}}\ and\ \bibinfo {author} {\bibfnamefont {G.~R.}\ \bibnamefont
  {Zemba}},\ }\href@noop {} {\bibfield  {journal} {\bibinfo  {journal} {Nucl.
  Phys. B}\ }\textbf {\bibinfo {volume} {490}},\ \bibinfo {pages} {595}
  (\bibinfo {year} {1997})}\BibitemShut {NoStop}%
\bibitem [{\citenamefont {Cappelli}\ \emph {et~al.}(2010)\citenamefont
  {Cappelli}, \citenamefont {Viola},\ and\ \citenamefont {Zemba}}]{Cappelli10}%
  \BibitemOpen
  \bibfield  {author} {\bibinfo {author} {\bibfnamefont {A.}~\bibnamefont
  {Cappelli}}, \bibinfo {author} {\bibfnamefont {G.}~\bibnamefont {Viola}}, \
  and\ \bibinfo {author} {\bibfnamefont {G.~R.}\ \bibnamefont {Zemba}},\
  }\href@noop {} {\bibfield  {journal} {\bibinfo  {journal} {Annals of
  Physics}\ }\textbf {\bibinfo {volume} {325}},\ \bibinfo {pages} {465}
  (\bibinfo {year} {2010})}\BibitemShut {NoStop}%
\bibitem [{\citenamefont {Cappelli}\ and\ \citenamefont
  {Viola}(2011)}]{Cappelli11}%
  \BibitemOpen
  \bibfield  {author} {\bibinfo {author} {\bibfnamefont {A.}~\bibnamefont
  {Cappelli}}\ and\ \bibinfo {author} {\bibfnamefont {G.}~\bibnamefont
  {Viola}},\ }\href@noop {} {\bibfield  {journal} {\bibinfo  {journal} {J.
  Phys. A}\ }\textbf {\bibinfo {volume} {44}},\ \bibinfo {pages} {075401}
  (\bibinfo {year} {2011})}\BibitemShut {NoStop}%
\bibitem [{\citenamefont {Ryu}\ and\ \citenamefont {Zhang}(2012)}]{Ryu2012}%
  \BibitemOpen
  \bibfield  {author} {\bibinfo {author} {\bibfnamefont {S.}~\bibnamefont
  {Ryu}}\ and\ \bibinfo {author} {\bibfnamefont {S.-C.}\ \bibnamefont
  {Zhang}},\ }\href@noop {} {\bibfield  {journal} {\bibinfo  {journal} {Phys.
  Rev. B}\ }\textbf {\bibinfo {volume} {85}},\ \bibinfo {pages} {245132}
  (\bibinfo {year} {2012})}\BibitemShut {NoStop}%
\bibitem [{\citenamefont {Sule}\ \emph {et~al.}(2013)\citenamefont {Sule},
  \citenamefont {Chen},\ and\ \citenamefont {Ryu}}]{Sule13}%
  \BibitemOpen
  \bibfield  {author} {\bibinfo {author} {\bibfnamefont {O.~M.}\ \bibnamefont
  {Sule}}, \bibinfo {author} {\bibfnamefont {X.}~\bibnamefont {Chen}}, \ and\
  \bibinfo {author} {\bibfnamefont {S.}~\bibnamefont {Ryu}},\ }\href@noop {}
  {\bibfield  {journal} {\bibinfo  {journal} {Phys.\ Rev.\ B}\ }\textbf
  {\bibinfo {volume} {88}},\ \bibinfo {pages} {075125} (\bibinfo {year}
  {2013})}\BibitemShut {NoStop}%
\bibitem [{\citenamefont {{Cho}}\ \emph {et~al.}(2014)\citenamefont {{Cho}},
  \citenamefont {{Teo}},\ and\ \citenamefont {{Ryu}}}]{Cho2014}%
  \BibitemOpen
  \bibfield  {author} {\bibinfo {author} {\bibfnamefont {G.~Y.}\ \bibnamefont
  {{Cho}}}, \bibinfo {author} {\bibfnamefont {J.~C.~Y.}\ \bibnamefont {{Teo}}},
  \ and\ \bibinfo {author} {\bibfnamefont {S.}~\bibnamefont {{Ryu}}},\ }\href
  {\doibase 10.1103/PhysRevB.89.235103} {\bibfield  {journal} {\bibinfo
  {journal} {\prb}\ }\textbf {\bibinfo {volume} {89}},\ \bibinfo {eid} {235103}
  (\bibinfo {year} {2014})},\ \Eprint {http://arxiv.org/abs/1403.2018}
  {arXiv:1403.2018 [cond-mat.str-el]} \BibitemShut {NoStop}%
\bibitem [{\citenamefont {{Hsieh}}\ \emph {et~al.}(2014)\citenamefont
  {{Hsieh}}, \citenamefont {{Sule}}, \citenamefont {{Cho}}, \citenamefont
  {{Ryu}},\ and\ \citenamefont {{Leigh}}}]{Hsieh2014a}%
  \BibitemOpen
  \bibfield  {author} {\bibinfo {author} {\bibfnamefont {C.-T.}\ \bibnamefont
  {{Hsieh}}}, \bibinfo {author} {\bibfnamefont {O.~M.}\ \bibnamefont {{Sule}}},
  \bibinfo {author} {\bibfnamefont {G.~Y.}\ \bibnamefont {{Cho}}}, \bibinfo
  {author} {\bibfnamefont {S.}~\bibnamefont {{Ryu}}}, \ and\ \bibinfo {author}
  {\bibfnamefont {R.~G.}\ \bibnamefont {{Leigh}}},\ }\href {\doibase
  10.1103/PhysRevB.90.165134} {\bibfield  {journal} {\bibinfo  {journal}
  {\prb}\ }\textbf {\bibinfo {volume} {90}},\ \bibinfo {eid} {165134} (\bibinfo
  {year} {2014})},\ \Eprint {http://arxiv.org/abs/1403.6902} {arXiv:1403.6902
  [cond-mat.str-el]} \BibitemShut {NoStop}%
\bibitem [{\citenamefont {Lu}\ and\ \citenamefont {Vishwanath}(2012)}]{Lu2012}%
  \BibitemOpen
  \bibfield  {author} {\bibinfo {author} {\bibfnamefont {Y.-M.}\ \bibnamefont
  {Lu}}\ and\ \bibinfo {author} {\bibfnamefont {A.}~\bibnamefont
  {Vishwanath}},\ }\href@noop {} {\bibfield  {journal} {\bibinfo  {journal}
  {Phys. Rev. B}\ }\textbf {\bibinfo {volume} {86}},\ \bibinfo {pages} {125119}
  (\bibinfo {year} {2012})}\BibitemShut {NoStop}%
\bibitem [{\citenamefont {Cappelli}\ and\ \citenamefont
  {Randellini}(2013)}]{Cappelli13}%
  \BibitemOpen
  \bibfield  {author} {\bibinfo {author} {\bibfnamefont {A.}~\bibnamefont
  {Cappelli}}\ and\ \bibinfo {author} {\bibfnamefont {E.}~\bibnamefont
  {Randellini}},\ }\href@noop {} {\bibfield  {journal} {\bibinfo  {journal}
  {JHEP}\ }\textbf {\bibinfo {volume} {12}},\ \bibinfo {pages} {101} (\bibinfo
  {year} {2013})}\BibitemShut {NoStop}%
\bibitem [{\citenamefont {Hsieh}\ \emph {et~al.}(2016)\citenamefont {Hsieh},
  \citenamefont {Cho},\ and\ \citenamefont {Ryu}}]{Hsieh2015}%
  \BibitemOpen
  \bibfield  {author} {\bibinfo {author} {\bibfnamefont {C.-T.}\ \bibnamefont
  {Hsieh}}, \bibinfo {author} {\bibfnamefont {G.~Y.}\ \bibnamefont {Cho}}, \
  and\ \bibinfo {author} {\bibfnamefont {S.}~\bibnamefont {Ryu}},\ }\href
  {\doibase 10.1103/PhysRevB.93.075135} {\bibfield  {journal} {\bibinfo
  {journal} {Phys. Rev. B}\ }\textbf {\bibinfo {volume} {93}},\ \bibinfo
  {pages} {075135} (\bibinfo {year} {2016})}\BibitemShut {NoStop}%
\bibitem [{\citenamefont {{Horowitz}}(1989)}]{Horowitz1989}%
  \BibitemOpen
  \bibfield  {author} {\bibinfo {author} {\bibfnamefont {G.~T.}\ \bibnamefont
  {{Horowitz}}},\ }\href {\doibase 10.1007/BF01218410} {\bibfield  {journal}
  {\bibinfo  {journal} {Communications in Mathematical Physics}\ }\textbf
  {\bibinfo {volume} {125}},\ \bibinfo {pages} {417} (\bibinfo {year}
  {1989})}\BibitemShut {NoStop}%
\bibitem [{\citenamefont {{Horowitz}}\ and\ \citenamefont
  {{Srednicki}}(1990)}]{Horowitz1990}%
  \BibitemOpen
  \bibfield  {author} {\bibinfo {author} {\bibfnamefont {G.~T.}\ \bibnamefont
  {{Horowitz}}}\ and\ \bibinfo {author} {\bibfnamefont {M.}~\bibnamefont
  {{Srednicki}}},\ }\href {\doibase 10.1007/BF02099875} {\bibfield  {journal}
  {\bibinfo  {journal} {Communications in Mathematical Physics}\ }\textbf
  {\bibinfo {volume} {130}},\ \bibinfo {pages} {83} (\bibinfo {year}
  {1990})}\BibitemShut {NoStop}%
\bibitem [{\citenamefont {Blau}\ and\ \citenamefont
  {Thompson}(1991)}]{Blau:1989bq}%
  \BibitemOpen
  \bibfield  {author} {\bibinfo {author} {\bibfnamefont {M.}~\bibnamefont
  {Blau}}\ and\ \bibinfo {author} {\bibfnamefont {G.}~\bibnamefont
  {Thompson}},\ }\href {\doibase 10.1016/0003-4916(91)90240-9} {\bibfield
  {journal} {\bibinfo  {journal} {Annals Phys.}\ }\textbf {\bibinfo {volume}
  {205}},\ \bibinfo {pages} {130} (\bibinfo {year} {1991})}\BibitemShut
  {NoStop}%
\bibitem [{\citenamefont {Blau}\ and\ \citenamefont
  {Thompson}(1989)}]{Blau:1989dh}%
  \BibitemOpen
  \bibfield  {author} {\bibinfo {author} {\bibfnamefont {M.}~\bibnamefont
  {Blau}}\ and\ \bibinfo {author} {\bibfnamefont {G.}~\bibnamefont
  {Thompson}},\ }\href {\doibase 10.1016/0370-2693(89)90526-1} {\bibfield
  {journal} {\bibinfo  {journal} {Phys. Lett.}\ }\textbf {\bibinfo {volume}
  {B228}},\ \bibinfo {pages} {64} (\bibinfo {year} {1989})}\BibitemShut
  {NoStop}%
\bibitem [{\citenamefont {Birmingham}\ \emph {et~al.}(1991)\citenamefont
  {Birmingham}, \citenamefont {Blau}, \citenamefont {Rakowski},\ and\
  \citenamefont {Thompson}}]{Birmingham1991}%
  \BibitemOpen
  \bibfield  {author} {\bibinfo {author} {\bibfnamefont {D.}~\bibnamefont
  {Birmingham}}, \bibinfo {author} {\bibfnamefont {M.}~\bibnamefont {Blau}},
  \bibinfo {author} {\bibfnamefont {M.}~\bibnamefont {Rakowski}}, \ and\
  \bibinfo {author} {\bibfnamefont {G.}~\bibnamefont {Thompson}},\ }\href
  {\doibase 10.1016/0370-1573(91)90117-5} {\bibfield  {journal} {\bibinfo
  {journal} {Phys. Rept.}\ }\textbf {\bibinfo {volume} {209}},\ \bibinfo
  {pages} {129} (\bibinfo {year} {1991})}\BibitemShut {NoStop}%
\bibitem [{\citenamefont {Oda}\ and\ \citenamefont
  {Yahikozawa}(1990)}]{Oda1989}%
  \BibitemOpen
  \bibfield  {author} {\bibinfo {author} {\bibfnamefont {I.}~\bibnamefont
  {Oda}}\ and\ \bibinfo {author} {\bibfnamefont {S.}~\bibnamefont
  {Yahikozawa}},\ }\href {\doibase 10.1016/0370-2693(90)91735-T} {\bibfield
  {journal} {\bibinfo  {journal} {Phys. Lett.}\ }\textbf {\bibinfo {volume}
  {B238}},\ \bibinfo {pages} {272} (\bibinfo {year} {1990})}\BibitemShut
  {NoStop}%
\bibitem [{\citenamefont {{Bergeron}}\ \emph {et~al.}(1995)\citenamefont
  {{Bergeron}}, \citenamefont {{Semenoff}},\ and\ \citenamefont
  {{Szabo}}}]{BergeronSemenoffSzabo1995}%
  \BibitemOpen
  \bibfield  {author} {\bibinfo {author} {\bibfnamefont {M.}~\bibnamefont
  {{Bergeron}}}, \bibinfo {author} {\bibfnamefont {G.~W.}\ \bibnamefont
  {{Semenoff}}}, \ and\ \bibinfo {author} {\bibfnamefont {R.~J.}\ \bibnamefont
  {{Szabo}}},\ }\href {\doibase 10.1016/0550-3213(94)00503-7} {\bibfield
  {journal} {\bibinfo  {journal} {Nuclear Physics B}\ }\textbf {\bibinfo
  {volume} {437}},\ \bibinfo {pages} {695} (\bibinfo {year} {1995})},\ \Eprint
  {http://arxiv.org/abs/hep-th/9407020} {hep-th/9407020} \BibitemShut {NoStop}%
\bibitem [{\citenamefont {{Hansson}}\ \emph {et~al.}(2004)\citenamefont
  {{Hansson}}, \citenamefont {{Oganesyan}},\ and\ \citenamefont
  {{Sondhi}}}]{HanssonOganesyanSondhi2004}%
  \BibitemOpen
  \bibfield  {author} {\bibinfo {author} {\bibfnamefont {T.~H.}\ \bibnamefont
  {{Hansson}}}, \bibinfo {author} {\bibfnamefont {V.}~\bibnamefont
  {{Oganesyan}}}, \ and\ \bibinfo {author} {\bibfnamefont {S.~L.}\ \bibnamefont
  {{Sondhi}}},\ }\href {\doibase 10.1016/j.aop.2004.05.006} {\bibfield
  {journal} {\bibinfo  {journal} {Annals of Physics}\ }\textbf {\bibinfo
  {volume} {313}},\ \bibinfo {pages} {497} (\bibinfo {year} {2004})},\ \Eprint
  {http://arxiv.org/abs/cond-mat/0404327} {cond-mat/0404327} \BibitemShut
  {NoStop}%
\bibitem [{\citenamefont {{Banks}}\ and\ \citenamefont
  {{Seiberg}}(2011)}]{Banks2011}%
  \BibitemOpen
  \bibfield  {author} {\bibinfo {author} {\bibfnamefont {T.}~\bibnamefont
  {{Banks}}}\ and\ \bibinfo {author} {\bibfnamefont {N.}~\bibnamefont
  {{Seiberg}}},\ }\href {\doibase 10.1103/PhysRevD.83.084019} {\bibfield
  {journal} {\bibinfo  {journal} {\prd}\ }\textbf {\bibinfo {volume} {83}},\
  \bibinfo {eid} {084019} (\bibinfo {year} {2011})},\ \Eprint
  {http://arxiv.org/abs/1011.5120} {arXiv:1011.5120 [hep-th]} \BibitemShut
  {NoStop}%
\bibitem [{\citenamefont {{Cho}}\ and\ \citenamefont
  {{Moore}}(2011)}]{ChoMoore2011}%
  \BibitemOpen
  \bibfield  {author} {\bibinfo {author} {\bibfnamefont {G.~Y.}\ \bibnamefont
  {{Cho}}}\ and\ \bibinfo {author} {\bibfnamefont {J.~E.}\ \bibnamefont
  {{Moore}}},\ }\href {\doibase 10.1016/j.aop.2010.12.011} {\bibfield
  {journal} {\bibinfo  {journal} {Annals of Physics}\ }\textbf {\bibinfo
  {volume} {326}},\ \bibinfo {pages} {1515} (\bibinfo {year} {2011})},\ \Eprint
  {http://arxiv.org/abs/1011.3485} {arXiv:1011.3485 [cond-mat.str-el]}
  \BibitemShut {NoStop}%
\bibitem [{\citenamefont {{Vishwanath}}\ and\ \citenamefont
  {{Senthil}}(2013)}]{Vishwanath2013}%
  \BibitemOpen
  \bibfield  {author} {\bibinfo {author} {\bibfnamefont {A.}~\bibnamefont
  {{Vishwanath}}}\ and\ \bibinfo {author} {\bibfnamefont {T.}~\bibnamefont
  {{Senthil}}},\ }\href {\doibase 10.1103/PhysRevX.3.011016} {\bibfield
  {journal} {\bibinfo  {journal} {Physical Review X}\ }\textbf {\bibinfo
  {volume} {3}},\ \bibinfo {eid} {011016} (\bibinfo {year} {2013})},\ \Eprint
  {http://arxiv.org/abs/1209.3058} {arXiv:1209.3058 [cond-mat.str-el]}
  \BibitemShut {NoStop}%
\bibitem [{\citenamefont {{Chan}}\ \emph {et~al.}(2013)\citenamefont {{Chan}},
  \citenamefont {{Hughes}}, \citenamefont {{Ryu}},\ and\ \citenamefont
  {{Fradkin}}}]{Chan2013}%
  \BibitemOpen
  \bibfield  {author} {\bibinfo {author} {\bibfnamefont {A.}~\bibnamefont
  {{Chan}}}, \bibinfo {author} {\bibfnamefont {T.~L.}\ \bibnamefont
  {{Hughes}}}, \bibinfo {author} {\bibfnamefont {S.}~\bibnamefont {{Ryu}}}, \
  and\ \bibinfo {author} {\bibfnamefont {E.}~\bibnamefont {{Fradkin}}},\ }\href
  {\doibase 10.1103/PhysRevB.87.085132} {\bibfield  {journal} {\bibinfo
  {journal} {\prb}\ }\textbf {\bibinfo {volume} {87}},\ \bibinfo {eid} {085132}
  (\bibinfo {year} {2013})},\ \Eprint {http://arxiv.org/abs/1210.4305}
  {arXiv:1210.4305 [cond-mat.str-el]} \BibitemShut {NoStop}%
\bibitem [{\citenamefont {{Tiwari}}\ \emph {et~al.}(2014)\citenamefont
  {{Tiwari}}, \citenamefont {{Chen}}, \citenamefont {{Neupert}}, \citenamefont
  {{Santos}}, \citenamefont {{Ryu}}, \citenamefont {{Chamon}},\ and\
  \citenamefont {{Mudry}}}]{Tiwari2014}%
  \BibitemOpen
  \bibfield  {author} {\bibinfo {author} {\bibfnamefont {A.}~\bibnamefont
  {{Tiwari}}}, \bibinfo {author} {\bibfnamefont {X.}~\bibnamefont {{Chen}}},
  \bibinfo {author} {\bibfnamefont {T.}~\bibnamefont {{Neupert}}}, \bibinfo
  {author} {\bibfnamefont {L.~H.}\ \bibnamefont {{Santos}}}, \bibinfo {author}
  {\bibfnamefont {S.}~\bibnamefont {{Ryu}}}, \bibinfo {author} {\bibfnamefont
  {C.}~\bibnamefont {{Chamon}}}, \ and\ \bibinfo {author} {\bibfnamefont
  {C.}~\bibnamefont {{Mudry}}},\ }\href {\doibase 10.1103/PhysRevB.90.235118}
  {\bibfield  {journal} {\bibinfo  {journal} {\prb}\ }\textbf {\bibinfo
  {volume} {90}},\ \bibinfo {eid} {235118} (\bibinfo {year} {2014})},\ \Eprint
  {http://arxiv.org/abs/1408.5417} {arXiv:1408.5417 [cond-mat.mes-hall]}
  \BibitemShut {NoStop}%
\bibitem [{\citenamefont {{Ye}}\ and\ \citenamefont
  {{Gu}}(2015{\natexlab{a}})}]{Ye2015}%
  \BibitemOpen
  \bibfield  {author} {\bibinfo {author} {\bibfnamefont {P.}~\bibnamefont
  {{Ye}}}\ and\ \bibinfo {author} {\bibfnamefont {Z.-C.}\ \bibnamefont
  {{Gu}}},\ }\href {\doibase 10.1103/PhysRevX.5.021029} {\bibfield  {journal}
  {\bibinfo  {journal} {Physical Review X}\ }\textbf {\bibinfo {volume} {5}},\
  \bibinfo {eid} {021029} (\bibinfo {year} {2015}{\natexlab{a}})},\ \Eprint
  {http://arxiv.org/abs/1410.2594} {arXiv:1410.2594 [cond-mat.str-el]}
  \BibitemShut {NoStop}%
\bibitem [{\citenamefont {{Gaiotto}}\ \emph {et~al.}(2015)\citenamefont
  {{Gaiotto}}, \citenamefont {{Kapustin}}, \citenamefont {{Seiberg}},\ and\
  \citenamefont {{Willett}}}]{Gaiotto2015}%
  \BibitemOpen
  \bibfield  {author} {\bibinfo {author} {\bibfnamefont {D.}~\bibnamefont
  {{Gaiotto}}}, \bibinfo {author} {\bibfnamefont {A.}~\bibnamefont
  {{Kapustin}}}, \bibinfo {author} {\bibfnamefont {N.}~\bibnamefont
  {{Seiberg}}}, \ and\ \bibinfo {author} {\bibfnamefont {B.}~\bibnamefont
  {{Willett}}},\ }\href {\doibase 10.1007/JHEP02(2015)172} {\bibfield
  {journal} {\bibinfo  {journal} {Journal of High Energy Physics}\ }\textbf
  {\bibinfo {volume} {2}},\ \bibinfo {pages} {172} (\bibinfo {year} {2015})},\
  \Eprint {http://arxiv.org/abs/1412.5148} {arXiv:1412.5148 [hep-th]}
  \BibitemShut {NoStop}%
\bibitem [{\citenamefont {Cirio}\ \emph {et~al.}(2014)\citenamefont {Cirio},
  \citenamefont {Palumbo},\ and\ \citenamefont {Pachos}}]{Cirio2014}%
  \BibitemOpen
  \bibfield  {author} {\bibinfo {author} {\bibfnamefont {M.}~\bibnamefont
  {Cirio}}, \bibinfo {author} {\bibfnamefont {G.}~\bibnamefont {Palumbo}}, \
  and\ \bibinfo {author} {\bibfnamefont {J.~K.}\ \bibnamefont {Pachos}},\
  }\href@noop {} {\bibfield  {journal} {\bibinfo  {journal} {Phys. Rev. B}\
  }\textbf {\bibinfo {volume} {90}},\ \bibinfo {pages} {085114} (\bibinfo
  {year} {2014})}\BibitemShut {NoStop}%
\bibitem [{\citenamefont {Francesco}\ \emph {et~al.}(1997)\citenamefont
  {Francesco}, \citenamefont {Mathieu},\ and\ \citenamefont
  {Senechal}}]{FMS-CFT}%
  \BibitemOpen
  \bibfield  {author} {\bibinfo {author} {\bibfnamefont {P.~D.}\ \bibnamefont
  {Francesco}}, \bibinfo {author} {\bibfnamefont {P.}~\bibnamefont {Mathieu}},
  \ and\ \bibinfo {author} {\bibfnamefont {D.}~\bibnamefont {Senechal}},\
  }\href@noop {} {\emph {\bibinfo {title} {{Conformal Field Theory}}}}\
  (\bibinfo  {publisher} {Springer-Verlag New York},\ \bibinfo {year}
  {1997})\BibitemShut {NoStop}%
\bibitem [{\citenamefont {{Moradi}}\ and\ \citenamefont
  {{Wen}}(2015)}]{MoradiWen2015}%
  \BibitemOpen
  \bibfield  {author} {\bibinfo {author} {\bibfnamefont {H.}~\bibnamefont
  {{Moradi}}}\ and\ \bibinfo {author} {\bibfnamefont {X.-G.}\ \bibnamefont
  {{Wen}}},\ }\href {\doibase 10.1103/PhysRevB.91.075114} {\bibfield  {journal}
  {\bibinfo  {journal} {\prb}\ }\textbf {\bibinfo {volume} {91}},\ \bibinfo
  {eid} {075114} (\bibinfo {year} {2015})},\ \Eprint
  {http://arxiv.org/abs/1404.4618} {arXiv:1404.4618 [cond-mat.str-el]}
  \BibitemShut {NoStop}%
\bibitem [{\citenamefont {{Jiang}}\ \emph {et~al.}(2014)\citenamefont
  {{Jiang}}, \citenamefont {{Mesaros}},\ and\ \citenamefont
  {{Ran}}}]{JiangMesarosRan2014}%
  \BibitemOpen
  \bibfield  {author} {\bibinfo {author} {\bibfnamefont {S.}~\bibnamefont
  {{Jiang}}}, \bibinfo {author} {\bibfnamefont {A.}~\bibnamefont {{Mesaros}}},
  \ and\ \bibinfo {author} {\bibfnamefont {Y.}~\bibnamefont {{Ran}}},\ }\href
  {\doibase 10.1103/PhysRevX.4.031048} {\bibfield  {journal} {\bibinfo
  {journal} {Physical Review X}\ }\textbf {\bibinfo {volume} {4}},\ \bibinfo
  {eid} {031048} (\bibinfo {year} {2014})},\ \Eprint
  {http://arxiv.org/abs/1404.1062} {arXiv:1404.1062 [cond-mat.str-el]}
  \BibitemShut {NoStop}%
\bibitem [{\citenamefont {{Wang}}\ and\ \citenamefont
  {{Levin}}(2014)}]{WangLevin2014}%
  \BibitemOpen
  \bibfield  {author} {\bibinfo {author} {\bibfnamefont {C.}~\bibnamefont
  {{Wang}}}\ and\ \bibinfo {author} {\bibfnamefont {M.}~\bibnamefont
  {{Levin}}},\ }\href {\doibase 10.1103/PhysRevLett.113.080403} {\bibfield
  {journal} {\bibinfo  {journal} {Physical Review Letters}\ }\textbf {\bibinfo
  {volume} {113}},\ \bibinfo {eid} {080403} (\bibinfo {year} {2014})},\ \Eprint
  {http://arxiv.org/abs/1403.7437} {arXiv:1403.7437 [cond-mat.str-el]}
  \BibitemShut {NoStop}%
\bibitem [{\citenamefont {{Wang}}\ and\ \citenamefont
  {{Wen}}(2015)}]{WangWen2015}%
  \BibitemOpen
  \bibfield  {author} {\bibinfo {author} {\bibfnamefont {J.~C.}\ \bibnamefont
  {{Wang}}}\ and\ \bibinfo {author} {\bibfnamefont {X.-G.}\ \bibnamefont
  {{Wen}}},\ }\href {\doibase 10.1103/PhysRevB.91.035134} {\bibfield  {journal}
  {\bibinfo  {journal} {\prb}\ }\textbf {\bibinfo {volume} {91}},\ \bibinfo
  {eid} {035134} (\bibinfo {year} {2015})},\ \Eprint
  {http://arxiv.org/abs/1404.7854} {arXiv:1404.7854 [cond-mat.str-el]}
  \BibitemShut {NoStop}%
\bibitem [{\citenamefont {{Jian}}\ and\ \citenamefont
  {{Qi}}(2014)}]{JianQi2014}%
  \BibitemOpen
  \bibfield  {author} {\bibinfo {author} {\bibfnamefont {C.-M.}\ \bibnamefont
  {{Jian}}}\ and\ \bibinfo {author} {\bibfnamefont {X.-L.}\ \bibnamefont
  {{Qi}}},\ }\href {\doibase 10.1103/PhysRevX.4.041043} {\bibfield  {journal}
  {\bibinfo  {journal} {Physical Review X}\ }\textbf {\bibinfo {volume} {4}},\
  \bibinfo {eid} {041043} (\bibinfo {year} {2014})},\ \Eprint
  {http://arxiv.org/abs/1405.6688} {arXiv:1405.6688 [cond-mat.str-el]}
  \BibitemShut {NoStop}%
\bibitem [{\citenamefont {{Wang}}\ and\ \citenamefont
  {{Levin}}(2015)}]{WangLevin2015}%
  \BibitemOpen
  \bibfield  {author} {\bibinfo {author} {\bibfnamefont {C.}~\bibnamefont
  {{Wang}}}\ and\ \bibinfo {author} {\bibfnamefont {M.}~\bibnamefont
  {{Levin}}},\ }\href {\doibase 10.1103/PhysRevB.91.165119} {\bibfield
  {journal} {\bibinfo  {journal} {\prb}\ }\textbf {\bibinfo {volume} {91}},\
  \bibinfo {eid} {165119} (\bibinfo {year} {2015})},\ \Eprint
  {http://arxiv.org/abs/1412.1781} {arXiv:1412.1781 [cond-mat.str-el]}
  \BibitemShut {NoStop}%
\bibitem [{\citenamefont {{Lin}}\ and\ \citenamefont
  {{Levin}}(2015)}]{LinLevin2015}%
  \BibitemOpen
  \bibfield  {author} {\bibinfo {author} {\bibfnamefont {C.-H.}\ \bibnamefont
  {{Lin}}}\ and\ \bibinfo {author} {\bibfnamefont {M.}~\bibnamefont
  {{Levin}}},\ }\href@noop {} {\bibfield  {journal} {\bibinfo  {journal} {ArXiv
  e-prints}\ } (\bibinfo {year} {2015})},\ \Eprint
  {http://arxiv.org/abs/1503.00142} {arXiv:1503.00142 [cond-mat.str-el]}
  \BibitemShut {NoStop}%
\bibitem [{\citenamefont {Wan}\ \emph {et~al.}(2015{\natexlab{a}})\citenamefont
  {Wan}, \citenamefont {Wang},\ and\ \citenamefont {He}}]{Wan2015}%
  \BibitemOpen
  \bibfield  {author} {\bibinfo {author} {\bibfnamefont {Y.}~\bibnamefont
  {Wan}}, \bibinfo {author} {\bibfnamefont {J.}~\bibnamefont {Wang}}, \ and\
  \bibinfo {author} {\bibfnamefont {H.}~\bibnamefont {He}},\ }\href@noop {}
  {\bibfield  {journal} {\bibinfo  {journal} {Phys. Rev. B}\ }\textbf {\bibinfo
  {volume} {92}},\ \bibinfo {pages} {045101} (\bibinfo {year}
  {2015}{\natexlab{a}})}\BibitemShut {NoStop}%
\bibitem [{\citenamefont {{Cano}}\ \emph {et~al.}(2014)\citenamefont {{Cano}},
  \citenamefont {{Cheng}}, \citenamefont {{Mulligan}}, \citenamefont {{Nayak}},
  \citenamefont {{Plamadeala}},\ and\ \citenamefont {{Yard}}}]{Cano2014}%
  \BibitemOpen
  \bibfield  {author} {\bibinfo {author} {\bibfnamefont {J.}~\bibnamefont
  {{Cano}}}, \bibinfo {author} {\bibfnamefont {M.}~\bibnamefont {{Cheng}}},
  \bibinfo {author} {\bibfnamefont {M.}~\bibnamefont {{Mulligan}}}, \bibinfo
  {author} {\bibfnamefont {C.}~\bibnamefont {{Nayak}}}, \bibinfo {author}
  {\bibfnamefont {E.}~\bibnamefont {{Plamadeala}}}, \ and\ \bibinfo {author}
  {\bibfnamefont {J.}~\bibnamefont {{Yard}}},\ }\href {\doibase
  10.1103/PhysRevB.89.115116} {\bibfield  {journal} {\bibinfo  {journal}
  {\prb}\ }\textbf {\bibinfo {volume} {89}},\ \bibinfo {eid} {115116} (\bibinfo
  {year} {2014})},\ \Eprint {http://arxiv.org/abs/1310.5708} {arXiv:1310.5708
  [cond-mat.str-el]} \BibitemShut {NoStop}%
\bibitem [{\citenamefont {Polyakov}(1975)}]{Polyakov1975}%
  \BibitemOpen
  \bibfield  {author} {\bibinfo {author} {\bibfnamefont {A.~M.}\ \bibnamefont
  {Polyakov}},\ }\href {\doibase 10.1016/0370-2693(75)90162-8} {\bibfield
  {journal} {\bibinfo  {journal} {Phys. Lett.}\ }\textbf {\bibinfo {volume}
  {B59}},\ \bibinfo {pages} {82} (\bibinfo {year} {1975})}\BibitemShut
  {NoStop}%
\bibitem [{\citenamefont {Polyakov}(1977)}]{Polyakov1976}%
  \BibitemOpen
  \bibfield  {author} {\bibinfo {author} {\bibfnamefont {A.~M.}\ \bibnamefont
  {Polyakov}},\ }\href {\doibase 10.1016/0550-3213(77)90086-4} {\bibfield
  {journal} {\bibinfo  {journal} {Nucl. Phys.}\ }\textbf {\bibinfo {volume}
  {B120}},\ \bibinfo {pages} {429} (\bibinfo {year} {1977})}\BibitemShut
  {NoStop}%
\bibitem [{\citenamefont {Coxeter}\ and\ \citenamefont
  {Moser}(1980)}]{Generators}%
  \BibitemOpen
  \bibfield  {author} {\bibinfo {author} {\bibfnamefont {H.~S.~M.}\
  \bibnamefont {Coxeter}}\ and\ \bibinfo {author} {\bibfnamefont {W.~O.~J.}\
  \bibnamefont {Moser}},\ }\href@noop {} {\emph {\bibinfo {title} {{Generators
  and Relations for Discrete Groups}}}}\ (\bibinfo  {publisher}
  {Springer-Verlag Berlin Heidelberg New York},\ \bibinfo {year}
  {1980})\BibitemShut {NoStop}%
\bibitem [{Note1()}]{Note1}%
  \BibitemOpen
  \bibinfo {note} {The Euclidean time coordinate $\tau $ should not be confused
  with the modular parameter.}\BibitemShut {Stop}%
\bibitem [{\citenamefont {{Wu}}(1991)}]{Wu1991}%
  \BibitemOpen
  \bibfield  {author} {\bibinfo {author} {\bibfnamefont {S.}~\bibnamefont
  {{Wu}}},\ }\href {\doibase 10.1007/BF02096795} {\bibfield  {journal}
  {\bibinfo  {journal} {Communications in Mathematical Physics}\ }\textbf
  {\bibinfo {volume} {136}},\ \bibinfo {pages} {157} (\bibinfo {year}
  {1991})}\BibitemShut {NoStop}%
\bibitem [{\citenamefont {{Balachandran}}\ and\ \citenamefont
  {{Teotonio-Sobrinho}}(1993)}]{Balachandran1993}%
  \BibitemOpen
  \bibfield  {author} {\bibinfo {author} {\bibfnamefont {A.~P.}\ \bibnamefont
  {{Balachandran}}}\ and\ \bibinfo {author} {\bibfnamefont {P.}~\bibnamefont
  {{Teotonio-Sobrinho}}},\ }\href {\doibase 10.1142/S0217751X9300028X}
  {\bibfield  {journal} {\bibinfo  {journal} {International Journal of Modern
  Physics A}\ }\textbf {\bibinfo {volume} {8}},\ \bibinfo {pages} {723}
  (\bibinfo {year} {1993})},\ \Eprint {http://arxiv.org/abs/hep-th/9205116}
  {hep-th/9205116} \BibitemShut {NoStop}%
\bibitem [{\citenamefont {{Amoretti}}\ \emph {et~al.}(2012)\citenamefont
  {{Amoretti}}, \citenamefont {{Blasi}}, \citenamefont {{Maggiore}},\ and\
  \citenamefont {{Magnoli}}}]{Amoretti2012}%
  \BibitemOpen
  \bibfield  {author} {\bibinfo {author} {\bibfnamefont {A.}~\bibnamefont
  {{Amoretti}}}, \bibinfo {author} {\bibfnamefont {A.}~\bibnamefont {{Blasi}}},
  \bibinfo {author} {\bibfnamefont {N.}~\bibnamefont {{Maggiore}}}, \ and\
  \bibinfo {author} {\bibfnamefont {N.}~\bibnamefont {{Magnoli}}},\ }\href
  {\doibase 10.1088/1367-2630/14/11/113014} {\bibfield  {journal} {\bibinfo
  {journal} {New Journal of Physics}\ }\textbf {\bibinfo {volume} {14}},\
  \bibinfo {eid} {113014} (\bibinfo {year} {2012})},\ \Eprint
  {http://arxiv.org/abs/1205.6156} {arXiv:1205.6156 [hep-th]} \BibitemShut
  {NoStop}%
\bibitem [{\citenamefont {{Kitaev}}\ and\ \citenamefont
  {{Preskill}}(2006)}]{KitaevPreskill2006}%
  \BibitemOpen
  \bibfield  {author} {\bibinfo {author} {\bibfnamefont {A.}~\bibnamefont
  {{Kitaev}}}\ and\ \bibinfo {author} {\bibfnamefont {J.}~\bibnamefont
  {{Preskill}}},\ }\href {\doibase 10.1103/PhysRevLett.96.110404} {\bibfield
  {journal} {\bibinfo  {journal} {Physical Review Letters}\ }\textbf {\bibinfo
  {volume} {96}},\ \bibinfo {eid} {110404} (\bibinfo {year} {2006})},\ \Eprint
  {http://arxiv.org/abs/hep-th/0510092} {hep-th/0510092} \BibitemShut {NoStop}%
\bibitem [{\citenamefont {{Fendley}}\ \emph {et~al.}(2007)\citenamefont
  {{Fendley}}, \citenamefont {{Fisher}},\ and\ \citenamefont
  {{Nayak}}}]{Fendley2007}%
  \BibitemOpen
  \bibfield  {author} {\bibinfo {author} {\bibfnamefont {P.}~\bibnamefont
  {{Fendley}}}, \bibinfo {author} {\bibfnamefont {M.~P.~A.}\ \bibnamefont
  {{Fisher}}}, \ and\ \bibinfo {author} {\bibfnamefont {C.}~\bibnamefont
  {{Nayak}}},\ }\href {\doibase 10.1007/s10955-006-9275-8} {\bibfield
  {journal} {\bibinfo  {journal} {Journal of Statistical Physics}\ }\textbf
  {\bibinfo {volume} {126}},\ \bibinfo {pages} {1111} (\bibinfo {year}
  {2007})},\ \Eprint {http://arxiv.org/abs/cond-mat/0609072} {cond-mat/0609072}
  \BibitemShut {NoStop}%
\bibitem [{\citenamefont {{Qi}}\ \emph {et~al.}(2012)\citenamefont {{Qi}},
  \citenamefont {{Katsura}},\ and\ \citenamefont
  {{Ludwig}}}]{QiKatsuraLudwig2012}%
  \BibitemOpen
  \bibfield  {author} {\bibinfo {author} {\bibfnamefont {X.-L.}\ \bibnamefont
  {{Qi}}}, \bibinfo {author} {\bibfnamefont {H.}~\bibnamefont {{Katsura}}}, \
  and\ \bibinfo {author} {\bibfnamefont {A.~W.~W.}\ \bibnamefont {{Ludwig}}},\
  }\href {\doibase 10.1103/PhysRevLett.108.196402} {\bibfield  {journal}
  {\bibinfo  {journal} {Physical Review Letters}\ }\textbf {\bibinfo {volume}
  {108}},\ \bibinfo {eid} {196402} (\bibinfo {year} {2012})},\ \Eprint
  {http://arxiv.org/abs/1103.5437} {arXiv:1103.5437 [cond-mat.mes-hall]}
  \BibitemShut {NoStop}%
\bibitem [{\citenamefont {Affleck}\ and\ \citenamefont
  {Ludwig}(1991)}]{Affleck:1991tk}%
  \BibitemOpen
  \bibfield  {author} {\bibinfo {author} {\bibfnamefont {I.}~\bibnamefont
  {Affleck}}\ and\ \bibinfo {author} {\bibfnamefont {A.~W.~W.}\ \bibnamefont
  {Ludwig}},\ }\href {\doibase 10.1103/PhysRevLett.67.161} {\bibfield
  {journal} {\bibinfo  {journal} {Phys. Rev. Lett.}\ }\textbf {\bibinfo
  {volume} {67}},\ \bibinfo {pages} {161} (\bibinfo {year} {1991})}\BibitemShut
  {NoStop}%
\bibitem [{\citenamefont {{Lopes}}\ \emph {et~al.}(2015)\citenamefont
  {{Lopes}}, \citenamefont {{Teo}},\ and\ \citenamefont {{Ryu}}}]{Lopes2015}%
  \BibitemOpen
  \bibfield  {author} {\bibinfo {author} {\bibfnamefont {P.~L.~e.~S.}\
  \bibnamefont {{Lopes}}}, \bibinfo {author} {\bibfnamefont {J.~C.~Y.}\
  \bibnamefont {{Teo}}}, \ and\ \bibinfo {author} {\bibfnamefont
  {S.}~\bibnamefont {{Ryu}}},\ }\href {\doibase 10.1103/PhysRevB.91.184111}
  {\bibfield  {journal} {\bibinfo  {journal} {\prb}\ }\textbf {\bibinfo
  {volume} {91}},\ \bibinfo {eid} {184111} (\bibinfo {year} {2015})},\ \Eprint
  {http://arxiv.org/abs/1501.04109} {arXiv:1501.04109 [cond-mat.other]}
  \BibitemShut {NoStop}%
\bibitem [{\citenamefont {{Ye}}\ and\ \citenamefont
  {{Gu}}(2015{\natexlab{b}})}]{YeGu2015}%
  \BibitemOpen
  \bibfield  {author} {\bibinfo {author} {\bibfnamefont {P.}~\bibnamefont
  {{Ye}}}\ and\ \bibinfo {author} {\bibfnamefont {Z.-C.}\ \bibnamefont
  {{Gu}}},\ }\href@noop {} {\bibfield  {journal} {\bibinfo  {journal} {ArXiv
  e-prints}\ } (\bibinfo {year} {2015}{\natexlab{b}})},\ \Eprint
  {http://arxiv.org/abs/1508.05689} {arXiv:1508.05689 [cond-mat.str-el]}
  \BibitemShut {NoStop}%
\bibitem [{\citenamefont {{Kapustin}}\ and\ \citenamefont
  {{Thorngren}}(2014)}]{Kapustin2014d}%
  \BibitemOpen
  \bibfield  {author} {\bibinfo {author} {\bibfnamefont {A.}~\bibnamefont
  {{Kapustin}}}\ and\ \bibinfo {author} {\bibfnamefont {R.}~\bibnamefont
  {{Thorngren}}},\ }\href@noop {} {\bibfield  {journal} {\bibinfo  {journal}
  {ArXiv e-prints}\ } (\bibinfo {year} {2014})},\ \Eprint
  {http://arxiv.org/abs/1404.3230} {1404.3230} \BibitemShut {NoStop}%
\bibitem [{\citenamefont {Wang}\ \emph {et~al.}(2015)\citenamefont {Wang},
  \citenamefont {Gu},\ and\ \citenamefont {Wen}}]{wang2015_prl}%
  \BibitemOpen
  \bibfield  {author} {\bibinfo {author} {\bibfnamefont {J.}~\bibnamefont
  {Wang}}, \bibinfo {author} {\bibfnamefont {Z.-C.}\ \bibnamefont {Gu}}, \ and\
  \bibinfo {author} {\bibfnamefont {X.-G.}\ \bibnamefont {Wen}},\ }\href@noop
  {} {\bibfield  {journal} {\bibinfo  {journal} {Phys. Rev. Lett.}\ }\textbf
  {\bibinfo {volume} {114}},\ \bibinfo {pages} {031601} (\bibinfo {year}
  {2015})}\BibitemShut {NoStop}%
\bibitem [{\citenamefont {Wan}\ \emph {et~al.}(2015{\natexlab{b}})\citenamefont
  {Wan}, \citenamefont {Wang},\ and\ \citenamefont {He}}]{wan2015twisted}%
  \BibitemOpen
  \bibfield  {author} {\bibinfo {author} {\bibfnamefont {Y.}~\bibnamefont
  {Wan}}, \bibinfo {author} {\bibfnamefont {J.~C.}\ \bibnamefont {Wang}}, \
  and\ \bibinfo {author} {\bibfnamefont {H.}~\bibnamefont {He}},\ }\href@noop
  {} {\bibfield  {journal} {\bibinfo  {journal} {Physical Review B}\ }\textbf
  {\bibinfo {volume} {92}},\ \bibinfo {pages} {045101} (\bibinfo {year}
  {2015}{\natexlab{b}})}\BibitemShut {NoStop}%
\bibitem [{\citenamefont {Dijkgraaf}\ and\ \citenamefont
  {Witten}(1990)}]{dijkgraaf1990topological}%
  \BibitemOpen
  \bibfield  {author} {\bibinfo {author} {\bibfnamefont {R.}~\bibnamefont
  {Dijkgraaf}}\ and\ \bibinfo {author} {\bibfnamefont {E.}~\bibnamefont
  {Witten}},\ }\href@noop {} {\bibfield  {journal} {\bibinfo  {journal}
  {Communications in Mathematical Physics}\ }\textbf {\bibinfo {volume}
  {129}},\ \bibinfo {pages} {393} (\bibinfo {year} {1990})}\BibitemShut
  {NoStop}%
\bibitem [{\citenamefont {Chen}\ \emph {et~al.}(2013)\citenamefont {Chen},
  \citenamefont {Gu}, \citenamefont {Liu},\ and\ \citenamefont
  {Wen}}]{Chen2013}%
  \BibitemOpen
  \bibfield  {author} {\bibinfo {author} {\bibfnamefont {X.}~\bibnamefont
  {Chen}}, \bibinfo {author} {\bibfnamefont {Z.-C.}\ \bibnamefont {Gu}},
  \bibinfo {author} {\bibfnamefont {Z.-X.}\ \bibnamefont {Liu}}, \ and\
  \bibinfo {author} {\bibfnamefont {X.-G.}\ \bibnamefont {Wen}},\ }\href@noop
  {} {\bibfield  {journal} {\bibinfo  {journal} {Phys. Rev. B}\ }\textbf
  {\bibinfo {volume} {87}},\ \bibinfo {pages} {155114} (\bibinfo {year}
  {2013})}\BibitemShut {NoStop}%
\end{thebibliography}%

\end{document}